\documentclass[11pt,a4paper]{article}
\usepackage{jheppub}
\makeatletter
\renewcommand{\@fpheader}{}
\makeatother           
\usepackage{graphicx}
\usepackage{amssymb}
\usepackage{amsmath}
\usepackage{color}
\hypersetup{hidelinks}
\usepackage{booktabs}
\usepackage{array}
\usepackage{amsthm}                        
\usepackage[labelformat=simple]{subcaption}

\usepackage{tabularx}
\usepackage[T1]{fontenc}
\usepackage{lmodern}
\usepackage{microtype}
\usepackage{needspace}
\allowdisplaybreaks[4]
\title{Higher-order semi-rational solutions of the
(2+1)-dimensional nonlocal multi-component
Yajima--Oikawa system}
\author[a]{Wenqiang Lu,}
\author[a,1]{Kelei Tian,\note{Corresponding author.}}
\author[a]{Ying Xu,}
\author[a]{Ge Yi}
\author[b]{and Xiaoming Zhu}
\affiliation[a]{School of Mathematics, Hefei University of Technology, Hefei, Anhui 230601, P. R. China}
\affiliation[b]{Department of Mathematics and Statistics, Hefei University, Hefei, Anhui 230601, P. R. China}
\emailAdd{kltian@hfut.edu.cn}
\abstract{Higher-order semi-rational solutions are constructed for the (2+1)-dimensional nonlocal multi-component Yajima-Oikawa system. This system is obtained from its local \(2M\)-component counterpart through a spatial-reflection reduction. Using Hirota's bilinear method and the Kadomtsev-Petviashvili hierarchy reduction, we derive a unified \(6\times6\) block Gram determinant. Its entries involve Schur polynomials and exponential functions. The complex-conjugation and spatial-reflection reductions are imposed through algebraic constraints on the spectral parameters and determinant blocks, with the former ensuring a real-valued denominator tau function. Separate spectral parameter sets are assigned to the rational, breather, and dark-soliton blocks. These solutions describe paired lumps or line rogue waves on breather, dark-soliton, or combined backgrounds. Pure rational, breather, and dark-soliton solutions follow by setting the appropriate block sizes to zero. Asymptotic analysis establishes the elastic interactions of the fundamental breather and dark-soliton pairs and determines their position and phase shifts.
}
\keywords{Nonlocal Yajima-Oikawa system;
Kadomtsev-Petviashvili hierarchy reduction;
Integrable Hierarchies; Integrable Field Theories.}
\begin{document}
\maketitle
\raggedbottom
\section{Introduction}
Resonant interactions between short and long waves arise in Bose-Einstein
condensates \cite{NFKM-2008}, plasma physics \cite{YO-1976,Z-1972},
nonlinear optics \cite{Kivshar-1992,CT-2008}, and water waves
\cite{B-1976,B-1977}. The long-wave-short-wave resonance (LSRI) occurs
when the group velocity of a short wave coincides with the phase
velocity of a long wave \cite{B-1976,B-1977}. This resonance mechanism
is described by a hierarchy of integrable evolution equations
\cite{YO-1976,M-1983}. In one dimension, it is governed by the
Yajima-Oikawa (YO) system \cite{YO-1976}. Its (2+1)-dimensional single-,
two-, and general \(M\)-component extensions were introduced in
\cite{OO-1989,OMO-2007,KVL-2014}. As a completely integrable model, the
YO system admits a Lax pair and a variety of exact solutions, including
lumps, line rogue waves \cite{YO3}, breathers, dark solitons and
semi-rational solutions. These solutions have been constructed by
Hirota's bilinear method, the Kadomtsev-Petviashvili (KP) hierarchy
reduction, and Darboux transformations
\cite{YO-1976,H-2004,Maccari-1996,OMO-2007,OY-2012,CCFM-2015,YO3,RPHK-2018,CGS-2014}.
The construction of higher-dimensional integrable systems has also
been pursued through conservation-law deformations \cite{LHJ-2023}.

Beyond the local LSRI systems discussed above, the emergence of
parity-time ($\mathcal{PT}$) symmetry has broadened the scope of
integrable nonlinear wave equations. It has led to a variety of
nonlocal integrable models. Non-Hermitian Hamiltonians with
$\mathcal{PT}$ symmetry may possess entirely real spectra
\cite{BB-1998,BBJ-2002}. This symmetry also enters soliton constructions
\cite{CF-2016}. Ablowitz and Musslimani proposed the nonlocal nonlinear
Schr\"odinger (NLS) equation \cite{AM-2013}, in which the evolution at
\(x\) depends on the field at \(-x\). Its inverse scattering transform and complete
integrability were subsequently established \cite{AM-2016,AM-2017}.
Fokas proposed multidimensional nonlocal integrable equations
\cite{Fokas-2016}. Subsequently, nonlocal extensions of several
(2+1)-dimensional integrable models were developed, including the
Davey--Stewartson equations \cite{RCH-2018} and the Maccari system
\cite{CCM-2021}. Nonlocal reductions have also been constructed for
the coupled bosonic massive Thirring model \cite{BS-2024}.
In addition, Yang introduced a nonlocal NLS equation through a
spatial-reflection reduction of the Manakov system \cite{Yang-2018}.
Following this framework, Lu, Tian and Zhu \cite{LTZ-2026} obtained
an \(M\)-component nonlocal YO system from the local \(2M\)-component
system proposed in \cite{KVL-2014}:
\begin{equation}\label{M-1}
\begin{aligned}
&\mathrm{i}(S_{\ell,t}+S_{\ell,y})-S_{\ell,xx}+LS_\ell=0,\quad \ell=1,\dots,2M,\\
&L_t=2\sum_{\ell=1}^{2M}\sigma_\ell(|S_\ell|^2)_x,\quad \sigma_\ell=\pm1,
\end{aligned}
\end{equation}
subject to the following Yang-type spatial-reflection reduction:
\begin{equation}\label{nonlocal-conditions}
\begin{aligned}
&\sigma_{M+\ell}=-\sigma_\ell,\\
&S_{M+\ell}(x,y,t)=S_\ell(-x,y,t),\\
&L(-x,y,t)=L(x,y,t),
\qquad \ell=1,\ldots,M.
\end{aligned}
\end{equation}
The evenness condition on \(L\) is required for compatibility of the
short-wave equations under spatial reflection. Moreover, the quantity
\(\sum_{\ell=1}^{M}\sigma_\ell
\bigl(|S_\ell(x,y,t)|^2-|S_\ell(-x,y,t)|^2\bigr)\)
is odd in \(x\). Its \(x\)-derivative is therefore even, so that the
evolution equation for \(L\) preserves even initial data.
The reduction therefore yields
\begin{equation}\label{M-2}
\begin{aligned}
&\mathrm{i}(S_{\ell,t}+S_{\ell,y})-S_{\ell,xx}+LS_\ell=0,\quad \ell=1,\dots,M,\\
&L_t=2\sum_{\ell=1}^{M}\bigl(\sigma_\ell|S_\ell(x,y,t)|^2-\sigma_\ell|S_\ell(-x,y,t)|^2\bigr)_x.
\end{aligned}
\end{equation}
This spatial-reflection reduction couples the wave field with its
spatially reflected counterpart. It introduces nonlocal interactions
absent from the corresponding local system.

Related semi-rational solutions and resonant interactions have been
studied in the local YO \cite{RPHK-2018,RJ1}, KP-I \cite{C1},
Davey--Stewartson I \cite{RHC-2022}, and Maccari \cite{CHC-2024}
systems. For the nonlocal system \eqref{M-2}, Lu, Tian and Zhu
\cite{LTZ-2026} constructed Schur-polynomial Gram determinants and
analyzed paired fission, bound-state and fusion processes between
lumps and dark solitons. Their construction also includes pure
rational and rogue-wave--dark-soliton solutions. In the partially
resonant states, the rational factors and dark-soliton phases share
the same spectral parameters.

To construct semi-rational solutions of \eqref{M-2}, we combine
exponential functions with Schur polynomials in a unified tau-function
framework and analyze the resulting solutions asymptotically.
The main results and features of this work are summarized
as follows:

\textbf{\textbullet} In the semi-rational construction of
\cite{LTZ-2026}, the rational and dark-soliton parts share
the same spectral parameters. Here, we construct a family
of higher-order semi-rational solutions with separate
spectral parameter sets for the rational, breather, and
dark-soliton sectors. The resulting family describes
rational waves on breather, dark-soliton, or combined
backgrounds. Pure solutions follow by setting the
appropriate block sizes to zero.

\textbf{\textbullet} Gram-determinant and Schur-polynomial
methods have been used to construct rational and
semi-rational solutions \cite{CCFM-2015,OY-2012,Z3}.
For the present nonlocal YO system, we obtain a unified
\(6\times6\) block Gram determinant and derive the algebraic
constraints required by the complex-conjugation and
spatial-reflection reductions. These constraints apply
to all blocks, including those coupling different types
of waves. They also ensure that the denominator tau
function is real-valued.

\textbf{\textbullet} Interactions among rational waves,
breathers, and solitons have been studied in several
integrable systems \cite{Z2,Z3}. For the present system,
we derive the asymptotic profiles and background limits
of the fundamental pure solutions. The spectral conditions
distinguish lump pairs from line rogue waves.
For the fundamental breather and dark-soliton pairs,
explicit asymptotic formulas determine the position and
phase shifts and establish their elastic interactions.
We also examine the dynamics of representative mixed states.

The remainder of this paper is organized as follows. Section~2 states
the main result and presents the Gram-determinant representation of
the higher-order semi-rational solutions. Section~3 classifies the
resulting solutions and analyzes their dynamical and asymptotic
behaviors. Section~4 derives the determinant formulas through the KP
hierarchy reduction and verifies the complex-conjugation and
spatial-reflection reductions. Section~5 summarizes the main results and
discusses possible extensions.

\section{Gram-determinant representation of the semi-rational solutions}
In this section, we present a Gram-determinant representation for a family of semi-rational solutions of \eqref{M-2}. The proof is given in Section~4.

Let \(N_1,N_2,N_3\) be nonnegative integers. We take \(h,\rho_\ell,k_\ell,\omega_\ell\) to be real and \(\sigma_\ell=\pm1\), where \(\ell=1,\ldots,2M\). The background parameters satisfy
\[
\sigma_{M+\ell}=-\sigma_\ell,\qquad
\rho_{M+\ell}=\rho_\ell,\qquad
k_{M+\ell}=-k_\ell,\qquad
\omega_{M+\ell}=\omega_\ell,
\]
for \(\ell=1,\ldots,M\). In the formulas below, we set \(x_1=x\), \(x_2=-\mathrm{i}y\), and \(r_\ell=\sigma_\ell\rho_\ell^2(t-y)\) for \(\ell=1,\ldots,2M\).

\noindent{\bf Theorem 1}.The nonlocal multi-component YO system
\eqref{M-2} admits a family of higher-order semi-rational solutions
with rational, breather, and dark-soliton block sizes \(N_1,N_2,N_3\):
\begin{equation}
\label{12d-M-semi-rational-solutions-1-1}
\begin{aligned}
&L^{[N_{1},N_{2},N_{3}]}=h-2(\ln{f^{[N_{1},N_{2},N_{3}]}})_{xx},\\
&S_{\ell}^{[N_{1},N_{2},N_{3}]}= \rho_{\ell}e^{\mathrm{i}[k_{\ell}x+\omega_{\ell}y+(h-\omega_{\ell} + k_{\ell}^{2}) t]}\frac{g_{\ell}^{[N_{1},N_{2},N_{3}]}}{f^{[N_{1},N_{2},N_{3}]}},
\end{aligned}
\end{equation}
where
\[
\mathbf{m}=(m_{1},m_{2},\ldots,m_{2M})\in\mathbb{Z}^{2M},
\qquad
\mathbf{0}=(0,0,\ldots,0),
\]
and \(\mathbf{e}_{\ell}\) denotes the \(\ell\)-th standard unit vector in
\(\mathbb{Z}^{2M}\). The tau functions are given by
\begin{equation}\label{12d-M-semi-rational-solutions-1-2}
f^{[N_{1},N_{2},N_{3}]}
=
\gamma_{\mathbf{0}},
\qquad
g_{\ell}^{[N_{1},N_{2},N_{3}]}
=
\gamma_{\mathbf{e}_{\ell}},
\qquad
\ell=1,\ldots,M.
\end{equation}
The tau function $\gamma_{\mathbf{m}}$ in \eqref{12d-M-semi-rational-solutions-1-2}
is given by the $6\times6$ block determinant
\begin{equation}\label{12d-M-semi-rational-solutions-1-3}
\renewcommand{\arraystretch}{1.5}
\gamma_{\mathbf{m}}=\mathrm{det} \left(
\begin{array}{cccccc}
\gamma_{\mathbf{m}}^{[1,1]} & \gamma_{\mathbf{m}}^{[1,2]} & \gamma_{\mathbf{m}}^{[1,3]} & \gamma_{\mathbf{m}}^{[1,4]} & \gamma_{\mathbf{m}}^{[1,5]} & \gamma_{\mathbf{m}}^{[1,6]} \\
\gamma_{-\mathbf{m}}^{\dag\,[1,2]} & \gamma_{\mathbf{m}}^{[2,2]} & \gamma_{\mathbf{m}}^{[2,3]} & \gamma_{\mathbf{m}}^{[2,4]} & \gamma_{\mathbf{m}}^{[2,5]} & \gamma_{\mathbf{m}}^{[2,6]} \\
\gamma_{-\mathbf{m}}^{\dag\,[1,3]} & \gamma_{-\mathbf{m}}^{\dag\,[2,3]} & \gamma_{\mathbf{m}}^{[3,3]} & \gamma_{\mathbf{m}}^{[3,4]} & \gamma_{\mathbf{m}}^{[3,5]} & \gamma_{\mathbf{m}}^{[3,6]} \\
\gamma_{-\mathbf{m}}^{\dag\,[1,4]} & \gamma_{-\mathbf{m}}^{\dag\,[2,4]} & \gamma_{-\mathbf{m}}^{\dag\,[3,4]} & \gamma_{\mathbf{m}}^{[4,4]} & \gamma_{\mathbf{m}}^{[4,5]} & \gamma_{\mathbf{m}}^{[4,6]} \\
\gamma_{-\mathbf{m}}^{\dag\,[1,5]} & \gamma_{-\mathbf{m}}^{\dag\,[2,5]} & \gamma_{-\mathbf{m}}^{\dag\,[3,5]} & \gamma_{-\mathbf{m}}^{\dag\,[4,5]} & \gamma_{\mathbf{m}}^{[5,5]} & \gamma_{\mathbf{m}}^{[5,6]} \\
\gamma_{-\mathbf{m}}^{\dag\,[1,6]} & \gamma_{-\mathbf{m}}^{\dag\,[2,6]} & \gamma_{-\mathbf{m}}^{\dag\,[3,6]} & \gamma_{-\mathbf{m}}^{\dag\,[4,6]} & \gamma_{-\mathbf{m}}^{\dag\,[5,6]} & \gamma_{\mathbf{m}}^{[6,6]}
\end{array}\right),
\end{equation}
with `$\dagger$' denoting complex conjugation followed by
transposition. The block sizes are specified by
\[
K_1=K_2=N_1,\qquad
K_3=K_4=N_2,\qquad
K_5=K_6=N_3.
\]
For the rational sectors, the Schur-polynomial indices are arbitrary
positive integers:
\[
n_i\in\mathbb Z_{>0},\qquad i=1,\ldots,N_1.
\]
Here \(N_1\) denotes the rational block size, while \(n_i\) specifies the order of spectral differentiation in its \(i\)-th row. The integers \(N_2\) and \(N_3\) count the breather and dark-soliton pairs, respectively. The full determinant has order \(2(N_1+N_2+N_3)\). Each block \(\gamma_{\mathbf{m}}^{[I,J]}\), \(I,J=1,\ldots,6\), is given by
\begin{equation}\label{12d-M-semi-rational-solutions-1-4-1}
\gamma_{\mathbf{m}}^{[I,J]}=\mathrm{mat}_{1\leq i\leq K_I,\;1\leq j\leq K_J} \left(m_{i,j}^{[I,J,\mathbf{m}]}\right).
\end{equation}
The entries are given by:
\begin{footnotesize}
\begin{align*}
m_{i,j}^{[1,1,\mathbf{m}]}=&\frac{1}{p_{i,1}+p_{j,1}^{\,*}}\prod_{\ell=1}^{2M}\left(-\frac{p_{i,1}-\mathrm{i}k_{\ell}}{p_{j,1}^{\,*}+\mathrm{i}k_{\ell}}\right)^{m_{\ell}}
\sum_{\upsilon=0}^{\mathrm{min}\left(n_{i},n_{j}\right)}\left[\frac{p_{i,1}p_{j,1}^{\,*}}{(p_{i,1}+p_{j,1}^{\,*})^{2}}\right]^{\upsilon}
S_{n_{i}-\upsilon}\left[\mathbf{x}_{i,\mathbf{m}}^{+}-\mathbf{d}_{ij}^{[1,1]}+\upsilon\mathbf{w}_{ij}^{[1,1]}\right]\\
& S_{n_{j}-\upsilon}\left[\mathbf{x}_{j,\mathbf{m}}^{-}-\bigl(\mathbf{d}_{ji}^{[1,1]}\bigr)^{*}+\upsilon\bigl(\mathbf{w}_{ji}^{[1,1]}\bigr)^{*}\right],\\
m_{i,j}^{[1,2,\mathbf{m}]}=&\frac{1}{p_{i,1}-p_{j,1}^{\,*}}\prod_{\ell=1}^{2M}\left(\frac{p_{i,1}-\mathrm{i}k_{\ell}}{p_{j,1}^{\,*}-\mathrm{i} k_{\ell}}\right)^{m_{\ell}}
\sum_{\upsilon=0}^{\mathrm{min}\left(n_{i},n_{j}\right)}\left[\frac{-p_{i,1}p_{j,1}^{\,*}}{(p_{i,1}-p_{j,1}^{\,*})^{2}}\right]^{\upsilon}
S_{n_{i}-\upsilon}\left[\mathbf{x}_{i,\mathbf{m}}^{+}-\mathbf{d}_{ij}^{[1,2]}+\upsilon\mathbf{w}_{ij}^{[1,2]}\right]\\
& S_{n_{j}-\upsilon}\left[\bar{\mathbf{x}}_{j,\mathbf{m}}^{-}-\bigl(\mathbf{d}_{ji}^{[1,2]}\bigr)^{*}+\upsilon\bigl(\mathbf{w}_{ji}^{[1,2]}\bigr)^{*}\right],\\
m_{i,j}^{[2,2,\mathbf{m}]}=&-\frac{1}{p_{i,1}+p_{j,1}^{\,*}}\prod_{\ell=1}^{2M}\left(-\frac{p_{i,1}+\mathrm{i} k_{\ell}}{p_{j,1}^{\,*}-\mathrm{i}k_{\ell}}\right)^{m_{\ell}}
\sum_{\upsilon=0}^{\mathrm{min}\left(n_{i},n_{j}\right)}\left[\frac{p_{i,1}p_{j,1}^{\,*}}{(p_{i,1}+p_{j,1}^{\,*})^{2}}\right]^{\upsilon}
S_{n_{i}-\upsilon}\left[\bar{\mathbf{x}}_{i,\mathbf{m}}^{+}-\mathbf{d}_{ij}^{[1,1]}+\upsilon\mathbf{w}_{ij}^{[1,1]}\right]\\
& S_{n_{j}-\upsilon}\left[\bar{\mathbf{x}}_{j,\mathbf{m}}^{-}-\bigl(\mathbf{d}_{ji}^{[1,1]}\bigr)^{*}+\upsilon\bigl(\mathbf{w}_{ji}^{[1,1]}\bigr)^{*}\right],\\
m_{i,j}^{[1,3,\mathbf{m}]}=&\sum_{\nu=1}^{2}\frac{1}{p_{i,1}+p_{j,2}^{*\,[\nu]}}\prod_{\ell=1}^{2M}\left(-\frac{p_{i,1}-\mathrm{i}k_{\ell}}{p_{j,2}^{*\,[\nu]}+\mathrm{i}k_{\ell}}\right)^{m_{\ell}}
e^{\xi_{j}^{*\,[2,2,\nu]}}
S_{n_{i}}\left[\mathbf{x}_{i,\mathbf{m}}^{+}-\mathbf{d}^{[1,3,\nu]}\right],\\
m_{i,j}^{[3,5,\mathbf{m}]}=&\sum_{\mu=1}^{2}
\frac{1}{p_{i,2}^{[\mu]}+p_{j,3}^{\,*}}\prod_{\ell=1}^{2M}\left(-\frac{p_{i,2}^{[\mu]}-\mathrm{i}k_{\ell}}{p_{j,3}^{\,*}+\mathrm{i}k_{\ell}}\right)^{m_{\ell}}e^{\xi_{i}^{[2,2,\mu]}+\xi_{j}^{*\,[3,3]}},\\
m_{i,j}^{[1,4,\mathbf{m}]}=&\sum_{\nu=1}^{2}\frac{1}{p_{i,1}-p_{j,2}^{*\,[\nu]}}\prod_{\ell=1}^{2M}\left(\frac{p_{i,1}-\mathrm{i}k_{\ell}}{p_{j,2}^{*\,[\nu]}-\mathrm{i}k_{\ell}}\right)^{m_{\ell}}
e^{\bar{\xi}_{j}^{*\,[2,2,\nu]}}
S_{n_{i}}\left[\mathbf{x}_{i,\mathbf{m}}^{+}-\mathbf{d}^{[1,4,\nu]}\right],\\
m_{i,j}^{[3,6,\mathbf{m}]}=&\sum_{\mu=1}^{2}
\frac{1}{p_{i,2}^{[\mu]}-p_{j,3}^{\,*}}\prod_{\ell=1}^{2M}\left(\frac{p_{i,2}^{[\mu]}-\mathrm{i}k_{\ell}}{p_{j,3}^{\,*}-\mathrm{i}k_{\ell}}\right)^{m_{\ell}}e^{\xi_{i}^{[2,2,\mu]}+\bar{\xi}_{j}^{*\,[3,3]}},\\
m_{i,j}^{[1,5,\mathbf{m}]}=&\frac{1}{p_{i,1}+p_{j,3}^{\,*}}\prod_{\ell=1}^{2M}\left(-\frac{p_{i,1}-\mathrm{i}k_{\ell}}{p_{j,3}^{\,*}+\mathrm{i}k_{\ell}}\right)^{m_{\ell}}e^{\xi_{j}^{*\,[3,3]}}
S_{n_{i}}\left[\mathbf{x}_{i,\mathbf{m}}^{+}-\mathbf{d}^{[1,5]}\right],\\
m_{i,j}^{[3,3,\mathbf{m}]}=&\sum_{\mu=1}^{2}\sum_{\nu=1}^{2}
\frac{1}{p_{i,2}^{[\mu]}+p_{j,2}^{*\,[\nu]}}\prod_{\ell=1}^{2M}\left(-\frac{p_{i,2}^{[\mu]}-\mathrm{i}k_{\ell}}{p_{j,2}^{*\,[\nu]}+\mathrm{i}k_{\ell}}\right)^{m_{\ell}}
e^{\xi_{i}^{[2,2,\mu]}+\xi_{j}^{*\,[2,2,\nu]}},\\
m_{i,j}^{[1,6,\mathbf{m}]}=&\frac{1}{p_{i,1}-p_{j,3}^{\,*}}\prod_{\ell=1}^{2M}\left(\frac{p_{i,1}-\mathrm{i}k_{\ell}}{p_{j,3}^{\,*}-\mathrm{i}k_{\ell}}\right)^{m_{\ell}}e^{\bar{\xi}_{j}^{*\,[3,3]}}
S_{n_{i}}\left[\mathbf{x}_{i,\mathbf{m}}^{+}-\mathbf{d}^{[1,6]}\right],\\
m_{i,j}^{[3,4,\mathbf{m}]}=&\sum_{\mu=1}^{2}\sum_{\nu=1}^{2}
\frac{1}{p_{i,2}^{[\mu]}-p_{j,2}^{*\,[\nu]}}\prod_{\ell=1}^{2M}\left(\frac{p_{i,2}^{[\mu]}-\mathrm{i}k_{\ell}}{p_{j,2}^{*\,[\nu]}-\mathrm{i}k_{\ell}}\right)^{m_{\ell}}
e^{\xi_{i}^{[2,2,\mu]}+\bar{\xi}_{j}^{*\,[2,2,\nu]}},\\
m_{i,j}^{[2,3,\mathbf{m}]}=&-\sum_{\nu=1}^{2}\frac{1}{p_{i,1}-p_{j,2}^{*\,[\nu]}}\prod_{\ell=1}^{2M}\left(\frac{p_{i,1}+\mathrm{i}k_{\ell}}{p_{j,2}^{*\,[\nu]}+\mathrm{i}k_{\ell}}\right)^{m_{\ell}}
e^{\xi_{j}^{*\,[2,2,\nu]}}
S_{n_{i}}\left[\bar{\mathbf{x}}_{i,\mathbf{m}}^{+}-\mathbf{d}^{[1,4,\nu]}\right],\\
m_{i,j}^{[4,5,\mathbf{m}]}=&-\sum_{\mu=1}^{2}
\frac{1}{p_{i,2}^{[\mu]}-p_{j,3}^{\,*}}\prod_{\ell=1}^{2M}\left(\frac{p_{i,2}^{[\mu]}+\mathrm{i}k_{\ell}}{p_{j,3}^{\,*}+\mathrm{i}k_{\ell}}\right)^{m_{\ell}}e^{\bar{\xi}_{i}^{[2,2,\mu]}+\xi_{j}^{*\,[3,3]}},\\
m_{i,j}^{[2,4,\mathbf{m}]}=&-\sum_{\nu=1}^{2}\frac{1}{p_{i,1}+p_{j,2}^{*\,[\nu]}}\prod_{\ell=1}^{2M}\left(-\frac{p_{i,1}+\mathrm{i}k_{\ell}}{p_{j,2}^{*\,[\nu]}-\mathrm{i}k_{\ell}}\right)^{m_{\ell}}
e^{\bar{\xi}_{j}^{*\,[2,2,\nu]}}
S_{n_{i}}\left[\bar{\mathbf{x}}_{i,\mathbf{m}}^{+}-\mathbf{d}^{[1,3,\nu]}\right],\\
m_{i,j}^{[4,6,\mathbf{m}]}=&-\sum_{\mu=1}^{2}
\frac{1}{p_{i,2}^{[\mu]}+p_{j,3}^{\,*}}\prod_{\ell=1}^{2M}\left(-\frac{p_{i,2}^{[\mu]}+\mathrm{i}k_{\ell}}{p_{j,3}^{\,*}-\mathrm{i}k_{\ell}}\right)^{m_{\ell}}e^{\bar{\xi}_{i}^{[2,2,\mu]}+\bar{\xi}_{j}^{*\,[3,3]}},\\
m_{i,j}^{[2,5,\mathbf{m}]}=&-\frac{1}{p_{i,1}-p_{j,3}^{\,*}}\prod_{\ell=1}^{2M}\left(\frac{p_{i,1}+\mathrm{i}k_{\ell}}{p_{j,3}^{\,*}+\mathrm{i}k_{\ell}}\right)^{m_{\ell}}e^{\xi_{j}^{*\,[3,3]}}
S_{n_{i}}\left[\bar{\mathbf{x}}_{i,\mathbf{m}}^{+}-\mathbf{d}^{[1,6]}\right],\\
m_{i,j}^{[5,5,\mathbf{m}]}=&\delta_{i,j}+\frac{1}{p_{i,3}+p_{j,3}^{\,*}}\prod_{\ell=1}^{2M}\left(-\frac{p_{i,3}-\mathrm{i}k_{\ell}}{p_{j,3}^{\,*}+\mathrm{i}k_{\ell}}\right)^{m_{\ell}}e^{\xi_{i}^{[3,3]}+\xi_{j}^{*\,[3,3]}},\\
m_{i,j}^{[2,6,\mathbf{m}]}=&-\frac{1}{p_{i,1}+p_{j,3}^{\,*}}\prod_{\ell=1}^{2M}\left(-\frac{p_{i,1}+\mathrm{i}k_{\ell}}{p_{j,3}^{\,*}-\mathrm{i}k_{\ell}}\right)^{m_{\ell}}e^{\bar{\xi}_{j}^{*\,[3,3]}}
S_{n_{i}}\left[\bar{\mathbf{x}}_{i,\mathbf{m}}^{+}-\mathbf{d}^{[1,5]}\right],\\
m_{i,j}^{[5,6,\mathbf{m}]}=&\frac{1}{p_{i,3}-p_{j,3}^{\,*}}\prod_{\ell=1}^{2M}\left(-\frac{p_{i,3}-\mathrm{i}k_{\ell}}{-p_{j,3}^{\,*}+\mathrm{i}k_{\ell}}\right)^{m_{\ell}}e^{\xi_{i}^{[3,3]}+\bar{\xi}_{j}^{*\,[3,3]}},\\
m_{i,j}^{[4,4,\mathbf{m}]}=&-\sum_{\mu=1}^{2}\sum_{\nu=1}^{2}
\frac{1}{p_{i,2}^{[\mu]}+p_{j,2}^{*\,[\nu]}}\prod_{\ell=1}^{2M}\left(-\frac{p_{i,2}^{[\mu]}+\mathrm{i}k_{\ell}}{p_{j,2}^{*\,[\nu]}-\mathrm{i}k_{\ell}}\right)^{m_{\ell}}
e^{\bar{\xi}_{i}^{[2,2,\mu]}+\bar{\xi}_{j}^{*\,[2,2,\nu]}},\\
m_{i,j}^{[6,6,\mathbf{m}]}=&-\delta_{i,j}-\frac{1}{p_{i,3}+p_{j,3}^{\,*}}\prod_{\ell=1}^{2M}\left(-\frac{p_{i,3}+\mathrm{i}k_{\ell}}{p_{j,3}^{\,*}-\mathrm{i}k_{\ell}}\right)^{m_{\ell}}e^{\bar{\xi}_{i}^{[3,3]}+\bar{\xi}_{j}^{*\,[3,3]}}.
\end{align*}
\end{footnotesize}
Here $\delta_{i,j}$ denotes the Kronecker delta. The auxiliary quantities 
appearing in the matrix elements are collected below for reference:
\begin{equation*}
\begin{aligned}
&\xi_{i}^{[2,2,\mu]}=\sum_{\ell=1}^{2M}\frac{1}{{p_{i,2}^{[\mu]}}-\mathrm{i}k_{\ell}}r_{\ell}+{p_{i,2}^{[\mu]}}x_{1}+({p_{i,2}^{[\mu]}})^2x_{2}+\xi_{i,0}^{[2,2,\mu]},\\
&\xi_{i}^{[3,3]}=\sum_{\ell=1}^{2M}\frac{1}{{p_{i,3}}-\mathrm{i}k_{\ell}}r_{\ell}+{p_{i,3}}x_{1}+p_{i,3}^2x_{2}+\xi_{i,0}^{[3,3]},\\
&\bar{\xi}_{i}^{[2,2,\mu]}=-\sum_{\ell=1}^{2M}\frac{1}{{p_{i,2}^{[\mu]}}+\mathrm{i}k_{\ell}}r_{\ell}-p_{i,2}^{[\mu]}x_{1}+({p_{i,2}^{[\mu]}})^2x_{2}+\xi_{i,0}^{[2,2,\mu]},\\
&\bar{\xi}_{i}^{[3,3]}=-\sum_{\ell=1}^{2M}\frac{1}{{p_{i,3}}+\mathrm{i}k_{\ell}}r_{\ell}-p_{i,3}x_{1}+p_{i,3}^2x_{2}+\xi_{i,0}^{[3,3]}.
\end{aligned}
\end{equation*}
Write \(\mathbf{x}_{i,\mathbf m}^{\pm}=(x_{r,i,\mathbf m}^{\pm})_{r\ge1}\)
and \(\bar{\mathbf{x}}_{i,\mathbf m}^{\pm}=(\bar x_{r,i,\mathbf m}^{\pm})_{r\ge1}\).
Their components are
\begin{equation*}
\begin{aligned}
x_{r,i,\mathbf m}^{+}
&=\sum_{\ell=1}^{2M}\theta_{r,\ell}(p_{i,1})r_\ell
+\frac{p_{i,1}}{r!}x_1+\frac{2^rp_{i,1}^2}{r!}x_2
+\sum_{\ell=1}^{2M}m_\ell h_{r,\ell}(p_{i,1})+a_{i,r},\\
x_{r,j,\mathbf m}^{-}
&=\sum_{\ell=1}^{2M}\theta_{r,\ell}(p_{j,1})^*r_\ell
+\frac{p_{j,1}^*}{r!}x_1-\frac{2^r(p_{j,1}^*)^2}{r!}x_2
-\sum_{\ell=1}^{2M}m_\ell h_{r,\ell}(p_{j,1})^*+a_{j,r}^*,\\
\bar x_{r,i,\mathbf m}^{+}
&=\sum_{\ell=1}^{2M}\bar\theta_{r,\ell}(p_{i,1})r_\ell
-\frac{p_{i,1}}{r!}x_1+\frac{2^rp_{i,1}^2}{r!}x_2
+\sum_{\ell=1}^{2M}m_\ell\bar h_{r,\ell}(p_{i,1})+a_{i,r},\\
\bar x_{r,j,\mathbf m}^{-}
&=\sum_{\ell=1}^{2M}\bar\theta_{r,\ell}(p_{j,1})^*r_\ell
-\frac{p_{j,1}^*}{r!}x_1-\frac{2^r(p_{j,1}^*)^2}{r!}x_2
-\sum_{\ell=1}^{2M}m_\ell\bar h_{r,\ell}(p_{j,1})^*+a_{j,r}^*.
\end{aligned}
\end{equation*}
The vectors \(\mathbf d_{ij}^{[I,J]}=(d_{r,ij}^{[I,J]})_{r\ge1}\)
and \(\mathbf w_{ij}^{[I,J]}=(w_{r,ij}^{[I,J]})_{r\ge1}\)
retain both spectral indices in the rational blocks. In particular,
\begin{equation*}
d_{1,ij}^{[1,1]}=\frac{p_{i,1}}{p_{i,1}+p_{j,1}^*},\qquad
\bigl(d_{1,ji}^{[1,1]}\bigr)^{*}=\frac{p_{j,1}^*}{p_{i,1}+p_{j,1}^*},\qquad
\bigl(d_{1,ji}^{[1,2]}\bigr)^{*}=\frac{-p_{j,1}^*}{p_{i,1}-p_{j,1}^*}.
\end{equation*}
Thus, the column Schur factors in the rational blocks use the complex conjugates of the \(d\)- and \(w\)-coefficients with \(i\) and \(j\) interchanged. In the mixed blocks, boldface symbols denote coefficient vectors, with the indices \(i,j\) understood from the corresponding matrix entry. The coefficients are defined by the following generating series:
\begin{align*}
&\frac{1}{pe^{\kappa}-\mathrm{i}k_{\ell}}-\frac{1}{p-\mathrm{i}k_{\ell}}=\sum_{r=1}^{\infty}\theta_{r,\ell}(p)\kappa^{r},
\quad
\frac{1}{p+\mathrm{i}k_{\ell}} - \frac{1}{pe^{\kappa}+\mathrm{i}k_{\ell}}=\sum_{r=1}^{\infty}\bar{\theta}_{r,\ell}(p)\kappa^{r},\\
&\mathrm{ln}\left[\frac{pe^{\kappa}-\mathrm{i}k_{\ell}}{p-\mathrm{i}k_{\ell}}\right]=\sum_{r=1}^{\infty}h_{r,\ell}(p)\kappa^{r},\\
&\mathrm{ln}\left[\frac{pe^{\kappa}+\mathrm{i}k_{\ell}}{p+\mathrm{i}k_{\ell}}\right]=\sum_{r=1}^{\infty}\bar{h}_{r,\ell}(p)\kappa^{r},
\quad
\mathrm{ln}\left[\frac{p_{i,1}e^{\kappa}+p_{j,1}^{\,*}}{p_{i,1}+p_{j,1}^{\,*}}\right]=\sum_{r=1}^{\infty}d_{r,ij}^{[1,1]}\kappa^{r},\\
&\mathrm{ln}\left[\left(\frac{p_{i,1}+p_{j,1}^{\,*}}{\kappa}\right)\left(\frac{e^{\kappa}-1}{p_{i,1}e^{\kappa}+p_{j,1}^{\,*}}\right)\right]=\sum_{r=1}^{\infty}w_{r,ij}^{[1,1]}\kappa^{r},\\
&\mathrm{ln}\left[\frac{p_{i,1}e^{\kappa}-p_{j,1}^{\,*}}{p_{i,1}-p_{j,1}^{\,*}}\right]=\sum_{r=1}^{\infty}d_{r,ij}^{[1,2]}\kappa^{r},
\quad
\mathrm{ln}\left[\frac{p_{i,1}e^{\kappa}+p_{j,2}^{*\,[\nu]}}{p_{i,1}+p_{j,2}^{*\,[\nu]}}\right]=\sum_{r=1}^{\infty}d_{r}^{[1,3,\nu]}\kappa^{r},\\
&\mathrm{ln}\left[\frac{p_{i,1}e^{\kappa}-p_{j,2}^{*\,[\nu]}}{p_{i,1}-p_{j,2}^{*\,[\nu]}}\right]=\sum_{r=1}^{\infty}d_{r}^{[1,4,\nu]}\kappa^{r},\\
&\mathrm{ln}\left[\frac{p_{i,1}e^{\kappa}+p_{j,3}^{\,*}}{p_{i,1}+p_{j,3}^{\,*}}\right]=\sum_{r=1}^{\infty}d_{r}^{[1,5]}\kappa^{r},
\quad 
\mathrm{ln}\left[\frac{p_{i,1}e^{\kappa}-p_{j,3}^{\,*}}{p_{i,1}-p_{j,3}^{\,*}}\right]=\sum_{r=1}^{\infty}d_{r}^{[1,6]}\kappa^{r},\\
&\mathrm{ln}\left[
\left(\frac{p_{i,1}-p_{j,1}^{\,*}}{\kappa}\right)
\left(
\frac{e^{\kappa}-1}
{p_{i,1}e^{\kappa}-p_{j,1}^{\,*}}
\right)
\right]
=
\sum_{r=1}^{\infty}w_{r,ij}^{[1,2]}\kappa^{r}.
\end{align*}
The spectral parameters $p_{i,1}$, $p_{i,2}^{[\mu]}$, and
$p_{i,3}$ lie off the imaginary axis. All Cauchy-type denominators
in the matrix elements and the generating relations above
are assumed to be nonzero.
The parameters $a_{i,r}$ ($i=1,\ldots,N_1$, $r=1,2,\ldots$)
are arbitrary complex constants.

\medskip

\noindent{\bf Remark}.
The complex-conjugation reduction in Section~4 gives
\(f^*=f\). With the Cauchy denominators nonzero, the solution
\eqref{12d-M-semi-rational-solutions-1-1} is regular wherever
\(f\ne0\).

\section{Dynamic behaviors of higher-order semi-rational solutions to the (2+1)-dimensional nonlocal  YO system}
In this section, we consider the one-component case \(M=1\). Thus \(\mathbf m=(m_1,m_2)\), and the spatial-reflection reduction gives
\[
\sigma_2=-\sigma_1,\qquad
\rho_2=\rho_1,\qquad
k_2=-k_1,\qquad
\omega_2=\omega_1,\qquad
r_2=-r_1.
\]
We first analyze lumps, line rogue waves, breathers, and dark solitons, and then discuss the mixed states listed in Table~1. In all figures, the rational rows have \(n_i=1\) for \(i=1,\ldots,N_1\). We write \(a_i=a_{i,1}\). The background parameters are chosen as
\[
\sigma_1=-\sigma_2=1,\qquad
\rho_1=\rho_2=1,\qquad
k_1=-k_2=1,\qquad
h=\omega_1=\omega_2=1.
\]
\begin{table}[htbp]
\centering
\caption{Pure and mixed solution families for different choices of the block sizes \(N_1,N_2,N_3\).}
\vspace{0.5em}
\begin{tabularx}{\textwidth}{@{}ll>{\raggedright\arraybackslash\leavevmode}X@{}}
\toprule
\textbf{Case} & \textbf{Conditions} & \textbf{Solution types} \\
\midrule
1a & $N_1>0,\ N_2=N_3=0$ & lumps \\
1b & $N_1>0,\ N_2=N_3=0$ & line rogue waves \\
2 & $N_2>0,\ N_1=N_3=0$ & breather solutions \\
3 & $N_3>0,\ N_1=N_2=0$ & dark-soliton solutions \\
4a & $N_1,N_2>0,\ N_3=0$ & lumps + breathers \\
4b & $N_1,N_2>0,\ N_3=0$ & line rogue waves + breathers \\
5a & $N_1,N_3>0,\ N_2=0$ & lumps + dark solitons \\
5b & $N_1,N_3>0,\ N_2=0$ & line rogue waves + dark solitons \\
6a & $N_1,N_2,N_3>0$ & lumps + breathers + dark solitons \\
6b & $N_1,N_2,N_3>0$ & line rogue waves + breathers + dark solitons \\
\bottomrule
\end{tabularx}
\end{table}

\subsection{Dynamic behaviors of lumps or line rogue waves}
The rational solutions in Theorem~1 correspond to
\(N_2=N_3=0\), with rational block size \(N_1\).
This rational family extends the zero-constant rational
construction in \cite{LTZ-2026}. The corresponding tau functions take the following \(2\times2\) block form:
\begin{equation}\label{1generalized-tau-Ls}
\renewcommand{\arraystretch}{1.5}
\gamma_{m_{1},m_{2}}=\mathrm{det} \left(
\begin{array}{cc}
\gamma_{m_{1},m_{2}}^{[1,1]} & \gamma_{m_{1},m_{2}}^{[1,2]} \\
\gamma_{-m_{1},-m_{2}}^{\,\dag\,[1,2]} & \gamma_{m_{1},m_{2}}^{[2,2]}
\end{array}
\right).
\end{equation}
For \(N_1=1\) and \(n_1=1\), the fundamental lump and
line-rogue-wave solutions have the common form
\begin{equation}\label{1L-soliton-1}
\begin{aligned}
S_{1}^{[1,0,0]}= \rho_{1}e^{\mathrm{i}[k_{1}x+\omega_{1}y+(h-\omega_{1} + k_{1}^{2}) t]}\frac{g_{1}^{[1,0,0]}}{f^{[1,0,0]}},
\end{aligned}
\end{equation}
where $f^{[1,0,0]}$ and $g_{1}^{[1,0,0]}$ are defined as:
\begin{equation}
\begin{aligned}
&f^{[1,0,0]}=\left|
\begin{array}{cc}
m_{1,1}^{[1,1,0,0]} & m_{1,1}^{[1,2,0,0]} \\
m_{1,1}^{\,*\,[1,2,0,0]} & m_{1,1}^{[2,2,0,0]}
\end{array}
\right|, \quad
g_{1}^{[1,0,0]}=\left|
\begin{array}{cc}
m_{1,1}^{[1,1,1,0]} & m_{1,1}^{[1,2,1,0]} \\
m_{1,1}^{\,*\,[1,2,-1,0]} & m_{1,1}^{[2,2,1,0]}
\end{array}
\right|,
\end{aligned}
\end{equation}
The matrix entries are
\begin{equation}
\label{LLL}
\begin{aligned}
m_{1,1}^{[1,1,m_{1},m_{2}]}=&\frac{1}{p_{1,1}+p_{1,1}^{\,*}}\left(-\frac{p_{1,1}-\mathrm{i}k_{1}}{p_{1,1}^{\,*}+\mathrm{i}k_{1}}\right)^{m_{1}}\left(-\frac{p_{1,1}+\mathrm{i}k_{1}}{p_{1,1}^{\,*}-\mathrm{i}k_{1}}\right)^{m_{2}}\\
& \left[\frac{p_{1,1}p_{1,1}^{\,*}}{(p_{1,1}+p_{1,1}^{\,*})^{2}}+\left(x_{1,m_{1},m_{2}}^{+}-d_{1}^{\,[1,1]}\right)\left(x_{1,m_{1},m_{2}}^{-}-d_{1}^{*\,[1,1]}\right)\right],\\
m_{1,1}^{[1,2,m_{1},m_{2}]}=&\frac{1}{p_{1,1}-p_{1,1}^{\,*}}\left(\frac{p_{1,1}-\mathrm{i}k_{1}}{p_{1,1}^{\,*}-\mathrm{i}k_{1}}\right)^{m_{1}}\left(\frac{p_{1,1}+\mathrm{i}k_{1}}{p_{1,1}^{\,*}+\mathrm{i}k_{1}}\right)^{m_{2}}\\
& \left[-\frac{p_{1,1}p_{1,1}^{\,*}}{(p_{1,1}-p_{1,1}^{\,*})^{2}}+\left(x_{1,m_{1},m_{2}}^{+}-d_{1}^{\,[1,2]}\right)\left(\bar{x}_{1,m_{1},m_{2}}^{-}-d_{1}^{*\,[1,2]}\right)\right],\\
m_{1,1}^{[2,2,m_{1},m_{2}]}=&-\frac{1}{p_{1,1}+p_{1,1}^{\,*}}\left(-\frac{p_{1,1}+\mathrm{i}k_{1}}{p_{1,1}^{\,*}-\mathrm{i}k_{1}}\right)^{m_{1}}\left(-\frac{p_{1,1}-\mathrm{i}k_{1}}{p_{1,1}^{\,*}+\mathrm{i}k_{1}}\right)^{m_{2}}\\
& \left[\frac{p_{1,1}p_{1,1}^{\,*}}{(p_{1,1}+p_{1,1}^{\,*})^{2}}+\left(\bar{x}_{1,m_{1},m_{2}}^{+}-d_{1}^{\,[1,1]}\right)\left(\bar{x}_{1,m_{1},m_{2}}^{-}-d_{1}^{*\,[1,1]}\right)\right],
\end{aligned}
\end{equation}
where
\begin{align*}
&x_{1,m_{1},m_{2}}^{\,+\,}=\theta_{1,1}r_{1}+\theta_{1,2}r_{2}+p_{1,1}x_{1}+2p_{1,1}^2x_{2}+m_{1}h_{1,1}+m_{2}h_{1,2}+a_{1},\\
&x_{1,m_{1},m_{2}}^{\,-\,}=\theta_{1,1}^{*}r_{1}+\theta_{1,2}^{*}r_{2}+p_{1,1}^{*}x_{1}-2p_{1,1}^{*\,2}x_{2}-m_{1}h_{1,1}^{*}-m_{2}h_{1,2}^{*}+a_{1}^{*},\\
&\bar{x}_{1,m_{1},m_{2}}^{\,+\,}=-\theta_{1,2}r_{1}-\theta_{1,1}r_{2}-p_{1,1}x_{1}+2p_{1,1}^2x_{2}+m_{1}h_{1,2}+m_{2}h_{1,1}+a_{1},\\
&\bar{x}_{1,m_{1},m_{2}}^{\,-\,}=-\theta_{1,2}^{*}r_{1}-\theta_{1,1}^{*}r_{2}-p_{1,1}^{\,*}x_{1}-2p_{1,1}^{*\,2}x_{2}-m_{1}h_{1,2}^{*}-m_{2}h_{1,1}^{*}+a_{1}^{\,*\,},\\
&\theta_{1,1}=\frac{-p_{1,1}}{(p_{1,1}-\mathrm{i}k_{1})^2},\quad
\theta_{1,2}=\frac{-p_{1,1}}{(p_{1,1}-\mathrm{i}k_{2})^2},
\quad h_{1,1}=\frac{p_{1,1}}{p_{1,1}-\mathrm{i}k_{1}},\\ 
&h_{1,2}=\frac{p_{1,1}}{p_{1,1}-\mathrm{i}k_{2}},\quad
d_{1}^{[1,1]}=\frac{p_{1,1}}{p_{1,1}+p_{1,1}^{\,*}},\quad
d_{1}^{[1,2]}=\frac{p_{1,1}}{p_{1,1}-p_{1,1}^{\,*}}.
\end{align*}
For the asymptotic analysis, we set
\[
\begin{aligned}
&A_{11}=\theta_{1,1}r_{1}+\theta_{1,2}r_{2}+p_{1,1}x_{1}+2p_{1,1}^{2}x_{2},\\
&A_{12}=-\theta_{1,2}r_{1}-\theta_{1,1}r_{2}-p_{1,1}x_{1}+2p_{1,1}^{2}x_{2}.
\end{aligned}
\]
The fundamental lump and line-rogue-wave regimes are distinguished by the spatial
dependence of \(A_{11}\). Writing \(p=p_{1,1}\), we have
\begin{equation*}
\frac{A_{11}}p=x+
\left[-2\mathrm{i}p+\frac{4\mathrm{i}\sigma_1\rho_1^2k_1p}{(p^2+k_1^2)^2}\right]y
-\frac{4\mathrm{i}\sigma_1\rho_1^2k_1p}{(p^2+k_1^2)^2}t.
\end{equation*}
\noindent\textbf{Lumps.} If the coefficient of \(y\) has nonzero imaginary
part, each separated rational profile is localized in both spatial
directions and gives a lump.

\noindent\textbf{Line rogue waves.} If
\begin{equation}\label{line-rogue-spectral-condition}
\operatorname{Re}\!\left[
p_{1,1}-\frac{2\sigma_1\rho_1^2k_1p_{1,1}}
{(p_{1,1}^2+k_1^2)^2}
\right]=0,
\end{equation}
the coefficient of \(y\) in \(A_{11}/p_{1,1}\) is real, whereas the coefficient of \(t\) has imaginary part \(-2\operatorname{Re}p_{1,1}\ne0\). Each separated profile is therefore localized transverse to a line and in time.

Since \(A_{11}-A_{12}=2p_{1,1}x\), we consider the following two asymptotic regimes as \(|x|\to\infty\):
\begin{equation*}
\begin{aligned}
&A_{11}=O(1),\qquad A_{12}=-2p_{1,1}x+O(1),\\
&A_{12}=O(1),\qquad A_{11}= 2p_{1,1}x+O(1).
\end{aligned}
\end{equation*}
The two regimes yield the separated lump or line-rogue-wave profiles in (a1) and (a2). When both \(|A_{11}|\) and \(|A_{12}|\) tend to infinity, the solution approaches the background given in (a3). Here \(S_1,S_2\), and later \(S_1^\pm,S_2^\pm\), denote asymptotic profiles of the same short-wave component.

\noindent{\bf (a1)} For $A_{11}=\mathcal{O}(1)$ and $|A_{12}|\rightarrow+\infty$, one obtains
\begin{equation*}
\begin{aligned}
S_{1}&={}\rho_{1}e^{\mathrm{i}[k_{1}x+\omega_{1}y+(h-\omega_{1} + k_{1}^{2}) t]} \frac{p_{1,1}^{2}+k_{1}^{2}}{p_{1,1}^{2\,*}+k_{1}^{2}}\\
&\quad \frac{\begin{aligned}&(p_{1,1} - p_{1,1}^*)^2 \left[ \frac{p_{1,1} p_{1,1}^*}{(p_{1,1}+p_{1,1}^*)^2} + (x_{1,1,0}^{+} - d_1^{[1,1]})(x_{1,1,0}^{-} - d_1^{*[1,1]}) \right]
\\[2pt]&{}- (p_{1,1} + p_{1,1}^*)^2 (x_{1,1,0}^{+} - d_1^{[1,2]})(x_{1,1,0}^{-} - d_1^{*[1,2]})\end{aligned}}{\begin{aligned}&(p_{1,1} - p_{1,1}^*)^2 \left[ \frac{p_{1,1} p_{1,1}^*}{(p_{1,1}+p_{1,1}^*)^2} + (x_{1,0,0}^{+} - d_1^{[1,1]})(x_{1,0,0}^{-} - d_1^{*[1,1]}) \right]
\\[2pt]&{}- (p_{1,1} + p_{1,1}^*)^2 (x_{1,0,0}^{+} - d_1^{[1,2]})(x_{1,0,0}^{-} - d_1^{*[1,2]})\end{aligned}}.
\end{aligned}
\end{equation*}
 
\begin{figure}[tbp]\centering
      \begin{subfigure}{0.24\textwidth}
          \centering
         \includegraphics[width=\textwidth]{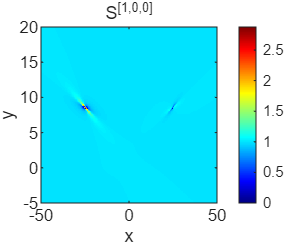}
         \caption*{t=-3}
        \end{subfigure}
      \begin{subfigure}{0.24\textwidth}
          \centering
          \includegraphics[width=\textwidth]{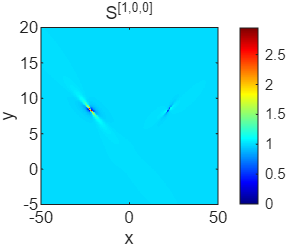}
       \caption*{t=-1}
        \end{subfigure}
      \begin{subfigure}{0.24\textwidth}
          \centering
         \includegraphics[width=\textwidth]{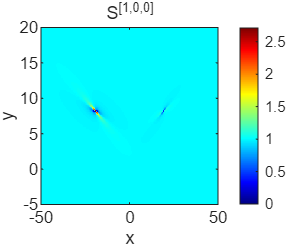}
      \caption*{t=1}
        \end{subfigure}
      \begin{subfigure}{0.24\textwidth}
          \centering
         \includegraphics[width=\textwidth]{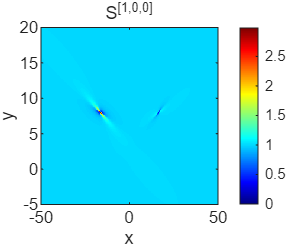}
      \caption*{t=3}
        \end{subfigure}
      \\
       \begin{subfigure}{0.24\textwidth}
          \centering
         \includegraphics[width=\textwidth]{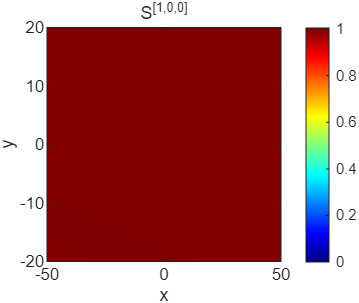}
         \caption*{t=-30}
        \end{subfigure}
      \begin{subfigure}{0.24\textwidth}
          \centering
          \includegraphics[width=\textwidth]{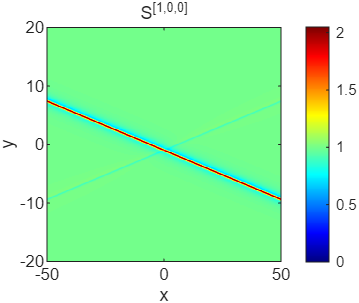}
       \caption*{t=-2}
        \end{subfigure}
      \begin{subfigure}{0.24\textwidth}
          \centering
         \includegraphics[width=\textwidth]{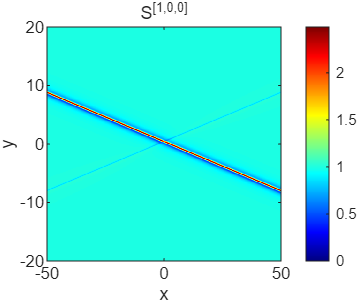}
      \caption*{t=0}
        \end{subfigure}
      \begin{subfigure}{0.24\textwidth}
          \centering
         \includegraphics[width=\textwidth]{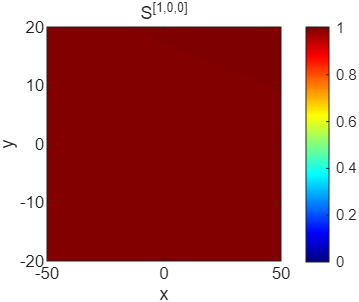}
      \caption*{t=30}
        \end{subfigure}
       \caption{Fundamental rational solutions \(S_1^{[1,0,0]}\).
Upper row: lumps for the parameters in \eqref{NL1}.
Lower row: line rogue waves for the parameters in \eqref{NL2}.}
\end{figure}

\noindent{\bf (a2)}
For $A_{12}=\mathcal{O}(1)$ and $|A_{11}|\rightarrow+\infty$, one obtains
\begin{equation*}
\begin{aligned}
S_{2}&={}\rho_{1}e^{\mathrm{i}[k_{1}x+\omega_{1}y+(h-\omega_{1} + k_{1}^{2}) t]} \frac{p_{1,1}^{2}+k_{1}^{2}}{p_{1,1}^{2\,*}+k_{1}^{2}}\\
&\quad \frac{\begin{aligned}&(p_{1,1} - p_{1,1}^*)^2 \left[ \frac{p_{1,1} p_{1,1}^*}{(p_{1,1}+p_{1,1}^*)^2} + (\bar{x}_{1,1,0}^{+} - d_1^{[1,1]})(\bar{x}_{1,1,0}^{-} - d_1^{*[1,1]}) \right]
\\[2pt]&{}- (p_{1,1} + p_{1,1}^*)^2 (\bar{x}_{1,1,0}^{+} - d_1^{[1,2]})(\bar{x}_{1,1,0}^{-} - d_1^{*[1,2]})\end{aligned}}{\begin{aligned}&(p_{1,1} - p_{1,1}^*)^2 \left[ \frac{p_{1,1} p_{1,1}^*}{(p_{1,1}+p_{1,1}^*)^2} + (\bar{x}_{1,0,0}^{+} - d_1^{[1,1]})(\bar{x}_{1,0,0}^{-} - d_1^{*[1,1]}) \right]
\\[2pt]&{}- (p_{1,1} + p_{1,1}^*)^2 (\bar{x}_{1,0,0}^{+} - d_1^{[1,2]})(\bar{x}_{1,0,0}^{-} - d_1^{*[1,2]})\end{aligned}}.
\end{aligned}
\end{equation*}
\noindent\textbf{(a3)} When both \(|A_{11}|\) and \(|A_{12}|\) tend to infinity, the leading terms of the tau functions \(f^{[1,0,0]}\) and \(g_1^{[1,0,0]}\) yield
\begin{equation}\label{lump-background-limit}
\frac{S_1^{[1,0,0]}}
{\rho_1e^{\mathrm{i}[k_1x+\omega_1y+(h-\omega_1+k_1^2)t]}}
\longrightarrow
\frac{p_{1,1}^2+k_1^2}{p_{1,1}^{*\,2}+k_1^2},
\qquad
|S_1^{[1,0,0]}|\longrightarrow|\rho_1|.
\end{equation}
Thus, the solution approaches the plane-wave background up to a constant phase factor.

For the fundamental lump pair, the parameters are chosen as
\begin{equation}
\label{NL1}
\begin{aligned}
&p_{1,1}=1+\mathrm{i},\qquad a_{1}=-20+10\mathrm{i}.
\end{aligned}
\end{equation}
The upper row of Fig.~1 shows the two lump profiles of
\eqref{1L-soliton-1} propagating on the plane-wave background.
For the lump configuration with \(N_1=2\), the parameters are chosen as
\begin{equation}\label{NLL1}
\begin{aligned}
&p_{1,1}=1+\mathrm{i},
\quad p_{2,1}=1+ \frac{3}{4}\mathrm{i},
\quad a_{1}=-20 + 10\mathrm{i},\quad a_{2}=20 - 10\mathrm{i}.
\end{aligned}
\end{equation}
The resulting solution \(S_1^{[2,0,0]}\), shown in the upper row
of Fig.~2, exhibits several localized peaks and depressions.

For the fundamental line rogue wave, the parameters are chosen as
\begin{equation}\label{NL2}
\begin{aligned} 
&p_{1,1}=0.5+0.9435789098\mathrm{i}, \quad a_{1}=-2\mathrm{i}.
\end{aligned}
\end{equation}
The lower row of Fig.~1 shows a line wave emerging from the
plane-wave background, reaching a large amplitude, and decaying back
to the background. For the configuration with \(N_1=2\),
the parameters are chosen as
\begin{equation}\label{NLL2}
\begin{aligned}
&p_{1,1}=0.5+0.9435789098\mathrm{i}, 
\quad p_{2,1}=0.7+0.7269065339\mathrm{i},
\quad a_{1}=-2\mathrm{i},\quad a_{2}= 2\mathrm{i}.
\end{aligned}
\end{equation}
The lower row of Fig.~2 shows the corresponding
\(S_1^{[2,0,0]}\), with intersecting elongated wave fronts at
intermediate times.

\begin{figure}[tbp]\centering
      \begin{subfigure}{0.24\textwidth}
          \centering
         \includegraphics[width=\textwidth]{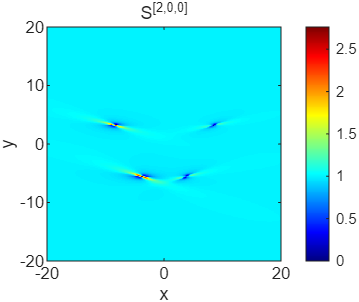}
         \caption*{t=-10}
        \end{subfigure}
      \begin{subfigure}{0.24\textwidth}
          \centering
          \includegraphics[width=\textwidth]{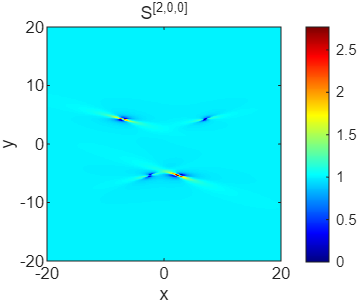}
       \caption*{t=-3}
        \end{subfigure}
      \begin{subfigure}{0.24\textwidth}
          \centering
         \includegraphics[width=\textwidth]{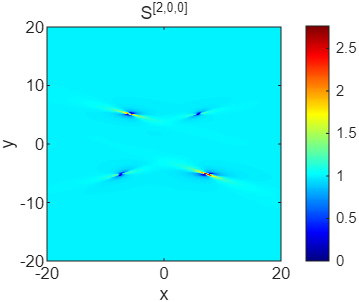}
      \caption*{t=3}
        \end{subfigure}
      \begin{subfigure}{0.24\textwidth}
          \centering
         \includegraphics[width=\textwidth]{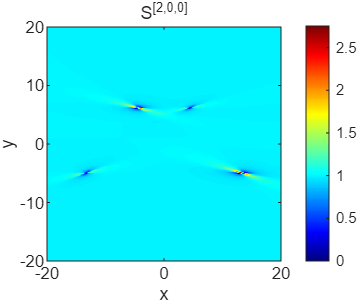}
      \caption*{t=10}
        \end{subfigure}
      \\
       \begin{subfigure}{0.24\textwidth}
          \centering
         \includegraphics[width=\textwidth]{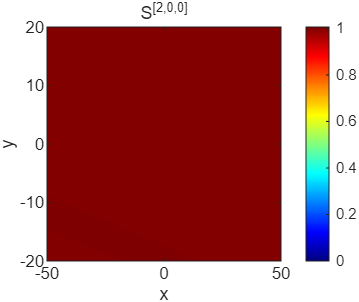}
         \caption*{t=-30}
        \end{subfigure}
      \begin{subfigure}{0.24\textwidth}
          \centering
          \includegraphics[width=\textwidth]{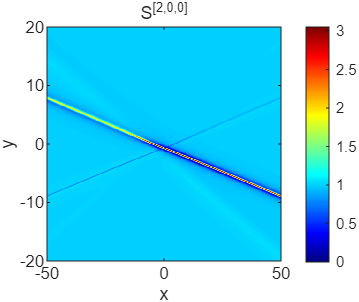}
       \caption*{t=-2}
        \end{subfigure}
      \begin{subfigure}{0.24\textwidth}
          \centering
         \includegraphics[width=\textwidth]{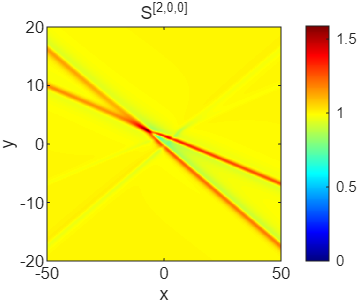}
      \caption*{t=1}
        \end{subfigure}
      \begin{subfigure}{0.24\textwidth}
          \centering
         \includegraphics[width=\textwidth]{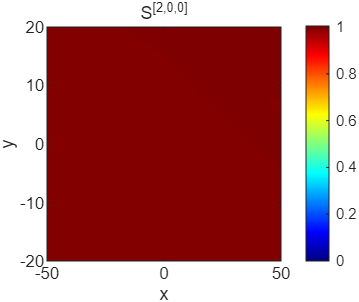}
      \caption*{t=30}
        \end{subfigure}
       \caption{Rational solutions \(S_1^{[2,0,0]}\) with \(n_1=n_2=1\).
Upper row: lumps for the parameters in \eqref{NLL1}.
Lower row: line rogue waves for the parameters in \eqref{NLL2}.}
\end{figure}

\subsection{Dynamic behaviors of breathers}
For \(N_2>0\) and \(N_1=N_3=0\), the solutions
\eqref{12d-M-semi-rational-solutions-1-1} reduce to the pure breather
family. The tau function reduces to a \(2\times2\) block determinant
with block size \(N_2\):
\begin{equation}\label{2generalized-tau-Ls}
\renewcommand{\arraystretch}{1.5}
\gamma_{m_1,m_{2}}=\mathrm{det} \left(
\begin{array}{cc}
\gamma_{m_1,m_{2}}^{[3,3]} & \gamma_{m_1,m_{2}}^{[3,4]} \\
\gamma_{-m_1,-m_{2}}^{\,\dag\,[3,4]} & \gamma_{m_1,m_{2}}^{[4,4]}
\end{array}
\right).
\end{equation}
For $N_{2}=1$, the fundamental breather solution is given by
\begin{equation}\label{2L-soliton-1}
\begin{aligned}
S_{1}^{[0,1,0]}
=
\rho_{1}e^{\mathrm{i}[k_{1}x+\omega_{1}y+
(h-\omega_{1}+k_{1}^{2})t]}
\frac{g_{1}^{[0,1,0]}}{f^{[0,1,0]}},
\end{aligned}
\end{equation}
where $f^{[0,1,0]}$ and $g_{1}^{[0,1,0]}$ are defined as:
\begin{equation}
\begin{aligned}
&f^{[0,1,0]}=\left|
\begin{array}{cc}
m_{1,1}^{[3,3,0,0]} & m_{1,1}^{[3,4,0,0]} \\
m_{1,1}^{\,*\,[3,4,0,0]} & m_{1,1}^{[4,4,0,0]}
\end{array}
\right|, \quad
g_{1}^{[0,1,0]}=\left|
\begin{array}{cc}
m_{1,1}^{[3,3,1,0]} & m_{1,1}^{[3,4,1,0]} \\
m_{1,1}^{\,*\,[3,4,-1,0]} & m_{1,1}^{[4,4,1,0]}
\end{array}
\right|,
\end{aligned}
\end{equation}
The matrix entries are
\begin{equation}
\label{BBB}
\begin{aligned}
&m_{1,1}^{[3,3,m_{1},m_{2}]}=\sum_{\mu=1}^{2}\sum_{\nu=1}^{2}
\frac{1}{p_{1,2}^{[\mu]}+p_{1,2}^{*\,[\nu]}}\left(-\frac{p_{1,2}^{[\mu]}-\mathrm{i}k_{1}}{p_{1,2}^{*\,[\nu]}+\mathrm{i}k_{1}}\right)^{m_{1}}\left(-\frac{p_{1,2}^{[\mu]}+\mathrm{i}k_{1}}{p_{1,2}^{*\,[\nu]}-\mathrm{i}k_{1}}\right)^{m_{2}} e^{\xi_{1}^{[2,2,\mu]}+\xi_{1}^{*\,[2,2,\nu]}},\\
&m_{1,1}^{[3,4,m_{1},m_{2}]}=\sum_{\mu=1}^{2}\sum_{\nu=1}^{2}
\frac{1}{p_{1,2}^{[\mu]}-p_{1,2}^{*\,[\nu]}}\left(\frac{p_{1,2}^{[\mu]}-\mathrm{i}k_{1}}{p_{1,2}^{*\,[\nu]}-\mathrm{i}k_{1}}\right)^{m_{1}}\left(\frac{p_{1,2}^{[\mu]}+\mathrm{i}k_{1}}{p_{1,2}^{*\,[\nu]}+\mathrm{i}k_{1}}\right)^{m_{2}}
e^{\xi_{1}^{[2,2,\mu]}+\bar{\xi}_{1}^{*\,[2,2,\nu]}},\\
&m_{1,1}^{[4,4,m_{1},m_{2}]}=-\sum_{\mu=1}^{2}\sum_{\nu=1}^{2}
\frac{1}{p_{1,2}^{[\mu]}+p_{1,2}^{*\,[\nu]}}\left(-\frac{p_{1,2}^{[\mu]}+\mathrm{i}k_{1}}{p_{1,2}^{*\,[\nu]}-\mathrm{i}k_{1}}\right)^{m_{1}}\left(-\frac{p_{1,2}^{[\mu]}-\mathrm{i}k_{1}}{p_{1,2}^{*\,[\nu]}+\mathrm{i}k_{1}}\right)^{m_{2}} e^{\bar{\xi}^{[2,2,\mu]}_{1}+\bar{\xi}_{1}^{*\,[2,2,\nu]}},
\end{aligned}
\end{equation}
where
\begin{equation*}
\begin{aligned}
&\xi_{1}^{[2,2,\mu]}=\sum_{\ell=1}^{2}\frac{1}{p_{1,2}^{[\mu]}-\mathrm{i}k_{\ell}}r_{\ell}+p_{1,2}^{[\mu]}x_{1}+p_{1,2}^{[\mu]\,2}x_{2}+\xi_{1,0}^{[2,2,\mu]},\\
&\bar{\xi}_{1}^{[2,2,\mu]}=-\sum_{\ell=1}^{2}\frac{1}{p_{1,2}^{[\mu]}+\mathrm{i}k_{\ell}}r_{\ell}-p_{1,2}^{[\mu]}x_{1}+p_{1,2}^{[\mu]\,2}x_{2}+\xi_{1,0}^{[2,2,\mu]}.
\end{aligned}
\end{equation*}
For the asymptotic analysis, write
\(p_\mu=p_{1,2}^{[\mu]}\),
\(\xi_\mu=\xi_1^{[2,2,\mu]}\), and
\(\bar\xi_\mu=\bar\xi_1^{[2,2,\mu]}\), where \(\mu=1,2\), and set
\[
\begin{aligned}
&B_{11}=\operatorname{Re}(\xi_1-\xi_2),\qquad
B_{12}=-\operatorname{Re}(\bar\xi_1-\bar\xi_2),\\
&B_{21}=\operatorname{Im}(\xi_1-\xi_2),\qquad
B_{22}=-\operatorname{Im}(\bar\xi_1-\bar\xi_2).
\end{aligned}
\]
Since \(B_{11}+B_{12}=2\operatorname{Re}(p_1-p_2)x\), the two
profiles separate for \(\operatorname{Re}(p_1-p_2)\ne0\).
Keeping one envelope variable bounded as \(|x|\to\infty\) gives
\begin{equation*}
\begin{aligned}
&S_1^+:\quad B_{11}=O(1),\quad B_{12}\to-\infty,
&&S_1^-:\quad B_{11}=O(1),\quad B_{12}\to+\infty,\\
&S_2^+:\quad B_{12}=O(1),\quad B_{11}\to+\infty,
&&S_2^-:\quad B_{12}=O(1),\quad B_{11}\to-\infty.
\end{aligned}
\end{equation*}
The upper and lower signs correspond to the dominant spectral modes
\(c=1\) and \(c=2\), respectively. Using the entries in \eqref{BBB}, we factor out the dominant exponential from the tau functions and obtain the coefficients
\begin{equation}\label{breather-coefficient-factorization}
\begin{aligned}
r_{\mu\nu}^{(c)}
&=-\frac{1}{(p_\mu+p_\nu^*)(p_c+p_c^*)}
+\frac{1}{(p_\mu-p_c^*)(p_c-p_\nu^*)}\\
&=-\frac{(p_\mu+p_c)(p_\nu^*+p_c^*)}
{(p_c+p_c^*)(p_\mu+p_\nu^*)(p_\mu-p_c^*)(p_\nu^*-p_c)},
\end{aligned}
\end{equation}
and
\begin{equation}\label{breather-numerator-coefficients}
\begin{aligned}
\tilde r_{1,\mu\nu}^{(c)}
&=\frac{(p_c+\mathrm{i}k_1)(p_\mu-\mathrm{i}k_1)}
{(p_c^*-\mathrm{i}k_1)(p_\nu^*+\mathrm{i}k_1)}r_{\mu\nu}^{(c)},\\
\tilde r_{2,\mu\nu}^{(c)}
&=\frac{(p_c-\mathrm{i}k_1)(p_\mu+\mathrm{i}k_1)}
{(p_c^*+\mathrm{i}k_1)(p_\nu^*-\mathrm{i}k_1)}r_{\mu\nu}^{(c)},
\qquad \mu,\nu,c=1,2.
\end{aligned}
\end{equation}
The separated profiles are therefore
\begin{equation}\label{breather-exponential-profile-1}
S_1^\pm=\rho_1e^{\mathrm{i}[k_1x+\omega_1y+(h-\omega_1+k_1^2)t]}
\frac{\displaystyle\sum_{\mu,\nu=1}^{2}
\tilde r_{1,\mu\nu}^{(c)}e^{\xi_\mu+\xi_\nu^*}}
{\displaystyle\sum_{\mu,\nu=1}^{2}
r_{\mu\nu}^{(c)}e^{\xi_\mu+\xi_\nu^*}},
\end{equation}
\begin{equation}\label{breather-exponential-profile-2}
S_2^\pm=\rho_1e^{\mathrm{i}[k_1x+\omega_1y+(h-\omega_1+k_1^2)t]}
\frac{\displaystyle\sum_{\mu,\nu=1}^{2}
\tilde r_{2,\mu\nu}^{(c)}e^{\bar\xi_\mu+\bar\xi_\nu^*}}
{\displaystyle\sum_{\mu,\nu=1}^{2}
r_{\mu\nu}^{(c)}e^{\bar\xi_\mu+\bar\xi_\nu^*}}.
\end{equation}
The four asymptotic limits can be expressed in terms of
hyperbolic and trigonometric functions.
For \(ab\ne0\), the branches of the square roots and logarithms
are chosen so that
\begin{equation*}
\begin{aligned}
\sqrt{ab}\cosh\!\left(z+\ln\sqrt{a/b}\right)
&=\tfrac12\left(ae^z+be^{-z}\right),\\
\sqrt{ab}\cos\!\left(z-\mathrm{i}\ln\sqrt{a/b}\right)
&=\tfrac12\left(ae^{\mathrm{i}z}+be^{-\mathrm{i}z}\right).
\end{aligned}
\end{equation*}
When either coefficient vanishes, the corresponding asymptotic
form follows directly from the exponential expressions
\eqref{breather-exponential-profile-1}--\eqref{breather-exponential-profile-2}.

\noindent\textbf{(b1)} When \(B_{11}=O(1)\), it follows from
\(B_{12}=2\operatorname{Re}(p_1-p_2)x-B_{11}\) that
\begin{equation*}
\begin{aligned}
\text{(b11)}\quad&B_{12}\to-\infty
&&\text{as }\operatorname{sgn}\!\bigl(\operatorname{Re}(p_1-p_2)\bigr)x\to-\infty,\\
\text{(b12)}\quad&B_{12}\to+\infty
&&\text{as }\operatorname{sgn}\!\bigl(\operatorname{Re}(p_1-p_2)\bigr)x\to+\infty.
\end{aligned}
\end{equation*}
In case (b11), \(\operatorname{Re}(\bar\xi_1-\bar\xi_2)\to+\infty\).
Using \eqref{BBB}, factoring out
\(e^{\bar\xi_1+\bar\xi_1^*}\) from both tau functions yields
the following asymptotic form of \eqref{2L-soliton-1}:
\begin{equation}\label{asymptotic-analysis-breather-1}
\begin{aligned}
S_{1}^{+} &={}\rho_{1}e^{\mathrm{i}[k_{1}x+\omega_{1}y+(h-\omega_{1} + k_{1}^{2}) t]}\\[2pt]
&\quad\frac{\sqrt{\tilde{r}_{1,+}^{[1,1]}\tilde{r}_{1,+}^{[2,1]}}\cosh\left(B_{11}+\ln\sqrt{\frac{\tilde{r}_{1,+}^{[1,1]}}{\tilde{r}_{1,+}^{[2,1]}}}\right)
+\sqrt{\tilde{r}_{1,+}^{[1,2]}\tilde{r}_{1,+}^{[2,2]}} \cos\left(B_{21}-\mathrm{i}\ln\sqrt{\frac{\tilde{r}_{1,+}^{[1,2]}}{\tilde{r}_{1,+}^{[2,2]}}}\right)}
{\sqrt{r_{1,+}^{[1,1]}r_{1,+}^{[2,1]}}\cosh\left(B_{11}+\ln\sqrt{\frac{r_{1,+}^{[1,1]}}{r_{1,+}^{[2,1]}}}\right)
+\sqrt{r_{1,+}^{[1,2]}r_{1,+}^{[2,2]}}\cos\left(B_{21}-\mathrm{i}\ln\sqrt{\frac{r_{1,+}^{[1,2]}}{r_{1,+}^{[2,2]}}}\right)},
\end{aligned}
\end{equation}
where the coefficients $r_{1,+}^{[I,J]}$ and $\tilde{r}_{1,+}^{[I,J]}$ $(I,J=1,2)$ are
\begin{equation}
\begin{aligned}
&r_{1,+}^{[1,1]}=-\frac{1}{p_1+p_1^*}\frac{1}{p_1+p_1^*}+\frac{1}{p_1-p_1^*}\frac{1}{p_1-p_1^*},
\qquad
\tilde{r}_{1,+}^{[1,1]}=\frac{p_1-\mathrm{i}k_{1}}{p_1^*-\mathrm{i}k_{1}}\frac{p_1+\mathrm{i}k_{1}}{p_1^*+\mathrm{i}k_{1}}r_{1,+}^{[1,1]},
\\
&r_{1,+}^{[2,1]}=-\frac{1}{p_1+p_1^*}\frac{1}{p_2+p_2^*}+\frac{1}{p_1-p_2^*}\frac{1}{p_2-p_1^*},
\qquad
\tilde{r}_{1,+}^{[2,1]}=\frac{p_1+\mathrm{i}k_{1}}{p_1^*-\mathrm{i}k_{1}}\frac{p_2-\mathrm{i}k_{1}}{p_2^*+\mathrm{i}k_{1}}r_{1,+}^{[2,1]},
\\
&r_{1,+}^{[1,2]}=-\frac{1}{p_1+p_1^*}\frac{1}{p_1+p_2^*}+\frac{1}{p_1-p_1^*}\frac{1}{p_1-p_2^*},
\qquad
\tilde{r}_{1,+}^{[1,2]}=\frac{p_1-\mathrm{i}k_{1}}{p_1^*-\mathrm{i}k_{1}}\frac{p_1+\mathrm{i}k_{1}}{p_2^*+\mathrm{i}k_{1}}r_{1,+}^{[1,2]},
\\
&r_{1,+}^{[2,2]}=-\frac{1}{p_1+p_1^*}\frac{1}{p_2+p_1^*}+\frac{1}{p_1-p_1^*}\frac{1}{p_2-p_1^*},
\qquad
\tilde{r}_{1,+}^{[2,2]}=\frac{p_1+\mathrm{i}k_{1}}{p_1^*+\mathrm{i}k_{1}}\frac{p_2-\mathrm{i}k_{1}}{p_1^*-\mathrm{i}k_{1}}r_{1,+}^{[2,2]}.
\end{aligned}
\end{equation}
In case (b12), \(\operatorname{Re}(\bar\xi_1-\bar\xi_2)\to-\infty\).
Using \eqref{BBB}, factoring out
\(e^{\bar\xi_2+\bar\xi_2^*}\) from both tau functions yields
the following asymptotic form:
\begin{equation}\label{asymptotic-analysis-breather-2}
\begin{aligned}
S_{1}^{-} &={}\rho_{1}e^{\mathrm{i}[k_{1}x+\omega_{1}y+(h-\omega_{1} + k_{1}^{2}) t]}\\[2pt]
&\quad\frac{\sqrt{\tilde{r}_{1,-}^{[1,1]}\tilde{r}_{1,-}^{[2,1]}}\cosh\left(B_{11}+\ln\sqrt{\frac{\tilde{r}_{1,-}^{[1,1]}}{\tilde{r}_{1,-}^{[2,1]}}}\right)
+\sqrt{\tilde{r}_{1,-}^{[1,2]}\tilde{r}_{1,-}^{[2,2]}} \cos\left(B_{21}-\mathrm{i}\ln\sqrt{\frac{\tilde{r}_{1,-}^{[1,2]}}{\tilde{r}_{1,-}^{[2,2]}}}\right)}
{\sqrt{r_{1,-}^{[1,1]}r_{1,-}^{[2,1]}}\cosh\left(B_{11}+\ln\sqrt{\frac{r_{1,-}^{[1,1]}}{r_{1,-}^{[2,1]}}}\right)
+\sqrt{r_{1,-}^{[1,2]}r_{1,-}^{[2,2]}}\cos\left(B_{21}-\mathrm{i}\ln\sqrt{\frac{r_{1,-}^{[1,2]}}{r_{1,-}^{[2,2]}}}\right)},
\end{aligned}
\end{equation}
where the coefficients $r_{1,-}^{[I,J]}$ and $\tilde{r}_{1,-}^{[I,J]}$ $(I,J=1,2)$ are
\begin{equation}
\begin{aligned}
&r_{1,-}^{[1,1]}=-\frac{1}{p_1+p_1^*}\frac{1}{p_2+p_2^*}+\frac{1}{p_1-p_2^*}\frac{1}{p_2-p_1^*},
\qquad
\tilde{r}_{1,-}^{[1,1]}=\frac{p_1-\mathrm{i}k_{1}}{p_1^*+\mathrm{i}k_{1}}\frac{p_2+\mathrm{i}k_{1}}{p_2^*-\mathrm{i}k_{1}}r_{1,-}^{[1,1]},
\\
&r_{1,-}^{[2,1]}=-\frac{1}{p_2+p_2^*}\frac{1}{p_2+p_2^*}+\frac{1}{p_2-p_2^*}\frac{1}{p_2-p_2^*},
\qquad
\tilde{r}_{1,-}^{[2,1]}=\frac{p_2+\mathrm{i}k_{1}}{p_2^*+\mathrm{i}k_{1}}\frac{p_2-\mathrm{i}k_{1}}{p_2^*-\mathrm{i}k_{1}}r_{1,-}^{[2,1]},
\\
&r_{1,-}^{[1,2]}=-\frac{1}{p_1+p_2^*}\frac{1}{p_2+p_2^*}+\frac{1}{p_1-p_2^*}\frac{1}{p_2-p_2^*},
\qquad
\tilde{r}_{1,-}^{[1,2]}=\frac{p_1-\mathrm{i}k_{1}}{p_2^*-\mathrm{i}k_{1}}\frac{p_2+\mathrm{i}k_{1}}{p_2^*+\mathrm{i}k_{1}}r_{1,-}^{[1,2]},
\\
&r_{1,-}^{[2,2]}=-\frac{1}{p_2+p_1^*}\frac{1}{p_2+p_2^*}+\frac{1}{p_2-p_1^*}\frac{1}{p_2-p_2^*},
\qquad
\tilde{r}_{1,-}^{[2,2]}=\frac{p_2+\mathrm{i}k_{1}}{p_1^*+\mathrm{i}k_{1}}\frac{p_2-\mathrm{i}k_{1}}{p_2^*-\mathrm{i}k_{1}}r_{1,-}^{[2,2]}.
\end{aligned}
\end{equation}
\noindent\textbf{(b2)} For \(B_{12}=O(1)\),
the identity \(B_{11}=2\operatorname{Re}(p_1-p_2)x-B_{12}\) gives
\begin{equation*}
\begin{aligned}
\text{(b21)}\quad&B_{11}\to+\infty
&&\text{as }\operatorname{sgn}\!\bigl(\operatorname{Re}(p_1-p_2)\bigr)x\to+\infty,\\
\text{(b22)}\quad&B_{11}\to-\infty
&&\text{as }\operatorname{sgn}\!\bigl(\operatorname{Re}(p_1-p_2)\bigr)x\to-\infty.
\end{aligned}
\end{equation*}
In case (b21), factoring out \(e^{\xi_1+\xi_1^*}\)
from both tau functions yields the following asymptotic form:
\begin{equation}\label{asymptotic-analysis-breather-3}
\begin{aligned}
S_{2}^{+} &={}\rho_{1}e^{\mathrm{i}[k_{1}x+\omega_{1}y+(h-\omega_{1} + k_{1}^{2}) t]}\\[2pt]
&\quad\frac{\sqrt{\tilde{r}_{2,+}^{[1,1]}\tilde{r}_{2,+}^{[2,1]}}\cosh\left(B_{12}-\ln\sqrt{\frac{\tilde{r}_{2,+}^{[1,1]}}{\tilde{r}_{2,+}^{[2,1]}}}\right)
+\sqrt{\tilde{r}_{2,+}^{[1,2]}\tilde{r}_{2,+}^{[2,2]}} \cos\left(B_{22}+\mathrm{i}\ln\sqrt{\frac{\tilde{r}_{2,+}^{[1,2]}}{\tilde{r}_{2,+}^{[2,2]}}}\right)}
{\sqrt{r_{2,+}^{[1,1]}r_{2,+}^{[2,1]}}\cosh\left(B_{12}-\ln\sqrt{\frac{r_{2,+}^{[1,1]}}{r_{2,+}^{[2,1]}}}\right)
+\sqrt{r_{2,+}^{[1,2]}r_{2,+}^{[2,2]}}\cos\left(B_{22}+\mathrm{i}\ln\sqrt{\frac{r_{2,+}^{[1,2]}}{r_{2,+}^{[2,2]}}}\right)},
\end{aligned}
\end{equation}
where the coefficients $r_{2,+}^{[I,J]}$ and $\tilde{r}_{2,+}^{[I,J]}$ $(I,J=1,2)$ are
\begin{equation}
\begin{aligned}
&r_{2,+}^{[1,1]}=-\frac{1}{p_1+p_1^*}\frac{1}{p_1+p_1^*}+\frac{1}{p_1-p_1^*}\frac{1}{p_1-p_1^*},
\qquad
\tilde{r}_{2,+}^{[1,1]}=\frac{p_1+\mathrm{i}k_{1}}{p_1^*+\mathrm{i}k_{1}}\frac{p_1-\mathrm{i}k_{1}}{p_1^*-\mathrm{i}k_{1}}r_{2,+}^{[1,1]},
\\
&r_{2,+}^{[2,1]}=-\frac{1}{p_1+p_1^*}\frac{1}{p_2+p_2^*}+\frac{1}{p_1-p_2^*}\frac{1}{p_2-p_1^*},
\qquad
\tilde{r}_{2,+}^{[2,1]}=\frac{p_1-\mathrm{i}k_{1}}{p_1^*+\mathrm{i}k_{1}}\frac{p_2+\mathrm{i}k_{1}}{p_2^*-\mathrm{i}k_{1}}r_{2,+}^{[2,1]},
\\
&r_{2,+}^{[1,2]}=-\frac{1}{p_1+p_1^*}\frac{1}{p_1+p_2^*}+\frac{1}{p_1-p_1^*}\frac{1}{p_1-p_2^*},
\qquad
\tilde{r}_{2,+}^{[1,2]}=\frac{p_1+\mathrm{i}k_{1}}{p_1^*+\mathrm{i}k_{1}}\frac{p_1-\mathrm{i}k_{1}}{p_2^*-\mathrm{i}k_{1}}r_{2,+}^{[1,2]},
\\
&r_{2,+}^{[2,2]}=-\frac{1}{p_1+p_1^*}\frac{1}{p_2+p_1^*}+\frac{1}{p_1-p_1^*}\frac{1}{p_2-p_1^*},
\qquad
\tilde{r}_{2,+}^{[2,2]}=\frac{p_1-\mathrm{i}k_{1}}{p_1^*-\mathrm{i}k_{1}}\frac{p_2+\mathrm{i}k_{1}}{p_1^*+\mathrm{i}k_{1}}r_{2,+}^{[2,2]}.
\end{aligned}
\end{equation}
In case (b22), factoring out \(e^{\xi_2+\xi_2^*}\)
from both tau functions yields the fourth asymptotic form:
\begin{equation}\label{asymptotic-analysis-breather-4}
\begin{aligned}
S_{2}^{-} &={}\rho_{1}e^{\mathrm{i}[k_{1}x+\omega_{1}y+(h-\omega_{1} + k_{1}^{2}) t]}\\[2pt]
&\quad\frac{\sqrt{\tilde{r}_{2,-}^{[1,1]}\tilde{r}_{2,-}^{[2,1]}}\cosh\left(B_{12}-\ln\sqrt{\frac{\tilde{r}_{2,-}^{[1,1]}}{\tilde{r}_{2,-}^{[2,1]}}}\right)
+\sqrt{\tilde{r}_{2,-}^{[1,2]}\tilde{r}_{2,-}^{[2,2]}} \cos\left(B_{22}+\mathrm{i}\ln\sqrt{\frac{\tilde{r}_{2,-}^{[1,2]}}{\tilde{r}_{2,-}^{[2,2]}}}\right)}
{\sqrt{r_{2,-}^{[1,1]}r_{2,-}^{[2,1]}}\cosh\left(B_{12}-\ln\sqrt{\frac{r_{2,-}^{[1,1]}}{r_{2,-}^{[2,1]}}}\right)
+\sqrt{r_{2,-}^{[1,2]}r_{2,-}^{[2,2]}}\cos\left(B_{22}+\mathrm{i}\ln\sqrt{\frac{r_{2,-}^{[1,2]}}{r_{2,-}^{[2,2]}}}\right)},
\end{aligned}
\end{equation}
where the coefficients $r_{2,-}^{[I,J]}$ and $\tilde{r}_{2,-}^{[I,J]}$ $(I,J=1,2)$ are
\begin{equation}
\begin{aligned}
&r_{2,-}^{[1,1]}=-\frac{1}{p_1+p_1^*}\frac{1}{p_2+p_2^*}+\frac{1}{p_1-p_2^*}\frac{1}{p_2-p_1^*},
\qquad
\tilde{r}_{2,-}^{[1,1]}=\frac{p_1+\mathrm{i}k_{1}}{p_1^*-\mathrm{i}k_{1}}\frac{p_2-\mathrm{i}k_{1}}{p_2^*+\mathrm{i}k_{1}}r_{2,-}^{[1,1]},
\\
&r_{2,-}^{[2,1]}=-\frac{1}{p_2+p_2^*}\frac{1}{p_2+p_2^*}+\frac{1}{p_2-p_2^*}\frac{1}{p_2-p_2^*},
\qquad
\tilde{r}_{2,-}^{[2,1]}=\frac{p_2+\mathrm{i}k_{1}}{p_2^*+\mathrm{i}k_{1}}\frac{p_2-\mathrm{i}k_{1}}{p_2^*-\mathrm{i}k_{1}}r_{2,-}^{[2,1]},
\\
&r_{2,-}^{[1,2]}=-\frac{1}{p_1+p_2^*}\frac{1}{p_2+p_2^*}+\frac{1}{p_1-p_2^*}\frac{1}{p_2-p_2^*},
\qquad
\tilde{r}_{2,-}^{[1,2]}=\frac{p_1+\mathrm{i}k_{1}}{p_2^*+\mathrm{i}k_{1}}\frac{p_2-\mathrm{i}k_{1}}{p_2^*-\mathrm{i}k_{1}}r_{2,-}^{[1,2]},
\\
&r_{2,-}^{[2,2]}=-\frac{1}{p_2+p_1^*}\frac{1}{p_2+p_2^*}+\frac{1}{p_2-p_1^*}\frac{1}{p_2-p_2^*},
\qquad
\tilde{r}_{2,-}^{[2,2]}=\frac{p_2+\mathrm{i}k_{1}}{p_2^*+\mathrm{i}k_{1}}\frac{p_2-\mathrm{i}k_{1}}{p_1^*-\mathrm{i}k_{1}}r_{2,-}^{[2,2]}.
\end{aligned}
\end{equation}
\begin{figure}[tbp]\centering
      \begin{subfigure}{0.24\textwidth}
          \centering
         \includegraphics[width=\textwidth]{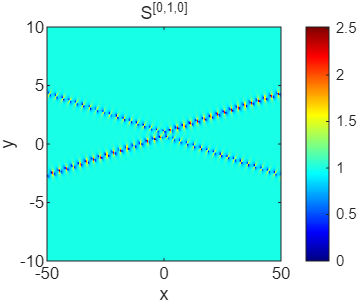}
         \caption*{t=-5}
        \end{subfigure}
      \begin{subfigure}{0.24\textwidth}
          \centering
          \includegraphics[width=\textwidth]{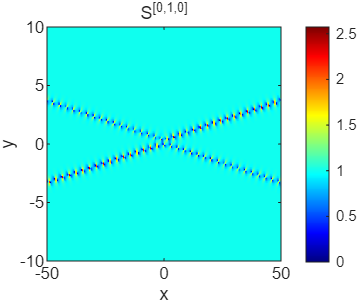}
       \caption*{t=-1}
        \end{subfigure}
      \begin{subfigure}{0.24\textwidth}
          \centering
         \includegraphics[width=\textwidth]{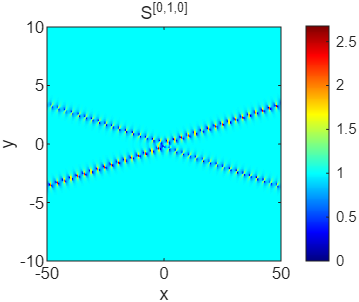}
      \caption*{t=1}
        \end{subfigure}
      \begin{subfigure}{0.24\textwidth}
          \centering
         \includegraphics[width=\textwidth]{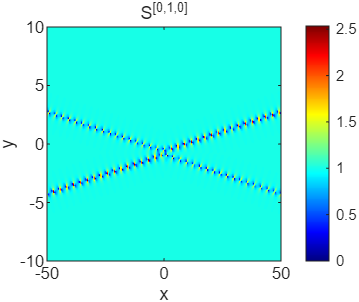}
      \caption*{t=5}
        \end{subfigure}
      \\
       \begin{subfigure}{0.24\textwidth}
          \centering
         \includegraphics[width=\textwidth]{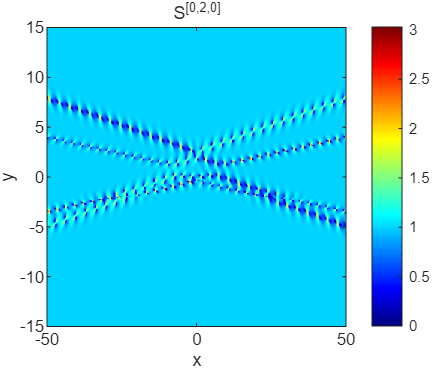}
         \caption*{t=-3}
        \end{subfigure}
      \begin{subfigure}{0.24\textwidth}
          \centering
          \includegraphics[width=\textwidth]{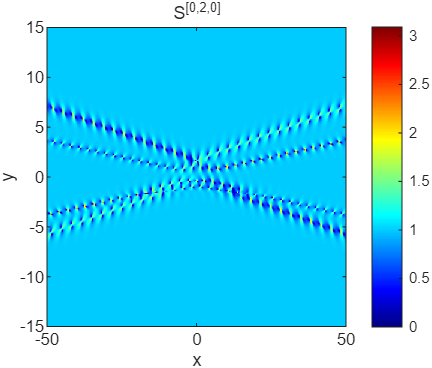}
       \caption*{t=-1}
        \end{subfigure}
      \begin{subfigure}{0.24\textwidth}
          \centering
         \includegraphics[width=\textwidth]{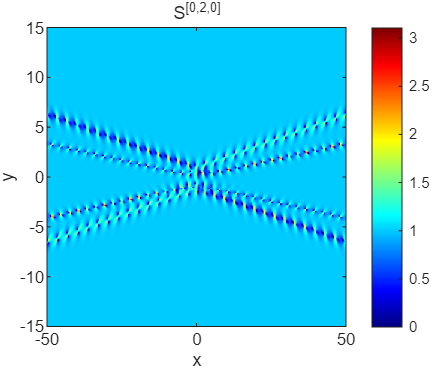}
      \caption*{t=1}
        \end{subfigure}
      \begin{subfigure}{0.24\textwidth}
          \centering
         \includegraphics[width=\textwidth]{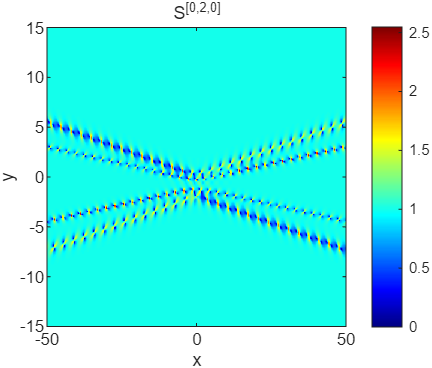}
      \caption*{t=3}
        \end{subfigure}
       \caption{Breather solutions with \(N_2=1\) and \(N_2=2\)
for the parameters in \eqref{NB1} and \eqref{NBB1}, respectively.}
\end{figure}
The coefficients above are related to those in
\eqref{breather-coefficient-factorization}--\eqref{breather-numerator-coefficients}
by
\begin{equation*}
\begin{aligned}
\left(r_{j,\pm}^{[1,1]},r_{j,\pm}^{[2,1]},
r_{j,\pm}^{[1,2]},r_{j,\pm}^{[2,2]}\right)
&=\left(r_{11}^{(c)},r_{22}^{(c)},r_{12}^{(c)},r_{21}^{(c)}\right),\\
\left(\tilde r_{j,\pm}^{[1,1]},\tilde r_{j,\pm}^{[2,1]},
\tilde r_{j,\pm}^{[1,2]},\tilde r_{j,\pm}^{[2,2]}\right)
&=\left(\tilde r_{j,11}^{(c)},\tilde r_{j,22}^{(c)},
\tilde r_{j,12}^{(c)},\tilde r_{j,21}^{(c)}\right),
\end{aligned}
\end{equation*}
where \(j=1,2\).
The upper and lower signs correspond to \(c=1\) and \(c=2\),
respectively.

\noindent\textbf{(b3)} Away from the two breather envelopes, both
\(|B_{11}|\) and \(|B_{12}|\) tend to infinity.
Let \(e^{\xi_a}\) and \(e^{\bar\xi_c}\) be the dominant
exponential terms. For \(r_{aa}^{(c)}\ne0\), the leading
terms in \eqref{BBB} yield
\begin{equation}\label{breather-background-limit}
\frac{S_1^{[0,1,0]}}
{\rho_1e^{\mathrm{i}[k_1x+\omega_1y+(h-\omega_1+k_1^2)t]}}
\longrightarrow
\frac{(p_c+\mathrm{i}k_1)(p_a-\mathrm{i}k_1)}
{(p_c^*-\mathrm{i}k_1)(p_a^*+\mathrm{i}k_1)}.
\end{equation}
Here \(a=1\) for \(B_{11}\to+\infty\) and \(a=2\) for
\(B_{11}\to-\infty\).
Similarly, \(c=1\) for \(B_{12}\to-\infty\) and \(c=2\)
for \(B_{12}\to+\infty\).
The limiting factor has unit modulus, so the far-field
intensity is \(\rho_1^2\).

For \(\partial_xB_{11}\ne0\) and \(\partial_tB_{11}\ne0\),
the first and second envelopes propagate at fixed \(y\)
with velocities
\(-(\partial_tB_{11})/(\partial_xB_{11})\) and
\((\partial_tB_{11})/(\partial_xB_{11})\), respectively.
Cases (b11)--(b22) give the corresponding asymptotic profiles
as \(t\to\pm\infty\) along these trajectories.

The factorization \eqref{breather-coefficient-factorization}
shows that the profiles \(S_1^\pm\) have the one-breather
form with each \(e^{\xi_\mu}\) replaced by
\((p_\mu+p_c)e^{\xi_\mu}/(p_\mu-p_c^*)\).
The additional factor
\(-(p_c+\mathrm{i}k_1)/(p_c^*-\mathrm{i}k_1)\)
has unit modulus and contributes only a constant phase shift.
The same argument applies to \(S_2^\pm\), with
\(\xi_\mu\) replaced by \(\bar\xi_\mu\) and \(k_1\) by \(-k_1\).
For each breather, the incoming and outgoing profiles differ
only by constant envelope and phase shifts.
The amplitudes, widths, and velocities are preserved,
so the interaction is elastic.

For one breather pair \((N_2=1)\), the parameters are chosen as
\begin{equation}\label{NB1}
\begin{aligned}
&p_{1,2}^{[1]}=\frac{4}{3}+\frac{4}{3}\mathrm{i},
\quad p_{1,2}^{[2]}=\frac{5}{3}- \frac{2}{3}\mathrm{i},
\quad \xi_{1,0}^{[2,2,1]}=10\mathrm{i},\quad \xi_{1,0}^{[2,2,2]}=10\mathrm{i}.
\end{aligned}
\end{equation}
For two breather pairs \((N_2=2)\), the parameters are chosen as
\begin{equation}
\label{NBB1}
\begin{aligned}
&p_{1,2}^{[1]}=\frac{4}{3}+\frac{4}{3}\mathrm{i},\quad
p_{1,2}^{[2]}=\frac{5}{3}- \frac{2}{3}\mathrm{i},\quad
p_{2,2}^{[1]}=\frac{9}{10}+\frac{2}{3}\mathrm{i},\\
&p_{2,2}^{[2]}=\frac{5}{4}- \frac{4}{3}\mathrm{i},
\quad \xi_{1,0}^{[2,2,1]}=10\mathrm{i},\quad \xi_{1,0}^{[2,2,2]}=10\mathrm{i},
\quad \xi_{2,0}^{[2,2,1]}=0,\quad
\xi_{2,0}^{[2,2,2]}=0.
\end{aligned}
\end{equation}
Figure~3 shows the one-pair and two-pair breather solutions
\(S_1^{[0,1,0]}\) and \(S_1^{[0,2,0]}\), respectively.

\Needspace{8\baselineskip}
\subsection{Dynamic behaviors of dark solitons}
For \(N_3>0\) and \(N_1=N_2=0\), the solutions
\eqref{12d-M-semi-rational-solutions-1-1} reduce to the pure
dark-soliton family. The corresponding tau function is a
\(2\times2\) block determinant with \(N_3\times N_3\) blocks:
\begin{equation}\label{3generalized-tau-Ls}
\renewcommand{\arraystretch}{1.5}
\gamma_{m_{1},m_{2}}=\mathrm{det} \left(
\begin{array}{cc}
\gamma_{m_{1},m_{2}}^{[5,5]} & \gamma_{m_{1},m_{2}}^{[5,6]} \\
\gamma_{-m_{1},-m_{2}}^{\,\dag\,[5,6]} & \gamma_{m_{1},m_{2}}^{[6,6]}
\end{array}
\right).
\end{equation}
For \(N_3=1\), the fundamental dark-soliton pair is given by
\begin{equation}\label{3L-soliton-1}
\begin{aligned}
S_{1}^{[0,0,1]}= \rho_{1}e^{\mathrm{i}[k_{1}x+\omega_{1}y+(h-\omega_{1} + k_{1}^{2}) t]}\frac{g_{1}^{[0,0,1]}}{f^{[0,0,1]}},
\end{aligned}
\end{equation}
where $f^{[0,0,1]}$ and $g_{1}^{[0,0,1]}$ are defined as:
\begin{equation}
\begin{aligned}
&f^{[0,0,1]}=\left|
\begin{array}{cc}
m_{1,1}^{[5,5,0,0]} & m_{1,1}^{[5,6,0,0]} \\
m_{1,1}^{\,*\,[5,6,0,0]} & m_{1,1}^{[6,6,0,0]}
\end{array}
\right|, \qquad
g_{1}^{[0,0,1]}=\left|
\begin{array}{cc}
m_{1,1}^{[5,5,1,0]} & m_{1,1}^{[5,6,1,0]} \\
m_{1,1}^{\,*\,[5,6,-1,0]} & m_{1,1}^{[6,6,1,0]}
\end{array}
\right|,
\end{aligned}
\end{equation}
The matrix entries are
\begin{equation}\label{SSS}
\begin{aligned}
&m_{1,1}^{[5,5,m_{1},m_{2}]}=1+
\frac{1}{p_{1,3}+p_{1,3}^{\,*}}\left(-\frac{p_{1,3}-\mathrm{i}k_{1}}{p_{1,3}^{\,*}+\mathrm{i}k_{1}}\right)^{m_{1}}\left(-\frac{p_{1,3}+\mathrm{i}k_{1}}{p_{1,3}^{\,*}-\mathrm{i}k_{1}}\right)^{m_{2}}
e^{\xi_{1}^{[3,3]}+\xi_{1}^{*\,[3,3]}},
\\
&m_{1,1}^{[5,6,m_{1},m_{2}]}=
\frac{1}{p_{1,3}-p_{1,3}^{\,*}}\left(\frac{p_{1,3}-\mathrm{i}k_{1}}{p_{1,3}^{\,*}-\mathrm{i}k_{1}}\right)^{m_{1}}\left(\frac{p_{1,3}+\mathrm{i}k_{1}}{p_{1,3}^{\,*}+\mathrm{i}k_{1}}\right)^{m_{2}}
e^{\xi_{1}^{[3,3]}+\bar{\xi}_{1}^{*\,[3,3]}}, \\
&m_{1,1}^{[6,6,m_{1},m_{2}]}=-1-
\frac{1}{p_{1,3}+p_{1,3}^{\,*}}\left(-\frac{p_{1,3}+\mathrm{i}k_{1}}{p_{1,3}^{\,*}-\mathrm{i}k_{1}}\right)^{m_{1}}\left(-\frac{p_{1,3}-\mathrm{i}k_{1}}{p_{1,3}^{\,*}+\mathrm{i}k_{1}}\right)^{m_{2}}
e^{\bar{\xi}_{1}^{[3,3]}+\bar{\xi}_{1}^{*\,[3,3]}},
\end{aligned}
\end{equation}
where
\begin{equation*}
\begin{aligned}
&\xi_{1}^{[3,3]}=\sum_{\ell=1}^{2}\frac{1}{p_{1,3}-\mathrm{i}k_{\ell}}r_{\ell}+p_{1,3}x_{1}+p_{1,3}^{2}x_{2}+\xi_{1,0}^{[3,3]},\\
&\bar{\xi}_{1}^{[3,3]}=-\sum_{\ell=1}^{2}\frac{1}{p_{1,3}+\mathrm{i}k_{\ell}}r_{\ell}-p_{1,3}x_{1}+p_{1,3}^{2}x_{2}+\xi_{1,0}^{[3,3]}.
\end{aligned}
\end{equation*}
\begin{figure}[tbp]\centering
      \begin{subfigure}{0.24\textwidth}
          \centering
         \includegraphics[width=\textwidth]{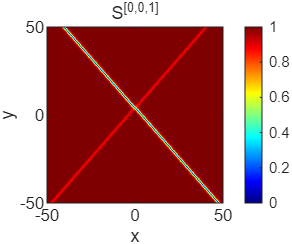}
         \caption*{t=-5}
        \end{subfigure}
      \begin{subfigure}{0.24\textwidth}
          \centering
          \includegraphics[width=\textwidth]{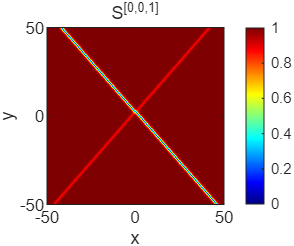}
       \caption*{t=-3}
        \end{subfigure}
      \begin{subfigure}{0.24\textwidth}
          \centering
         \includegraphics[width=\textwidth]{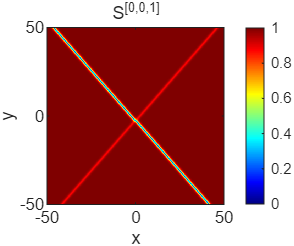}
      \caption*{t=3}
        \end{subfigure}
      \begin{subfigure}{0.24\textwidth}
          \centering
         \includegraphics[width=\textwidth]{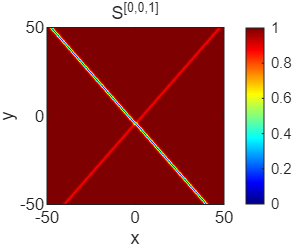}
      \caption*{t=5}
        \end{subfigure}
      \\
       \begin{subfigure}{0.24\textwidth}
          \centering
         \includegraphics[width=\textwidth]{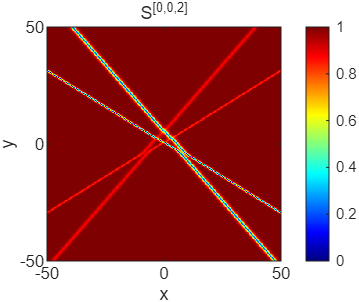}
         \caption*{t=-5}
        \end{subfigure}
      \begin{subfigure}{0.24\textwidth}
          \centering
          \includegraphics[width=\textwidth]{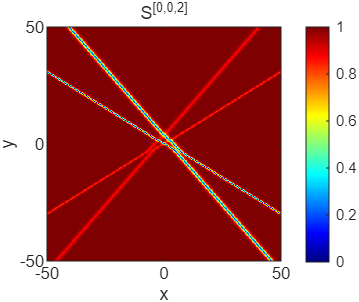}
       \caption*{t=-3}
        \end{subfigure}
      \begin{subfigure}{0.24\textwidth}
          \centering
         \includegraphics[width=\textwidth]{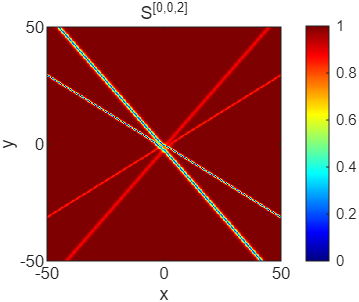}
      \caption*{t=3}
        \end{subfigure}
      \begin{subfigure}{0.24\textwidth}
          \centering
         \includegraphics[width=\textwidth]{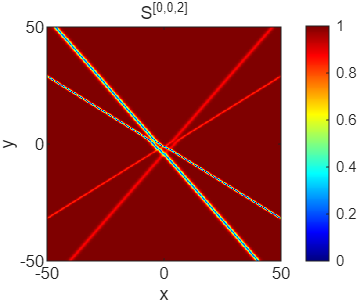}
      \caption*{t=5}
        \end{subfigure}
       \caption{Dark-soliton solutions with \(N_3=1\) and \(N_3=2\), for the parameters in \eqref{ND1} and \eqref{NDD1}, respectively.}
\end{figure}
For the asymptotic analysis, let
\(C_{11}=\xi_1^{[3,3]}+\xi_1^{*\,[3,3]}\) and
\(C_{12}=\bar\xi_1^{[3,3]}+\bar\xi_1^{*\,[3,3]}\).
Since \(C_{11}-C_{12}=4\operatorname{Re}(p_{1,3})x\),
the two asymptotic dark-soliton profiles follow by keeping
either \(C_{11}\) or \(C_{12}\) bounded as \(|x|\to\infty\).
The denominator tau function takes the form
\begin{equation*}
f^{[0,0,1]}=-1-\frac{e^{C_{11}}+e^{C_{12}}}{2\operatorname{Re}p_{1,3}}
-\frac{|p_{1,3}|^2e^{C_{11}+C_{12}}}
{4(\operatorname{Re}p_{1,3})^2(\operatorname{Im}p_{1,3})^2}.
\end{equation*}
The limiting intensities are given by

\textbf{(c1)} For \(C_{11}=O(1)\),
\begin{equation*}
\begin{aligned}
\text{(c11)}\quad&C_{12}\to-\infty
&&\text{as }\operatorname{sgn}(\operatorname{Re}p_{1,3})x\to+\infty,\\
\text{(c12)}\quad&C_{12}\to+\infty
&&\text{as }\operatorname{sgn}(\operatorname{Re}p_{1,3})x\to-\infty.
\end{aligned}
\end{equation*}
Taking these limits in \eqref{3L-soliton-1} yields
the following asymptotic profile:
\begin{equation}\label{nonlocal-dark-soliton-1-1}
|S_{1}^{\pm}|^{2}=\rho_{1}^{2}\left[1-
\frac{2\mathrm{Re}^{2}(p_{1,3})}
{[\mathrm{Re}^{2}(p_{1,3})+(\mathrm{Im}(p_{1,3})-k_{1})^{2}][\textrm{sgn}(\mathrm{Re}(p_{1,3}))\cosh{(C_{11}+\Delta_{1}^{\pm})+1}]}
\right],
\end{equation}
where \(\Delta_{1}^{+}=\ln\frac{1}{2|\mathrm{Re}(p_{1,3})|}\) and \(\Delta_{1}^{-}=\ln\frac{|p_{1,3}|^{2}}{2|\mathrm{Re}(p_{1,3})|\mathrm{Im}^{2}(p_{1,3})}\). Here $\Delta_{1}^{+}$ corresponds to the asymptotic regime
$C_{12}\to-\infty$ in (c11), whereas $\Delta_{1}^{-}$
corresponds to $C_{12}\to+\infty$ in (c12).
\textbf{(c2)} For \(C_{12}=O(1)\),
\begin{equation*}
\begin{aligned}
\text{(c21)}\quad&C_{11}\to+\infty
&&\text{as }\operatorname{sgn}(\operatorname{Re}p_{1,3})x\to+\infty,\\
\text{(c22)}\quad&C_{11}\to-\infty
&&\text{as }\operatorname{sgn}(\operatorname{Re}p_{1,3})x\to-\infty.
\end{aligned}
\end{equation*}
Taking these limits in \eqref{3L-soliton-1} yields
the following asymptotic form:
\begin{equation}\label{nonlocal-dark-soliton-1-2}
|S_{2}^{\pm}|^{2}=\rho_{1}^{2}\left[1-
\frac{2\mathrm{Re}^{2}(p_{1,3})}
{[\mathrm{Re}^{2}(p_{1,3})+(\mathrm{Im}(p_{1,3})+k_{1})^{2}][\textrm{sgn}(\mathrm{Re}(p_{1,3}))\cosh{(C_{12}-\Delta_{2}^{\pm})+1}]}
\right],
\end{equation}
where \(\Delta_{2}^{+}=\ln\frac{2|\mathrm{Re}(p_{1,3})|\mathrm{Im}^{2}(p_{1,3})}{|p_{1,3}|^{2}}\) and \(\Delta_{2}^{-}=\ln{2|\mathrm{Re}(p_{1,3})|}\). The upper sign corresponds to \(C_{11}\to+\infty\) in (c21), and the lower sign to \(C_{11}\to-\infty\) in (c22).
\medskip

\noindent\textbf{(c3)} When both \(|C_{11}|\) and \(|C_{12}|\)
tend to infinity, the leading terms in \eqref{SSS} yield
the four plane-wave limits
\begin{equation}\label{dark-background-limits}
\frac{S_1^{[0,0,1]}}
{\rho_1e^{\mathrm{i}[k_1x+\omega_1y+(h-\omega_1+k_1^2)t]}}
\longrightarrow
\begin{cases}
1, & C_{11},C_{12}\to-\infty,\\[3pt]
-\dfrac{p_{1,3}-\mathrm{i}k_1}{p_{1,3}^*+\mathrm{i}k_1},
& C_{11}\to+\infty,\ C_{12}\to-\infty,\\[6pt]
-\dfrac{p_{1,3}+\mathrm{i}k_1}{p_{1,3}^*-\mathrm{i}k_1},
& C_{11}\to-\infty,\ C_{12}\to+\infty,\\[6pt]
\dfrac{p_{1,3}^2+k_1^2}{p_{1,3}^{*\,2}+k_1^2},
& C_{11},C_{12}\to+\infty.
\end{cases}
\end{equation}
All four multipliers have unit modulus, so
\(|S_1^{[0,0,1]}|\to|\rho_1|\).
These limits also cover the case in which both \(|C_{11}|\)
and \(|C_{12}|\) grow linearly with \(|x|\).

For \(\operatorname{Re}p_{1,3}>0\), each profile in
\eqref{nonlocal-dark-soliton-1-1} and
\eqref{nonlocal-dark-soliton-1-2} retains its depth and width
in the two asymptotic limits. The shift parameters satisfy
\[
\Delta_1^--\Delta_1^+=\Delta_2^--\Delta_2^+
=\ln\frac{|p_{1,3}|^2}{\operatorname{Im}^2(p_{1,3})}.
\]
The corresponding center lines are
\(C_{11}+\Delta_1^\pm=0\) and \(C_{12}-\Delta_2^\pm=0\),
which determine the position shifts.
At fixed \(y\), the first and second solitons have velocities
\(-(\partial_tC_{11})/[2\operatorname{Re}(p_{1,3})]\) and
\((\partial_tC_{11})/[2\operatorname{Re}(p_{1,3})]\), respectively.
When \(\partial_tC_{11}\ne0\), cases (c11)--(c22) describe
the limits \(t\to\pm\infty\) along the two trajectories.
Thus, the fundamental dark-soliton pair undergoes an elastic
interaction.

For one dark-soliton pair \((N_3=1)\), the parameters are chosen as
\begin{equation}
\label{ND1}
\begin{aligned}
p_{1,3}=1-\frac{4}{5}\mathrm{i},\quad
\xi_{1,0}^{[3,3]}=0.
\end{aligned}
\end{equation}
For two dark-soliton pairs \((N_3=2)\), the parameters are chosen as
\begin{equation}\label{NDD1}
\begin{aligned}
&p_{1,3}=1-\frac{4}{5}\mathrm{i}, \quad
p_{2,3}=\frac{4}{3}+\mathrm{i},
\quad \xi_{1,0}^{[3,3]}=0, \quad \xi_{2,0}^{[3,3]}=0.
\end{aligned}
\end{equation}
Figure~4 shows the one-pair and two-pair dark-soliton
solutions \(S_1^{[0,0,1]}\) and \(S_1^{[0,0,2]}\), respectively.

\begin{figure}[tbp]\centering
      \begin{subfigure}{0.24\textwidth}
          \centering
         \includegraphics[width=\textwidth]{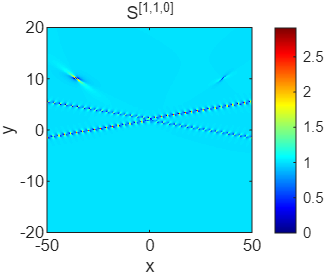}
         \caption*{t=-10}
        \end{subfigure}
      \begin{subfigure}{0.24\textwidth}
          \centering
          \includegraphics[width=\textwidth]{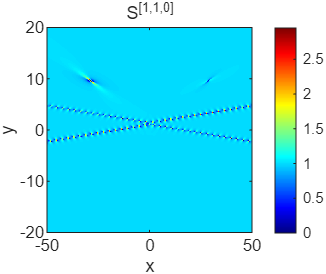}
       \caption*{t=-5}
        \end{subfigure}
      \begin{subfigure}{0.24\textwidth}
          \centering
         \includegraphics[width=\textwidth]{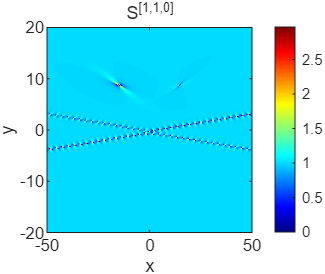}
      \caption*{t=5}
        \end{subfigure}
      \begin{subfigure}{0.24\textwidth}
          \centering
         \includegraphics[width=\textwidth]{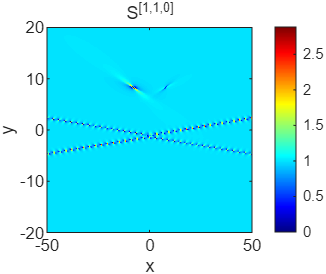}
      \caption*{t=10}
        \end{subfigure}
      \\
       \begin{subfigure}{0.24\textwidth}
          \centering
         \includegraphics[width=\textwidth]{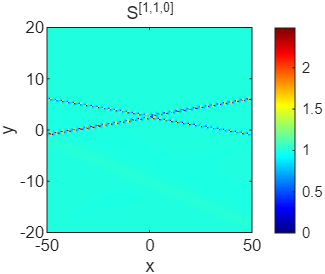}
         \caption*{t=-15}
        \end{subfigure}
      \begin{subfigure}{0.24\textwidth}
          \centering
          \includegraphics[width=\textwidth]{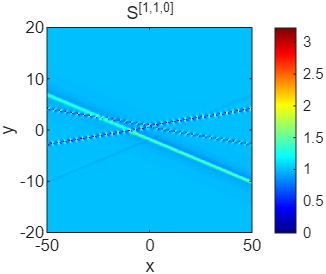}
       \caption*{t=-3}
        \end{subfigure}
      \begin{subfigure}{0.24\textwidth}
          \centering
         \includegraphics[width=\textwidth]{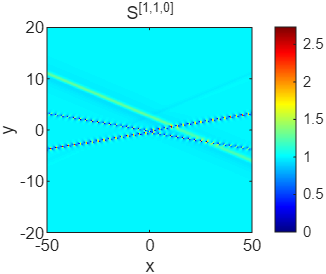}
      \caption*{t=3}
        \end{subfigure}
      \begin{subfigure}{0.24\textwidth}
          \centering
         \includegraphics[width=\textwidth]{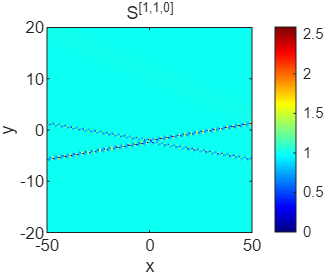}
      \caption*{t=15}
        \end{subfigure}
       \caption{Semi-rational solutions with \((N_1,N_2,N_3)=(1,1,0)\).
Upper row: lumps and breathers for the parameters in
\eqref{NL1} and \eqref{NB1}.
Lower row: line rogue waves and breathers for the parameters in
\eqref{NL2} and \eqref{NB1}.}
\end{figure}
\subsection{Dynamic behaviors of lumps or line rogue waves with backgrounds of breathers}
When \(N_3=0\) and \(N_1,N_2>0\), Theorem~1 yields lumps
or line rogue waves interacting with \(N_2\) breather pairs.
The corresponding tau function takes the \(4\times4\) block form
\begin{equation}\label{5generalized-tau-Ls}
\renewcommand{\arraystretch}{1.5}
\gamma_{m_{1},m_{2}}=\mathrm{det} \left(
\begin{array}{cccc}
\gamma_{m_{1},m_{2}}^{[1,1]} & \gamma_{m_{1},m_{2}}^{[1,2]} & \gamma_{m_{1},m_{2}}^{[1,3]} & \gamma_{m_{1},m_{2}}^{[1,4]}\\
\gamma_{-m_{1},-m_{2}}^{\,\dag\,[1,2]} & \gamma_{m_{1},m_{2}}^{[2,2]} & \gamma_{m_{1},m_{2}}^{[2,3]}& \gamma_{m_{1},m_{2}}^{[2,4]}\\
\gamma_{-m_{1},-m_{2}}^{\,\dag\,[1,3]} & \gamma_{-m_{1},-m_{2}}^{\,\dag\,[2,3]} & \gamma_{m_{1},m_{2}}^{[3,3]} & \gamma_{m_{1},m_{2}}^{[3,4]}\\
\gamma_{-m_{1},-m_{2}}^{\,\dag\,[1,4]} & \gamma_{-m_{1},-m_{2}}^{\,\dag\,[2,4]} & \gamma_{-m_{1},-m_{2}}^{\,\dag\,[3,4]} & \gamma_{m_{1},m_{2}}^{[4,4]}\\
\end{array}
\right).
\end{equation}
The fundamental semi-rational solutions with
\(N_1=N_2=1\) and \(N_3=0\) take the form
\begin{equation}\label{5L-soliton-1}
\begin{aligned}
S_{1}^{[1,1,0]}= \rho_{1}e^{\mathrm{i}[k_{1}x+\omega_{1}y+(h-\omega_{1} + k_{1}^{2}) t]}\frac{g_{1}^{[1,1,0]}}{f^{[1,1,0]}} ,
\end{aligned}
\end{equation}
Here \(f^{[1,1,0]}=\gamma_{0,0}\) and \(g_1^{[1,1,0]}=\gamma_{1,0}\), where
\(\gamma_{m_1,m_2}\) is defined by \eqref{5generalized-tau-Ls}.
\begin{figure}[tbp]\centering
      \begin{subfigure}{0.24\textwidth}
          \centering
         \includegraphics[width=\textwidth]{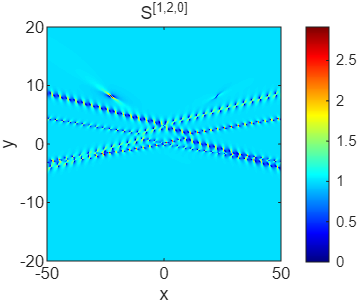}
         \caption*{t=-3}
        \end{subfigure}
      \begin{subfigure}{0.24\textwidth}
          \centering
          \includegraphics[width=\textwidth]{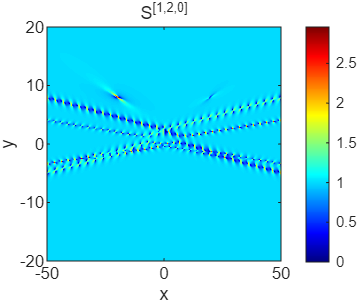}
       \caption*{t=-1}
        \end{subfigure}
      \begin{subfigure}{0.24\textwidth}
          \centering
         \includegraphics[width=\textwidth]{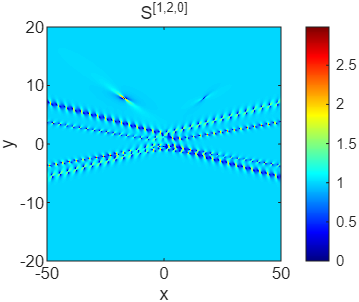}
      \caption*{t=1}
        \end{subfigure}
      \begin{subfigure}{0.24\textwidth}
          \centering
         \includegraphics[width=\textwidth]{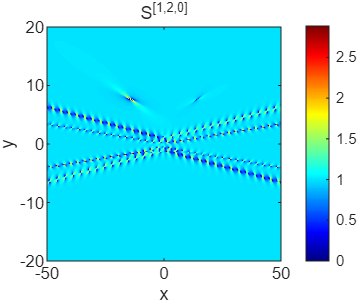}
      \caption*{t=3}
        \end{subfigure}
      \\
       \begin{subfigure}{0.24\textwidth}
          \centering
         \includegraphics[width=\textwidth]{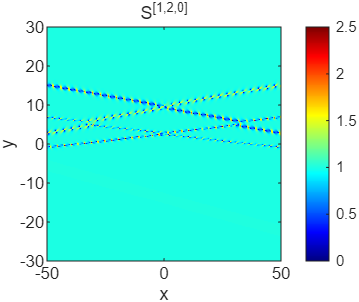}
         \caption*{t=-20}
        \end{subfigure}
      \begin{subfigure}{0.24\textwidth}
          \centering
          \includegraphics[width=\textwidth]{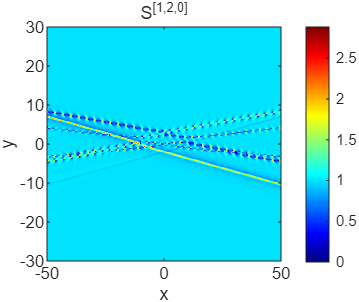}
       \caption*{t=-3}
        \end{subfigure}
      \begin{subfigure}{0.24\textwidth}
          \centering
         \includegraphics[width=\textwidth]{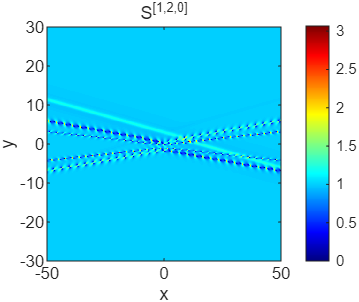}
      \caption*{t=3}
        \end{subfigure}
      \begin{subfigure}{0.24\textwidth}
          \centering
         \includegraphics[width=\textwidth]{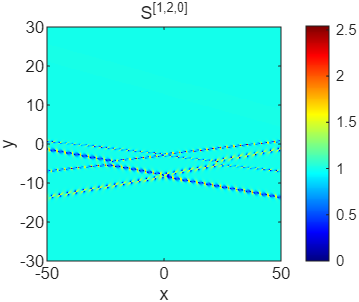}
      \caption*{t=20}
        \end{subfigure}
       \caption{Semi-rational solutions with \((N_1,N_2,N_3)=(1,2,0)\).
Upper row: lumps and two breather pairs for the parameters in
\eqref{NL1} and \eqref{NBB1}.
Lower row: line rogue waves and two breather pairs for the parameters in
\eqref{NL2} and \eqref{NBB1}.}
\end{figure}

The matrix entries are
\begin{equation}
\label{5LS1}
\begin{aligned}
m_{1,1}^{[1,3,m_1,m_{2}]}=&\sum_{\nu=1}^{2}\frac{1}{p_{1,1}+p_{1,2}^{*\,[\nu]}}\left(-\frac{p_{1,1}-\mathrm{i}k_{1}}{p_{1,2}^{*\,[\nu]}+\mathrm{i}k_{1}}\right)^{m_{1}}\left(-\frac{p_{1,1}+\mathrm{i}k_{1}}{p_{1,2}^{*\,[\nu]}-\mathrm{i}k_{1}}\right)^{m_{2}}
e^{\xi_{1}^{*\,[2,2,\nu]}}\\
& \left(x_{1,m_{1},m_{2}}^{+}-d_{1}^{[1,3,\nu]}\right),\\
m_{1,1}^{[1,4,m_1,m_{2}]}=&\sum_{\nu=1}^{2}\frac{1}{p_{1,1}-p_{1,2}^{*\,[\nu]}}\left(\frac{p_{1,1}-\mathrm{i}k_{1}}{p_{1,2}^{*\,[\nu]}-\mathrm{i}k_{1}}\right)^{m_{1}}\left(\frac{p_{1,1}+\mathrm{i}k_{1}}{p_{1,2}^{*\,[\nu]}+\mathrm{i}k_{1}}\right)^{m_{2}}
e^{\bar{\xi}_{1}^{*\,[2,2,\nu]}}\\
& \left(x_{1,m_{1},m_{2}}^{+}-d_{1}^{[1,4,\nu]}\right),\\
m_{1,1}^{[2,3,m_1,m_{2}]}=&-\sum_{\nu=1}^{2}\frac{1}{p_{1,1}-p_{1,2}^{*\,[\nu]}}\left(\frac{p_{1,1}+\mathrm{i}k_{1}}{p_{1,2}^{*\,[\nu]}+\mathrm{i}k_{1}}\right)^{m_{1}}\left(\frac{p_{1,1}-\mathrm{i}k_{1}}{p_{1,2}^{*\,[\nu]}-\mathrm{i}k_{1}}\right)^{m_{2}}
e^{\xi_{1}^{*\,[2,2,\nu]}}\\
& \left(\bar{x}_{1,m_{1},m_{2}}^{+}-d_{1}^{[1,4,\nu]}\right),\\
m_{1,1}^{[2,4,m_1,m_{2}]}=&-\sum_{\nu=1}^{2}\frac{1}{p_{1,1}+p_{1,2}^{*\,[\nu]}}\left(-\frac{p_{1,1}+\mathrm{i}k_{1}}{p_{1,2}^{*\,[\nu]}-\mathrm{i}k_{1}}\right)^{m_{1}}\left(-\frac{p_{1,1}-\mathrm{i}k_{1}}{p_{1,2}^{*\,[\nu]}+\mathrm{i}k_{1}}\right)^{m_{2}}
e^{\bar{\xi}_{1}^{*\,[2,2,\nu]}}\\
& \left(\bar{x}_{1,m_{1},m_{2}}^{+}-d_{1}^{[1,3,\nu]}\right),
\end{aligned}
\end{equation}
where
\begin{equation*}
\begin{aligned}
d_{1}^{[1,3,\nu]}=\frac{p_{1,1}}{p_{1,1}+p_{1,2}^{*\,[\nu]}},
\quad
d_{1}^{[1,4,\nu]}=\frac{p_{1,1}}{p_{1,1}-p_{1,2}^{*\,[\nu]}},
\end{aligned}
\end{equation*}
and the entries
\(m_{1,1}^{[1,1,m_1,m_2]}\),
\(m_{1,1}^{[1,2,m_1,m_2]}\), and
\(m_{1,1}^{[2,2,m_1,m_2]}\)
are given in \eqref{LLL}, while
\(m_{1,1}^{[3,3,m_1,m_2]}\),
\(m_{1,1}^{[3,4,m_1,m_2]}\), and
\(m_{1,1}^{[4,4,m_1,m_2]}\)
are given in \eqref{BBB}.

\noindent\textbf{Lumps and breathers.}
The upper rows of Figs.~5 and 6 show a lump pair interacting
with one and two breather pairs, respectively.
The lump parameters are given in \eqref{NL1}, and the corresponding
breather parameters in \eqref{NB1} and \eqref{NBB1}.
The upper row of Fig.~7 shows the solution \(S_1^{[2,1,0]}\)
for the parameters in \eqref{NLL1} and \eqref{NB1}.
The localized lump profiles intersect the oscillatory breather bands.

\noindent\textbf{Line rogue waves and breathers.}
The lower rows of Figs.~5 and 6 show transient line rogue waves
on backgrounds containing one and two breather pairs, respectively.
Here the line-rogue-wave parameters are given in \eqref{NL2},
and the corresponding breather parameters in \eqref{NB1}
and \eqref{NBB1}.
With the parameters in \eqref{NLL2} and \eqref{NB1},
the solution \(S_1^{[2,1,0]}\) is shown in the lower row of Fig.~7.
Its elongated wave fronts cross the breather bands at intermediate times.

\begin{figure}[tbp]\centering
      \begin{subfigure}{0.24\textwidth}
          \centering
         \includegraphics[width=\textwidth]{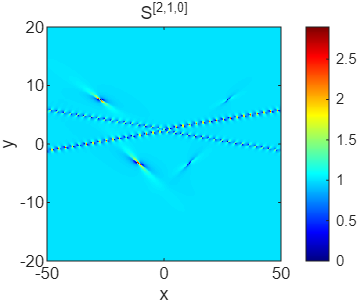}
         \caption*{t=-10}
        \end{subfigure}
      \begin{subfigure}{0.24\textwidth}
          \centering
          \includegraphics[width=\textwidth]{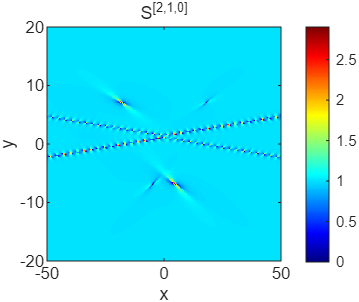}
       \caption*{t=-3}
        \end{subfigure}
      \begin{subfigure}{0.24\textwidth}
          \centering
         \includegraphics[width=\textwidth]{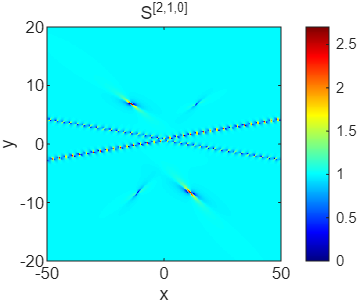}
      \caption*{t=0}
        \end{subfigure}
      \begin{subfigure}{0.24\textwidth}
          \centering
         \includegraphics[width=\textwidth]{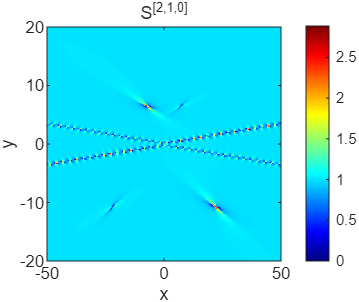}
      \caption*{t=5}
        \end{subfigure}
      \\
       \begin{subfigure}{0.24\textwidth}
          \centering
         \includegraphics[width=\textwidth]{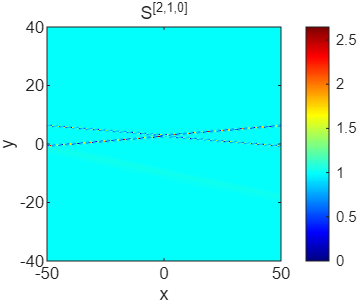}
         \caption*{t=-15}
        \end{subfigure}
      \begin{subfigure}{0.24\textwidth}
          \centering
          \includegraphics[width=\textwidth]{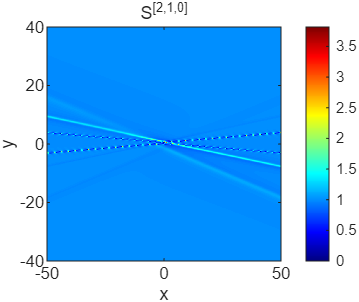}
       \caption*{t=0}
        \end{subfigure}
      \begin{subfigure}{0.24\textwidth}
          \centering
         \includegraphics[width=\textwidth]{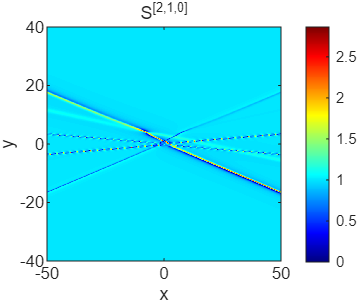}
      \caption*{t=3}
        \end{subfigure}
      \begin{subfigure}{0.24\textwidth}
          \centering
         \includegraphics[width=\textwidth]{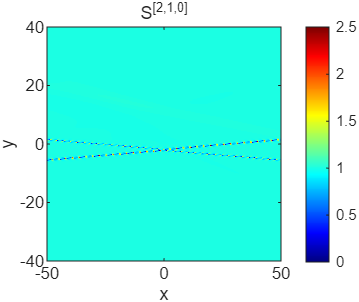}
      \caption*{t=15}
        \end{subfigure}
       \caption{Semi-rational solutions with \((N_1,N_2,N_3)=(2,1,0)\)
and \(n_1=n_2=1\).
Upper row: lumps and a breather pair for the parameters in
\eqref{NLL1} and \eqref{NB1}.
Lower row: line rogue waves and a breather pair for the parameters in
\eqref{NLL2} and \eqref{NB1}.}
\end{figure}

\subsection{Dynamic behaviors of lumps or line rogue waves with backgrounds of dark solitons}
The solutions in Theorem~1 with \(N_2=0\) and \(N_1,N_3>0\)
describe lumps or line rogue waves interacting with
\(N_3\) dark-soliton pairs.
The corresponding tau function is given by the \(4\times4\)
block determinant
\begin{equation}\label{4generalized-tau-Ls}
\renewcommand{\arraystretch}{1.5}
\gamma_{m_1,m_{2}}=\mathrm{det} \left(
\begin{array}{cccc}
\gamma_{m_1,m_{2}}^{[1,1]} & \gamma_{m_1,m_{2}}^{[1,2]} & \gamma_{m_1,m_{2}}^{[1,5]} & \gamma_{m_1,m_{2}}^{[1,6]}\\
\gamma_{-m_1,-m_{2}}^{\,\dag\,[1,2]} & \gamma_{m_1,m_{2}}^{[2,2]} & \gamma_{m_1,m_{2}}^{[2,5]}& \gamma_{m_1,m_{2}}^{[2,6]}\\
\gamma_{-m_1,-m_{2}}^{\,\dag\,[1,5]} & \gamma_{-m_1,-m_{2}}^{\,\dag\,[2,5]} & \gamma_{m_1,m_{2}}^{[5,5]} & \gamma_{m_1,m_{2}}^{[5,6]}\\
\gamma_{-m_1,-m_{2}}^{\,\dag\,[1,6]} & \gamma_{-m_1,-m_{2}}^{\,\dag\,[2,6]} & \gamma_{-m_1,-m_{2}}^{\,\dag\,[5,6]} & \gamma_{m_1,m_{2}}^{[6,6]}\\
\end{array}
\right).
\end{equation}
The fundamental semi-rational solutions with
\(N_1=N_3=1\) take the form
\begin{equation}\label{4L-soliton-1}
\begin{aligned}
S_{1}^{[1,0,1]}= \rho_{1}e^{\mathrm{i}[k_{1}x+\omega_{1}y+(h-\omega_{1} + k_{1}^{2}) t]}\frac{g_{1}^{[1,0,1]}}{f^{[1,0,1]}} ,
\end{aligned}
\end{equation}
Here \(f^{[1,0,1]}=\gamma_{0,0}\) and
\(g_1^{[1,0,1]}=\gamma_{1,0}\), where
\(\gamma_{m_1,m_2}\) is defined by \eqref{4generalized-tau-Ls}.
The matrix entries are given by
\begin{figure}[tbp]\centering
      \begin{subfigure}{0.24\textwidth}
          \centering
         \includegraphics[width=\textwidth]{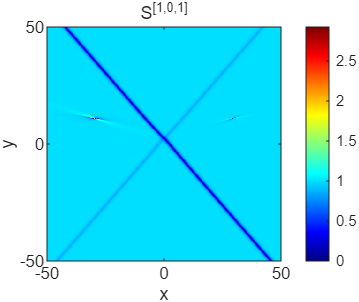}
         \caption*{t=0}
        \end{subfigure}
      \begin{subfigure}{0.24\textwidth}
          \centering
          \includegraphics[width=\textwidth]{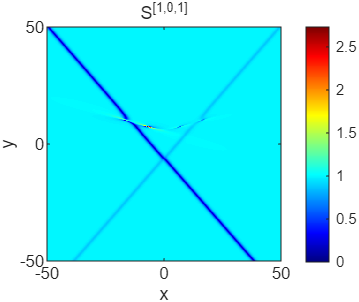}
       \caption*{t=10}
        \end{subfigure}
      \begin{subfigure}{0.24\textwidth}
          \centering
         \includegraphics[width=\textwidth]{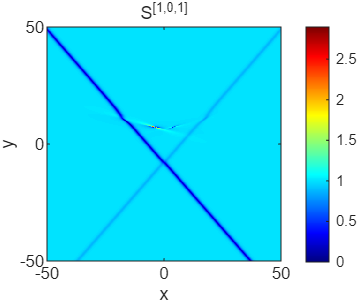}
      \caption*{t=12}
        \end{subfigure}
      \begin{subfigure}{0.24\textwidth}
          \centering
         \includegraphics[width=\textwidth]{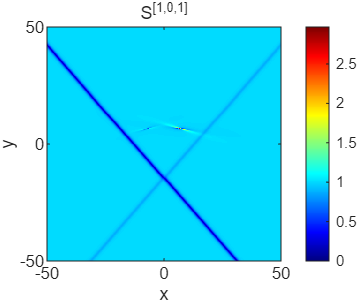}
      \caption*{t=20}
        \end{subfigure}
      \\
       \begin{subfigure}{0.24\textwidth}
          \centering
         \includegraphics[width=\textwidth]{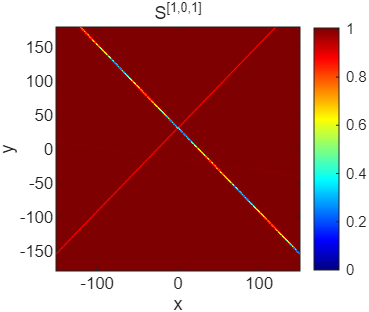}
         \caption*{t=-35}
        \end{subfigure}
      \begin{subfigure}{0.24\textwidth}
          \centering
          \includegraphics[width=\textwidth]{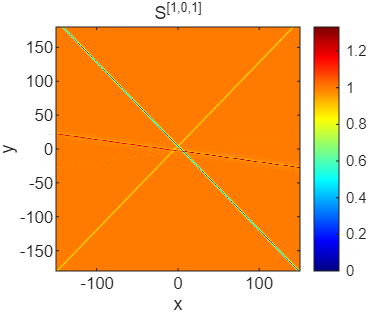}
       \caption*{t=-4}
        \end{subfigure}
      \begin{subfigure}{0.24\textwidth}
          \centering
         \includegraphics[width=\textwidth]{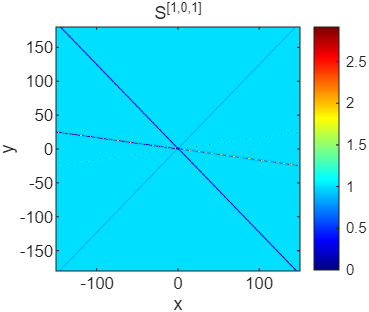}
      \caption*{t=0}
        \end{subfigure}
      \begin{subfigure}{0.24\textwidth}
          \centering
         \includegraphics[width=\textwidth]{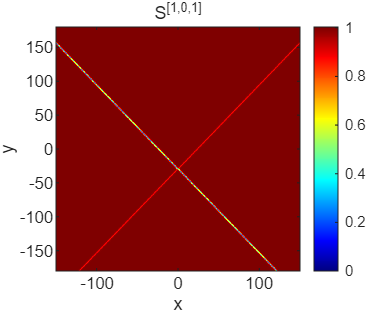}
      \caption*{t=35}
        \end{subfigure}
       \caption{\((N_1,N_2,N_3)=(1,0,1)\). Upper row: lumps and dark solitons with \eqref{NL1}, \eqref{ND1}. Lower row: line rogue waves and dark solitons with \eqref{NL2}, \eqref{ND1}.}
\end{figure} 
\begin{equation}
\label{4LS1}
\begin{aligned}
&m_{1,1}^{[1,5,m_{1},m_{2}]}=\frac{1}{p_{1,1}+p_{1,3}^{\,*}}\left(-\frac{p_{1,1}-\mathrm{i}k_{1}}{p_{1,3}^{\,*}+\mathrm{i}k_{1}}\right)^{m_1}\left(-\frac{p_{1,1}+\mathrm{i}k_{1}}{p_{1,3}^{\,*}-\mathrm{i}k_{1}}\right)^{m_{2}}
e^{\xi_{1}^{*\,[3,3]}} \left(x_{1,m_{1},m_{2}}^{+}-d_{1}^{[1,5]}\right), \\
&m_{1,1}^{[1,6,m_{1},m_{2}]}=\frac{1}{p_{1,1}-p_{1,3}^{\,*}}\left(\frac{p_{1,1}-\mathrm{i}k_{1}}{p_{1,3}^{\,*}-\mathrm{i}k_{1}}\right)^{m_{1}}\left(\frac{p_{1,1}+\mathrm{i}k_{1}}{p_{1,3}^{\,*}+\mathrm{i}k_{1}}\right)^{m_{2}}
e^{\bar{\xi}_{1}^{*\,[3,3]}}\left(x_{1,m_{1},m_{2}}^{+}-d_{1}^{[1,6]}\right),\\
&m_{1,1}^{[2,5,m_{1},m_{2}]}=-\frac{1}{p_{1,1}-p_{1,3}^{\,*}}\left(\frac{p_{1,1}+\mathrm{i}k_{1}}{p_{1,3}^{\,*}+\mathrm{i}k_{1}}\right)^{m_{1}}\left(\frac{p_{1,1}-\mathrm{i}k_{1}}{p_{1,3}^{\,*}-\mathrm{i}k_{1}}\right)^{m_{2}}
e^{\xi_{1}^{*\,[3,3]}}\left(\bar{x}_{1,m_{1},m_{2}}^{+}-d_{1}^{[1,6]}\right),\\
&m_{1,1}^{[2,6,m_{1},m_{2}]}=-\frac{1}{p_{1,1}+p_{1,3}^{\,*}}\left(-\frac{p_{1,1}+\mathrm{i}k_{1}}{p_{1,3}^{\,*}-\mathrm{i}k_{1}}\right)^{m_{1}}\left(-\frac{p_{1,1}-\mathrm{i}k_{1}}{p_{1,3}^{\,*}+\mathrm{i}k_{1}}\right)^{m_{2}}
e^{\bar{\xi}_{1}^{*\,[3,3]}}\left(\bar{x}_{1,m_{1},m_{2}}^{+}-d_{1}^{[1,5]}\right),\\
&d_{1}^{[1,6]}=\frac{p_{1,1}}{p_{1,1}-p_{1,3}^{\,*}},
\quad d_{1}^{[1,5]}=\frac{p_{1,1}}{p_{1,1}+p_{1,3}^{\,*}},
\end{aligned}
\end{equation}
and the entries
\(m_{1,1}^{[1,1,m_1,m_2]}\),
\(m_{1,1}^{[1,2,m_1,m_2]}\), and
\(m_{1,1}^{[2,2,m_1,m_2]}\)
are given in \eqref{LLL}, while
\(m_{1,1}^{[5,5,m_1,m_2]}\),
\(m_{1,1}^{[5,6,m_1,m_2]}\), and
\(m_{1,1}^{[6,6,m_1,m_2]}\)
are given in \eqref{SSS}.

\noindent\textbf{Lumps and dark solitons.}
The upper rows of Figs.~8 and 9 show a lump pair interacting
with one and two dark-soliton pairs, respectively.
Both configurations use the lump parameters in \eqref{NL1},
with the respective dark-soliton parameters given in
\eqref{ND1} and \eqref{NDD1}.
With the parameters in \eqref{NLL1} and \eqref{ND1},
the solution \(S_1^{[2,0,1]}\) is shown in the upper row of Fig.~10.
In these examples, localized lump profiles intersect
the dark-soliton troughs.

\noindent\textbf{Line rogue waves and dark solitons.}
The lower rows of Figs.~8 and 9 show the solutions obtained
by combining the parameters in \eqref{NL2} with those in
\eqref{ND1} and \eqref{NDD1}, respectively.
The line rogue waves grow and then decay on the dark-soliton background.
With the parameters in \eqref{NLL2} and \eqref{ND1},
the solution \(S_1^{[2,0,1]}\) is shown in the lower row of Fig.~10.
At intermediate times, it exhibits elongated wave fronts.

\begin{figure}[tbp]\centering
      \begin{subfigure}{0.24\textwidth}
          \centering
         \includegraphics[width=\textwidth]{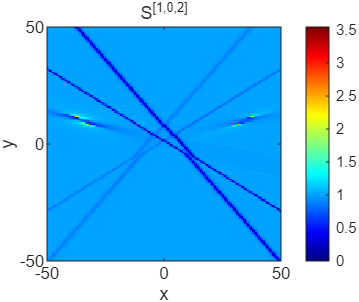}
         \caption*{t=-5}
        \end{subfigure}
      \begin{subfigure}{0.24\textwidth}
          \centering
          \includegraphics[width=\textwidth]{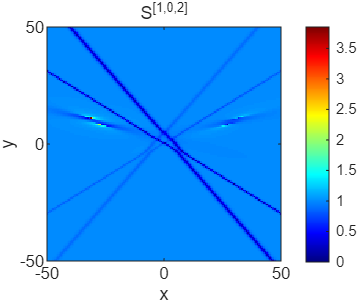}
       \caption*{t=-1}
        \end{subfigure}
      \begin{subfigure}{0.24\textwidth}
          \centering
         \includegraphics[width=\textwidth]{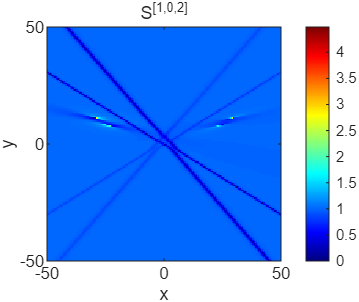}
      \caption*{t=1}
        \end{subfigure}
      \begin{subfigure}{0.24\textwidth}
          \centering
         \includegraphics[width=\textwidth]{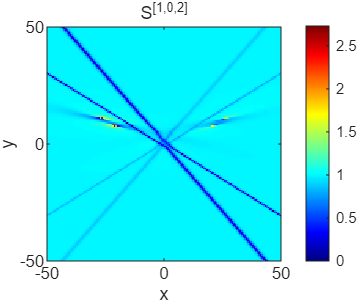}
      \caption*{t=3}
        \end{subfigure}
      \\
       \begin{subfigure}{0.24\textwidth}
          \centering
         \includegraphics[width=\textwidth]{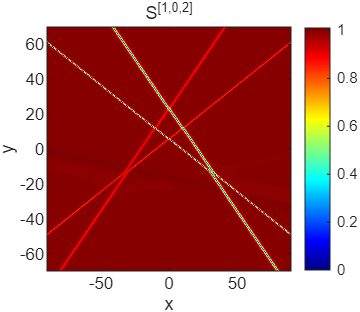}
         \caption*{t=-25}
        \end{subfigure}
      \begin{subfigure}{0.24\textwidth}
          \centering
          \includegraphics[width=\textwidth]{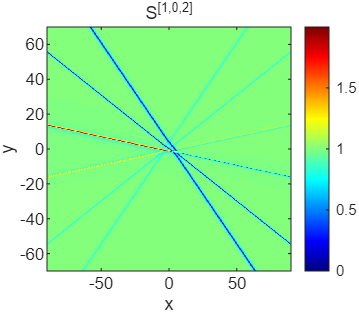}
       \caption*{t=-2}
        \end{subfigure}
      \begin{subfigure}{0.24\textwidth}
          \centering
         \includegraphics[width=\textwidth]{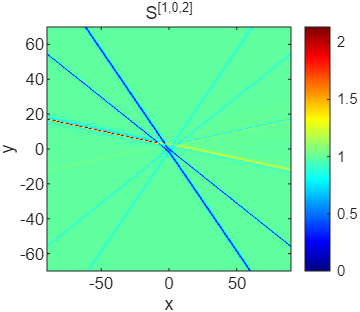}
      \caption*{t=0}
        \end{subfigure}
      \begin{subfigure}{0.24\textwidth}
          \centering
         \includegraphics[width=\textwidth]{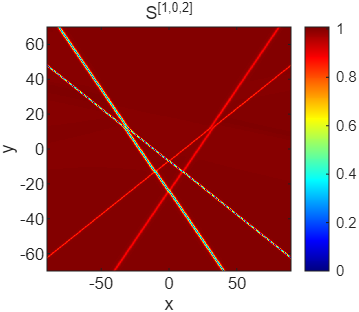}
      \caption*{t=30}
        \end{subfigure}
       \caption{\((N_1,N_2,N_3)=(1,0,2)\). Upper row: lumps and two dark-soliton pairs with \eqref{NL1}, \eqref{NDD1}. Lower row: line rogue waves and two dark-soliton pairs with \eqref{NL2}, \eqref{NDD1}.}
\end{figure}

\subsection{Dynamic behaviors of lumps or line rogue waves with backgrounds of breathers and dark solitons}
When \(N_1,N_2,N_3>0\), the full \(6\times6\) block determinant
\eqref{12d-M-semi-rational-solutions-1-3} yields lumps or
line rogue waves on a mixed background of breathers and dark solitons.
The fundamental semi-rational solutions with
\(N_1=N_2=N_3=1\) take the form
\begin{equation}\label{6L-soliton-1}
\begin{aligned}
S_{1}^{[1,1,1]}= \rho_{1}e^{\mathrm{i}[k_{1}x+\omega_{1}y+(h-\omega_{1} + k_{1}^{2}) t]}\frac{g_{1}^{[1,1,1]}}{f^{[1,1,1]}},
\end{aligned}
\end{equation}
Here \(f^{[1,1,1]}=\gamma_{0,0}\) and
\(g_1^{[1,1,1]}=\gamma_{1,0}\), where
\(\gamma_{m_1,m_2}\) is given by
\eqref{12d-M-semi-rational-solutions-1-3}
with \(N_1=N_2=N_3=1\).
The matrix entries are given by
\begin{figure}[tbp]\centering
      \begin{subfigure}{0.24\textwidth}
          \centering
         \includegraphics[width=\textwidth]{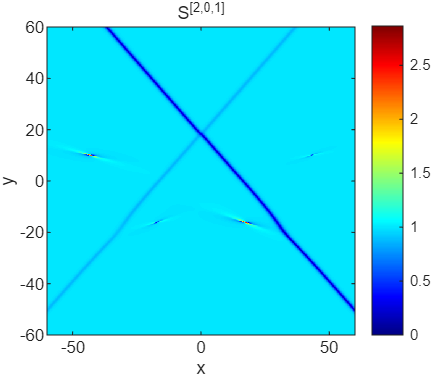}
         \caption*{t=-15}
        \end{subfigure}
      \begin{subfigure}{0.24\textwidth}
          \centering
          \includegraphics[width=\textwidth]{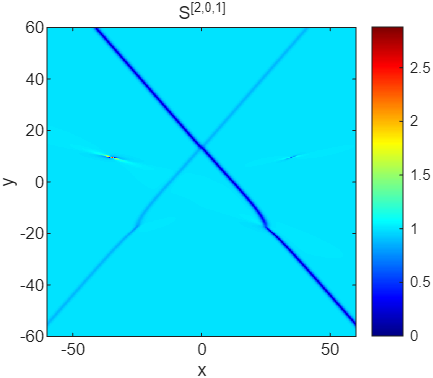}
       \caption*{t=-9}
        \end{subfigure}
      \begin{subfigure}{0.24\textwidth}
          \centering
         \includegraphics[width=\textwidth]{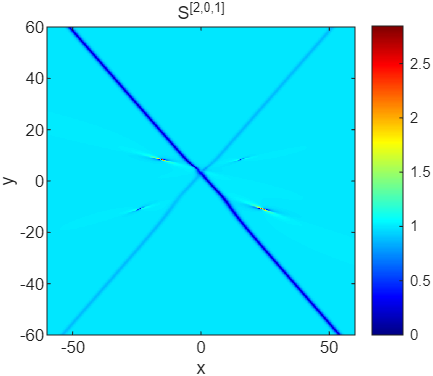}
      \caption*{t=5}
        \end{subfigure}
      \begin{subfigure}{0.24\textwidth}
          \centering
         \includegraphics[width=\textwidth]{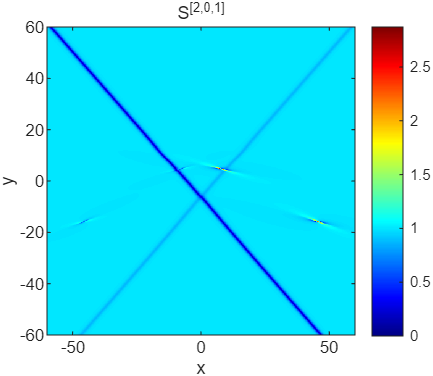}
      \caption*{t=15}
        \end{subfigure}
      \\
       \begin{subfigure}{0.24\textwidth}
          \centering
         \includegraphics[width=\textwidth]{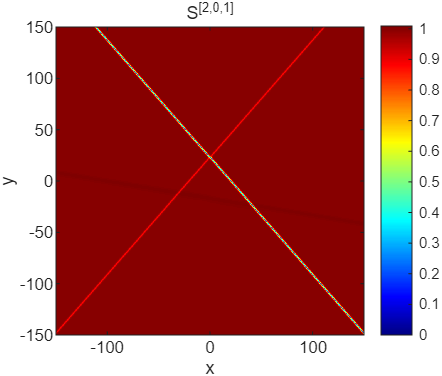}
         \caption*{t=-25}
        \end{subfigure}
      \begin{subfigure}{0.24\textwidth}
          \centering
          \includegraphics[width=\textwidth]{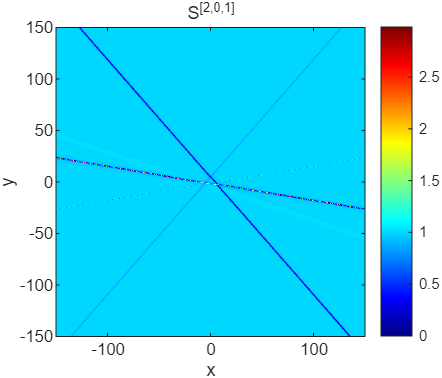}
       \caption*{t=-3}
        \end{subfigure}
      \begin{subfigure}{0.24\textwidth}
          \centering
         \includegraphics[width=\textwidth]{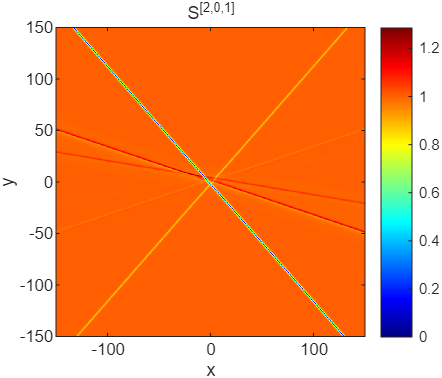}
      \caption*{t=5}
        \end{subfigure}
      \begin{subfigure}{0.24\textwidth}
          \centering
         \includegraphics[width=\textwidth]{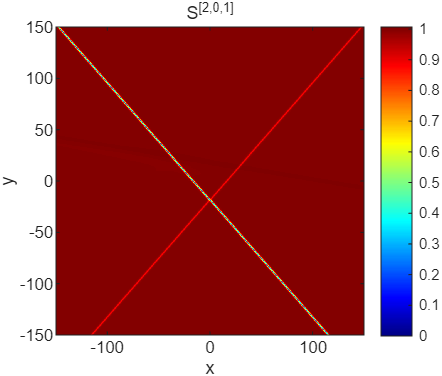}
      \caption*{t=25}
        \end{subfigure}
       \caption{Semi-rational solutions with one dark-soliton pair, \((N_1,N_2,N_3)=(2,0,1)\), and \(n_1=n_2=1\). Upper row: lump parameters \eqref{NLL1}, \eqref{ND1}. Lower row: line-rogue-wave spectral parameters \eqref{NLL2}, \eqref{ND1}.}
\end{figure}
\begin{equation}
\begin{aligned}
&m_{1,1}^{[3,5,m_{1},m_{2}]}=\sum_{\mu=1}^{2}
\frac{1}{p_{1,2}^{[\mu]}+p_{1,3}^{\,*}}\left(-\frac{p_{1,2}^{[\mu]}-\mathrm{i}k_{1}}{p_{1,3}^{\,*}+\mathrm{i}k_{1}}\right)^{m_{1}}\left(-\frac{p_{1,2}^{[\mu]}+\mathrm{i}k_{1}}{p_{1,3}^{\,*}-\mathrm{i}k_{1}}\right)^{m_{2}} e^{\xi_{1}^{[2,2,\mu]}+\xi_{1}^{*\,[3,3]}},
\\
&m_{1,1}^{[3,6,m_{1},m_{2}]}=\sum_{\mu=1}^{2}
\frac{1}{p_{1,2}^{[\mu]}-p_{1,3}^{\,*}}\left(\frac{p_{1,2}^{[\mu]}-\mathrm{i}k_{1}}{p_{1,3}^{\,*}-\mathrm{i}k_{1}}\right)^{m_{1}}\left(\frac{p_{1,2}^{[\mu]}+\mathrm{i}k_{1}}{p_{1,3}^{\,*}+\mathrm{i}k_{1}}\right)^{m_{2}} e^{\xi_{1}^{[2,2,\mu]}+\bar{\xi}_{1}^{*\,[3,3]}},
\\
&m_{1,1}^{[4,5,m_{1},m_{2}]}=-\sum_{\mu=1}^{2}
\frac{1}{p_{1,2}^{[\mu]}-p_{1,3}^{\,*}}\left(\frac{p_{1,2}^{[\mu]}+\mathrm{i}k_{1}}{p_{1,3}^{\,*}+\mathrm{i}k_{1}}\right)^{m_{1}}\left(\frac{p_{1,2}^{[\mu]}-\mathrm{i}k_{1}}{p_{1,3}^{\,*}-\mathrm{i}k_{1}}\right)^{m_{2}}e^{\bar{\xi}_{1}^{[2,2,\mu]}+\xi_{1}^{*\,[3,3]}},\\
&m_{1,1}^{[4,6,m_{1},m_{2}]}=-\sum_{\mu=1}^{2}
\frac{1}{p_{1,2}^{[\mu]}+p_{1,3}^{\,*}}\left(-\frac{p_{1,2}^{[\mu]}+\mathrm{i}k_{1}}{p_{1,3}^{\,*}-\mathrm{i}k_{1}}\right)^{m_{1}}\left(-\frac{p_{1,2}^{[\mu]}-\mathrm{i}k_{1}}{p_{1,3}^{\,*}+\mathrm{i}k_{1}}\right)^{m_{2}}e^{\bar{\xi}_{1}^{[2,2,\mu]}+\bar{\xi}_{1}^{*\,[3,3]}}.
\end{aligned}
\end{equation}
The remaining matrix entries are given in
\eqref{LLL}, \eqref{BBB}, \eqref{SSS}, \eqref{5LS1}, and \eqref{4LS1}.

\noindent\textbf{Lumps with breathers and dark solitons.}
The upper row of Fig.~11 shows the solution \(S_1^{[1,1,1]}\)
for the parameters in \eqref{NL1}, \eqref{NB1}, and \eqref{ND1}.
The localized lump profiles interact with the oscillatory
breather bands and dark-soliton troughs.

\noindent\textbf{Line rogue waves with breathers and dark solitons.}
The lower row of Fig.~11 shows the solution \(S_1^{[1,1,1]}\)
for the parameters in \eqref{NL2}, \eqref{NB1}, and \eqref{ND1}.
The line rogue waves grow and then decay on the mixed background.
In both rows, the background consists of one breather pair
and one dark-soliton pair.
\begin{figure}[tbp]\centering
      \begin{subfigure}{0.24\textwidth}
          \centering
         \includegraphics[width=\textwidth]{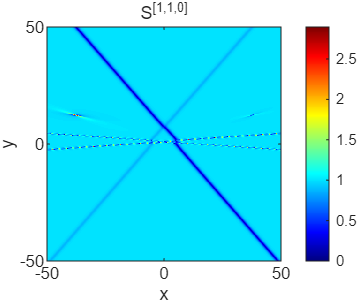}
         \caption*{t=-5}
        \end{subfigure}
      \begin{subfigure}{0.24\textwidth}
          \centering
          \includegraphics[width=\textwidth]{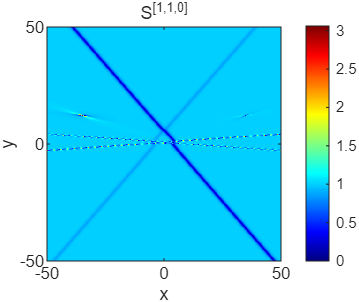}
       \caption*{t=-3}
        \end{subfigure}
      \begin{subfigure}{0.24\textwidth}
          \centering
         \includegraphics[width=\textwidth]{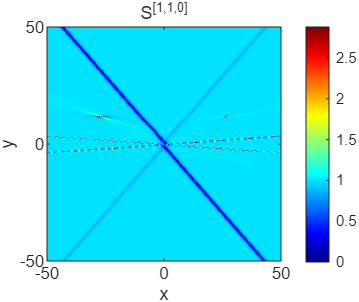}
      \caption*{t=3}
        \end{subfigure}
      \begin{subfigure}{0.24\textwidth}
          \centering
         \includegraphics[width=\textwidth]{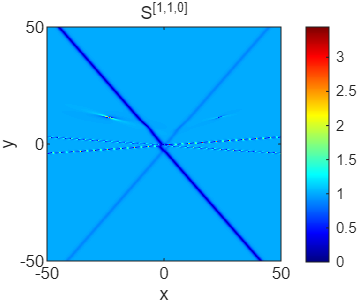}
      \caption*{t=5}
        \end{subfigure}
      \\
       \begin{subfigure}{0.24\textwidth}
          \centering
         \includegraphics[width=\textwidth]{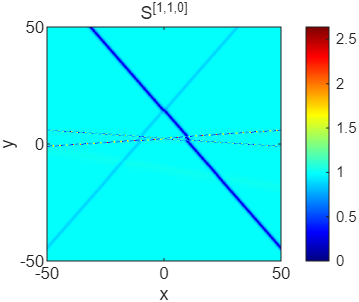}
         \caption*{t=-15}
        \end{subfigure}
      \begin{subfigure}{0.24\textwidth}
          \centering
          \includegraphics[width=\textwidth]{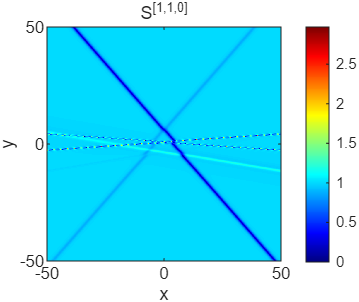}
       \caption*{t=-5}
        \end{subfigure}
      \begin{subfigure}{0.24\textwidth}
          \centering
         \includegraphics[width=\textwidth]{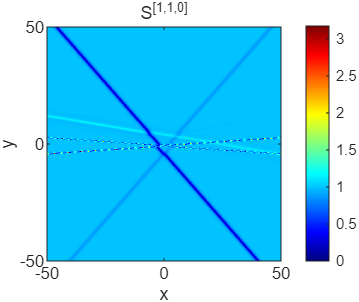}
      \caption*{t=5}
        \end{subfigure}
      \begin{subfigure}{0.24\textwidth}
          \centering
         \includegraphics[width=\textwidth]{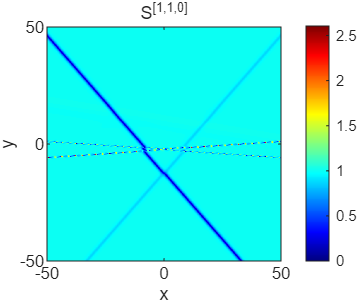}
      \caption*{t=15}
        \end{subfigure}
       \caption{Semi-rational solutions with \((N_1,N_2,N_3)=(1,1,1)\).
Upper row: lumps, breathers, and dark solitons for the parameters
in \eqref{NL1}, \eqref{NB1}, and \eqref{ND1}.
Lower row: line rogue waves, breathers, and dark solitons
for the parameters in \eqref{NL2}, \eqref{NB1}, and \eqref{ND1}.}
\end{figure}

\section{Derivation of the higher-order semi-rational solutions}
The proof of Theorem~1 is based on the KP hierarchy reduction
and Schur-polynomial construction in \cite{LTZ-2026}.
We verify the complex-conjugation and spatial-reflection reductions
for the rational, breather, dark-soliton, and mixed blocks.

\subsection{Bilinear forms and tau functions}
Under the transformation
\eqref{12d-M-semi-rational-solutions-1-1}, extended to
\(\ell=1,\ldots,2M\), the local \(2M\)-component YO system
\eqref{M-1} takes the bilinear form:
\begin{equation}
\label{2d-twris-b-2}
\begin{aligned}
&\left[\mathrm{i}(D_t+D_y)-D_x^2-2\mathrm{i}k_\ell D_x\right]
g_\ell\cdot f=0,\\
&D_t D_x f \cdot f - 2 \sum_{\ell=1}^{2M} \sigma_\ell \rho_\ell^2 f^2 + 2 \sum_{\ell=1}^{2M} \sigma_\ell \rho_\ell^2 g_{\ell} g_{\ell}^{*} = 0,
\end{aligned}
\end{equation}
Here \(f\) is real, \(g_\ell\) is complex, and \(D\) denotes
the Hirota bilinear operator \cite{H-2004}.
We use the change of variables
\begin{equation}
x_{1}=x,
\quad
x_{2}=-\mathrm{i} y,
\quad
r_{\ell}=\sigma_{\ell} \rho_{\ell}^2 (t-y),
\quad
z_{\ell}=\mathrm{i} k_{\ell},
\end{equation}
to reduce the following KP-hierarchy bilinear equations to
\eqref{2d-twris-b-2}:
\begin{equation}\label{2d-twris-b-4}
\begin{aligned}
&(D_{x_2}-D_{x_1}^{2}-2\mathrm{i}k_{\ell}D_{x_1})
\tilde{\gamma}_{\mathbf{m}+\mathbf{e}_{\ell}}
\cdot
\tilde{\gamma}_{\mathbf{m}}
=0,
\\
&\left(\frac{1}{2}D_{x_1}D_{r_{\ell}}-1\right)
\tilde{\gamma}_{\mathbf{m}}
\cdot
\tilde{\gamma}_{\mathbf{m}}
=
-\tilde{\gamma}_{\mathbf{m}+\mathbf{e}_{\ell}}
\tilde{\gamma}_{\mathbf{m}-\mathbf{e}_{\ell}},
\qquad
\ell=1,\ldots,2M.
\end{aligned}
\end{equation}
Here \(\mathbf{m}=(m_1,\ldots,m_{2M})\in\mathbb{Z}^{2M}\),
and \(\mathbf{e}_\ell\) denotes the \(\ell\)-th standard unit vector.
The change of variables gives
\(\partial_t+\partial_y=-\mathrm{i}\partial_{x_2}\) and
\(\partial_t=\sum_{\ell=1}^{2M}\sigma_\ell\rho_\ell^2
\partial_{r_\ell}\).
Summing the second equation in \eqref{2d-twris-b-4}
with weights \(2\sigma_\ell\rho_\ell^2\) yields
the long-wave bilinear equation in \eqref{2d-twris-b-2}.

The bilinear equations \eqref{2d-twris-b-4} admit
Gram-determinant solutions \cite{CCFM-2015,YO3,YY2021,YY2022}
of the form
\begin{equation}\label{2-tau-1}
\tilde{\gamma}_{\mathbf{m}}=\underset{1 \leq i,j \leq N}{\mathrm{det}}\left(\tilde{\gamma}_{i,j}^{[\mathbf{m}]}\right),
\end{equation}
where the matrix elements
\(\tilde{\gamma}_{i,j}^{[\mathbf m]}\)
satisfy the difference and differential relations given in \cite{CCFM-2015,YO3}. The tau functions are constrained by:
\begin{equation}
f=\tilde{\gamma}_{\mathbf{0}},\qquad
g_{\ell}=\tilde{\gamma}_{\mathbf{e}_{\ell}},\qquad
g_{\ell}^{*}=\tilde{\gamma}_{-\mathbf{e}_{\ell}}.
\end{equation}
Different choices of the matrix elements and determinant order
yield various families of solutions to the local
\(2M\)-component YO system.

The nonlocal YO system \eqref{M-2} is obtained from \eqref{M-1}
through the Yang-type spatial-reflection reduction
\eqref{nonlocal-conditions}.
With the parameter relations
\begin{equation}
\sigma_{\ell}=-\sigma_{M+\ell}, \qquad \rho_{\ell}=\rho_{M+\ell}, \qquad k_{\ell}=-k_{M+\ell},  \qquad \omega_{\ell}=\omega_{M+\ell}
\end{equation}
the short-wave reflection constraint becomes 
\begin{equation}\label{nonlocal-constraint}
\frac{g_{\ell}(-x,y,t)}{f(-x,y,t)}
=
\frac{g_{M+\ell}(x,y,t)}{f(x,y,t)}
\quad\Longleftrightarrow\quad
\frac{\tilde{\gamma}_{\mathbf{e}_{\ell}}(-x,y,t)}
{\tilde{\gamma}_{\mathbf{0}}(-x,y,t)}
=
\frac{\tilde{\gamma}_{\mathbf{e}_{M+\ell}}(x,y,t)}
{\tilde{\gamma}_{\mathbf{0}}(x,y,t)},
\qquad
\ell=1,\ldots,M.
\end{equation}

\subsection{Block determinant tau functions}
Following \cite{Z1,Z2}, we construct the tau functions
as a \(6\times6\) block Gram determinant with the block sizes
specified in Theorem~1:
\begin{equation}\label{tau-4-1}
\tilde\gamma_{\mathbf m}
=\det\!\left(\tilde\gamma_{\mathbf m}^{[I,J]}\right)_{I,J=1}^{6},
\end{equation}
where the block matrices $\tilde{\gamma}_{\mathbf{m}}^{[I,J]}$ for $I,J=1,2,3,4,5,6$ are specified as:
\begin{equation}\label{tau-ij-4-1}
\tilde{\gamma}_{\mathbf{m}}^{[I,J]}=\mathrm{mat}_{1\leq i\leq K_I,\;1\leq j\leq K_J}
\left(\tilde{m}_{i,j}^{[I,J,\mathbf{m}]}\right).
\end{equation}
For the rational blocks \(I,J=1,2\), we use the positive integers
\(n_i\) specified in Theorem~1.
We introduce the coefficient vectors
\(\mathbf a_i^{[I]}=(a_{i,r}^{[I]})_{r\ge1}\) and
\(\mathbf b_j^{[J]}=(b_{j,r}^{[J]})_{r\ge1}\).
The Schur polynomials are defined by the generating function
\(\sum_{n\ge0}S_n(\mathbf a)\kappa^n
=\exp(\sum_{r\ge1}a_r\kappa^r)\).
The matrix entries are given by
\begin{small}
\begin{align*}
&\tilde{m}_{i,j}^{[I,J,\mathbf{m}]}=\mathcal{A}_{i}^{[I]}\mathcal{B}_{j}^{[J]}\frac{1}{p+q}\prod_{\ell=1}^{2M}\left(-\frac{p-\mathrm{i}k_{\ell}}{q+\mathrm{i}k_{\ell}}\right)^{m_{\ell}}
e^{\xi^{[I,I]}+\eta^{[J,J]}}\Big|^{p=p_{i,1}^{[I]}}_{q=q_{j,1}^{[J]}},
\\
&\mathcal A_i^{[I]}=\sum_{s=0}^{n_i}
S_{n_i-s}(\mathbf a_i^{[I]})\frac{(p\partial_p)^s}{s!},\qquad
\mathcal B_j^{[J]}=\sum_{s=0}^{n_j}
S_{n_j-s}(\mathbf b_j^{[J]})\frac{(q\partial_q)^s}{s!},\\
&\tilde{m}_{i,j}^{[I,3,\mathbf{m}]}=\mathcal{A}_{i}^{[I]}\sum_{\nu=1}^{2}\frac{1}{p+q_{j,2}^{[\nu]}}\prod_{\ell=1}^{2M}\left(-\frac{p-\mathrm{i}k_{\ell}}{q_{j,2}^{[\nu]}+\mathrm{i}k_{\ell}}\right)^{m_{\ell}}
e^{\xi^{[I,I]}+\eta_{j}^{[2,2,\nu]}}\Big|_{p=p_{i,1}^{[I]}},\\
&\tilde{m}_{i,j}^{[I,5,\mathbf{m}]}=\mathcal{A}_{i}^{[I]}\frac{1}{p+q_{j,3}}\prod_{\ell=1}^{2M}\left(-\frac{p-\mathrm{i}k_{\ell}}{q_{j,3}+\mathrm{i}k_{\ell}}\right)^{m_{\ell}}
e^{\xi^{[I,I]}+\eta_{j}^{[3,3]}}\Big|_{p=p_{i,1}^{[I]}},\\
&\tilde{m}_{i,j}^{[I,4,\mathbf{m}]}=\mathcal{A}_{i}^{[I]}\sum_{\nu=1}^{2}\frac{1}{p+\bar{q}_{j,2}^{[\nu]}}\prod_{\ell=1}^{2M}\left(-\frac{p-\mathrm{i}k_{\ell}}{\bar{q}_{j,2}^{[\nu]}+\mathrm{i}k_{\ell}}\right)^{m_{\ell}}
e^{\xi^{[I,I]}+\bar{\eta}_{j}^{[2,2,\nu]}}\Big|_{p=p_{i,1}^{[I]}},\\
&\tilde{m}_{i,j}^{[I,6,\mathbf{m}]}=\mathcal{A}_{i}^{[I]}\frac{1}{p+\bar{q}_{j,3}}\prod_{\ell=1}^{2M}\left(-\frac{p-\mathrm{i}k_{\ell}}{\bar{q}_{j,3}+\mathrm{i}k_{\ell}}\right)^{m_{\ell}}
e^{\xi^{[I,I]}+\bar{\eta}_{j}^{[3,3]}}\Big|_{p=p_{i,1}^{[I]}},\\
&\tilde{m}_{i,j}^{[3,J,\mathbf{m}]}=\mathcal{B}_{j}^{[J]}\sum_{\mu=1}^{2}\frac{1}{p_{i,2}^{[\mu]}+q}\prod_{\ell=1}^{2M}\left(-\frac{p_{i,2}^{[\mu]}-\mathrm{i}k_{\ell}}{q+\mathrm{i}k_{\ell}}\right)^{m_{\ell}}
e^{\xi_{i}^{[2,2,\mu]}+\eta^{[J,J]}}\Big|_{q=q_{j,1}^{[J]}},\\
&\tilde{m}_{i,j}^{[5,J,\mathbf{m}]}=\mathcal{B}_{j}^{[J]}\frac{1}{p_{i,3}+q}\prod_{\ell=1}^{2M}\left(-\frac{p_{i,3}-\mathrm{i}k_{\ell}}{q+\mathrm{i}k_{\ell}}\right)^{m_{\ell}}
e^{\xi_{i}^{[3,3]}+\eta^{[J,J]}}\Big|_{q=q_{j,1}^{[J]}},\\
&\tilde{m}_{i,j}^{[4,J,\mathbf{m}]}=\mathcal{B}_{j}^{[J]}\sum_{\mu=1}^{2}\frac{1}{\bar{p}_{i,2}^{[\mu]}+q}\prod_{\ell=1}^{2M}\left(-\frac{\bar{p}_{i,2}^{[\mu]}-\mathrm{i}k_{\ell}}{q+\mathrm{i}k_{\ell}}\right)^{m_{\ell}}
e^{\bar{\xi}_{i}^{[2,2,\mu]}+\eta^{[J,J]}}\Big|_{q=q_{j,1}^{[J]}},\\
&\tilde{m}_{i,j}^{[6,J,\mathbf{m}]}=\mathcal{B}_{j}^{[J]}\frac{1}{\bar{p}_{i,3}+q}\prod_{\ell=1}^{2M}\left(-\frac{\bar{p}_{i,3}-\mathrm{i}k_{\ell}}{q+\mathrm{i}k_{\ell}}\right)^{m_{\ell}}
e^{\bar{\xi}_{i}^{[3,3]}+\eta^{[J,J]}}\Big|_{q=q_{j,1}^{[J]}}.
\end{align*}
\end{small}
For \(I=3,4\) and \(J=3,4,5,6\), the matrix elements
\(\tilde m_{i,j}^{[I,J,\mathbf m]}\) are given by:
\begin{align*}
&\tilde{m}_{i,j}^{[3,3,\mathbf{m}]}=\sum_{\mu=1}^{2}\sum_{\nu=1}^{2}
\frac{1}{p_{i,2}^{[\mu]}+q_{j,2}^{[\nu]}}\prod_{\ell=1}^{2M}\left(-\frac{p_{i,2}^{[\mu]}-\mathrm{i}k_{\ell}}{q_{j,2}^{[\nu]}+\mathrm{i}k_{\ell}}\right)^{m_{\ell}}
e^{\xi_{i}^{[2,2,\mu]}+\eta_{j}^{[2,2,\nu]}},\\
&\tilde{m}_{i,j}^{[3,5,\mathbf{m}]}=\sum_{\mu=1}^{2}\frac{1}{p_{i,2}^{[\mu]}+q_{j,3}}\prod_{\ell=1}^{2M}\left(-\frac{p_{i,2}^{[\mu]}-\mathrm{i}k_{\ell}}{q_{j,3}+\mathrm{i}k_{\ell}}\right)^{m_{\ell}}
e^{\xi_{i}^{[2,2,\mu]}+\eta_{j}^{[3,3]}},\\
&\tilde{m}_{i,j}^{[3,4,\mathbf{m}]}=\sum_{\mu=1}^{2}\sum_{\nu=1}^{2}
\frac{1}{p_{i,2}^{[\mu]}+\bar{q}_{j,2}^{[\nu]}}\prod_{\ell=1}^{2M}\left(-\frac{p_{i,2}^{[\mu]}-\mathrm{i}k_{\ell}}{\bar{q}_{j,2}^{[\nu]}+\mathrm{i}k_{\ell}}\right)^{m_{\ell}}
e^{\xi_{i}^{[2,2,\mu]}+\bar{\eta}_{j}^{[2,2,\nu]}},\\
&\tilde{m}_{i,j}^{[3,6,\mathbf{m}]}=\sum_{\mu=1}^{2}\frac{1}{p_{i,2}^{[\mu]}+\bar{q}_{j,3}}\prod_{\ell=1}^{2M}\left(-\frac{p_{i,2}^{[\mu]}-\mathrm{i}k_{\ell}}{\bar{q}_{j,3}+\mathrm{i}k_{\ell}}\right)^{m_{\ell}}
e^{\xi_{i}^{[2,2,\mu]}+\bar{\eta}_{j}^{[3,3]}},\\
&\tilde{m}_{i,j}^{[4,3,\mathbf{m}]}=\sum_{\mu=1}^{2}\sum_{\nu=1}^{2}
\frac{1}{\bar{p}_{i,2}^{[\mu]}+q_{j,2}^{[\nu]}}\prod_{\ell=1}^{2M}\left(-\frac{\bar{p}_{i,2}^{[\mu]}-\mathrm{i}k_{\ell}}{q_{j,2}^{[\nu]}+\mathrm{i}k_{\ell}}\right)^{m_{\ell}}
e^{\bar{\xi}_{i}^{[2,2,\mu]}+\eta_{j}^{[2,2,\nu]}},\\
&\tilde{m}_{i,j}^{[4,5,\mathbf{m}]}=\sum_{\mu=1}^{2}\frac{1}{\bar{p}_{i,2}^{[\mu]}+q_{j,3}}\prod_{\ell=1}^{2M}\left(-\frac{\bar{p}_{i,2}^{[\mu]}-\mathrm{i}k_{\ell}}{q_{j,3}+\mathrm{i}k_{\ell}}\right)^{m_{\ell}}
e^{\bar{\xi}_{i}^{[2,2,\mu]}+\eta_{j}^{[3,3]}},\\
&\tilde{m}_{i,j}^{[4,4,\mathbf{m}]}=\sum_{\mu=1}^{2}\sum_{\nu=1}^{2}
\frac{1}{\bar{p}_{i,2}^{[\mu]}+\bar{q}_{j,2}^{[\nu]}}\prod_{\ell=1}^{2M}\left(-\frac{\bar{p}_{i,2}^{[\mu]}-\mathrm{i}k_{\ell}}{\bar{q}_{j,2}^{[\nu]}+\mathrm{i}k_{\ell}}\right)^{m_{\ell}}
e^{\bar{\xi}_{i}^{[2,2,\mu]}+\bar{\eta}_{j}^{[2,2,\nu]}},\\
&\tilde{m}_{i,j}^{[4,6,\mathbf{m}]}=\sum_{\mu=1}^{2}\frac{1}{\bar{p}_{i,2}^{[\mu]}+\bar{q}_{j,3}}\prod_{\ell=1}^{2M}\left(-\frac{\bar{p}_{i,2}^{[\mu]}-\mathrm{i}k_{\ell}}{\bar{q}_{j,3}+\mathrm{i}k_{\ell}}\right)^{m_{\ell}}
e^{\bar{\xi}_{i}^{[2,2,\mu]}+\bar{\eta}_{j}^{[3,3]}}.
\end{align*}
For \(I=5,6\) and \(J=3,4,5,6\), the matrix elements
\(\tilde m_{i,j}^{[I,J,\mathbf m]}\) are defined by:
\begin{align*}
&\tilde{m}_{i,j}^{[5,3,\mathbf{m}]}=\sum_{\nu=1}^{2}\frac{1}{p_{i,3}+q_{j,2}^{[\nu]}}\prod_{\ell=1}^{2M}\left(-\frac{p_{i,3}-\mathrm{i}k_{\ell}}{q_{j,2}^{[\nu]}+\mathrm{i}k_{\ell}}\right)^{m_{\ell}}
e^{\xi_{i}^{[3,3]}+\eta_{j}^{[2,2,\nu]}},\\
&\tilde{m}_{i,j}^{[5,4,\mathbf{m}]}=\sum_{\nu=1}^{2}\frac{1}{p_{i,3}+\bar{q}_{j,2}^{[\nu]}}\prod_{\ell=1}^{2M}\left(-\frac{p_{i,3}-\mathrm{i}k_{\ell}}{\bar{q}_{j,2}^{[\nu]}+\mathrm{i}k_{\ell}}\right)^{m_{\ell}}
e^{\xi_{i}^{[3,3]}+\bar{\eta}_{j}^{[2,2,\nu]}},\\
&\tilde{m}_{i,j}^{[5,5,\mathbf{m}]}=\delta_{i,j}+\frac{1}{p_{i,3}+q_{j,3}}\prod_{\ell=1}^{2M}\left(-\frac{p_{i,3}-\mathrm{i}k_{\ell}}{q_{j,3}+\mathrm{i}k_{\ell}}\right)^{m_{\ell}}
e^{\xi_{i}^{[3,3]}+\eta_{j}^{[3,3]}},\\
&\tilde{m}_{i,j}^{[5,6,\mathbf{m}]}=\frac{1}{p_{i,3}+\bar{q}_{j,3}}\prod_{\ell=1}^{2M}\left(-\frac{p_{i,3}-\mathrm{i}k_{\ell}}{\bar{q}_{j,3}+\mathrm{i}k_{\ell}}\right)^{m_{\ell}}
e^{\xi_{i}^{[3,3]}+\bar{\eta}_{j}^{[3,3]}},\\
&\tilde{m}_{i,j}^{[6,3,\mathbf{m}]}=\sum_{\nu=1}^{2}\frac{1}{\bar{p}_{i,3}+q_{j,2}^{[\nu]}}\prod_{\ell=1}^{2M}\left(-\frac{\bar{p}_{i,3}-\mathrm{i}k_{\ell}}{q_{j,2}^{[\nu]}+\mathrm{i}k_{\ell}}\right)^{m_{\ell}}
e^{\bar{\xi}_{i}^{[3,3]}+\eta_{j}^{[2,2,\nu]}},\\
&\tilde{m}_{i,j}^{[6,4,\mathbf{m}]}=\sum_{\nu=1}^{2}\frac{1}{\bar{p}_{i,3}+\bar{q}_{j,2}^{[\nu]}}\prod_{\ell=1}^{2M}\left(-\frac{\bar{p}_{i,3}- \mathrm{i}k_{\ell}}{\bar{q}_{j,2}^{[\nu]} + \mathrm{i}k_{\ell}}\right)^{m_{\ell}}
e^{\bar{\xi}_{i}^{[3,3]}+\bar{\eta}_{j}^{[2,2,\nu]}},\\
&\tilde{m}_{i,j}^{[6,5,\mathbf{m}]}=\frac{1}{\bar{p}_{i,3}+q_{j,3}}\prod_{\ell=1}^{2M}\left(-\frac{\bar{p}_{i,3}-\mathrm{i}k_{\ell}}{q_{j,3}+\mathrm{i}k_{\ell}}\right)^{m_{\ell}}
e^{\bar{\xi}_{i}^{[3,3]}+\eta_{j}^{[3,3]}},\\
&\tilde{m}_{i,j}^{[6,6,\mathbf{m}]}=-\delta_{i,j}+\frac{1}{\bar{p}_{i,3}+\bar{q}_{j,3}}\prod_{\ell=1}^{2M}\left(-\frac{\bar{p}_{i,3}-\mathrm{i}k_{\ell}}{\bar{q}_{j,3}+\mathrm{i}k_{\ell}}\right)^{m_{\ell}}
e^{\bar{\xi}_{i}^{[3,3]}+\bar{\eta}_{j}^{[3,3]}}.
\end{align*}
The phase functions $\xi^{[I,I]}$, $\xi_{i}^{[2,2,\mu]}$, $\bar{\xi}_{i}^{[2,2,\mu]}$, $\xi_{i}^{[3,3]}$, $\bar{\xi}_{i}^{[3,3]}$, $\eta^{[J,J]}$, $\eta_{j}^{[2,2,\nu]}$, $\bar{\eta}_{j}^{[2,2,\nu]}$, $\eta_{j}^{[3,3]}$ and $\bar{\eta}_{j}^{[3,3]}$ are specified as follows:
\begin{align*}
&\xi^{[I,I]}=\sum_{\ell=1}^{2M}\frac{1}{p-\mathrm{i}k_{\ell}}r_{\ell}+px_{1}+p^{2}x_{2}+\xi_{0}^{[I,I]},\\
&\eta^{[J,J]}=\sum_{\ell=1}^{2M}\frac{1}{q+\mathrm{i}k_{\ell}}r_{\ell}+qx_{1}-q^{2}x_{2}+\eta_{0}^{[J,J]},\\
&\xi_{i}^{[2,2,\mu]}=\sum_{\ell=1}^{2M}\frac{1}{p_{i,2}^{[\mu]}-\mathrm{i}k_{\ell}}r_{\ell}+p_{i,2}^{[\mu]}x_{1}+p_{i,2}^{[\mu]\,2}x_{2}+\xi_{i,0}^{[2,2,\mu]},\\
&\xi_{i}^{[3,3]}=\sum_{\ell=1}^{2M}\frac{1}{p_{i,3}-\mathrm{i}k_{\ell}}r_{\ell}+p_{i,3}x_{1}+p_{i,3}^{2}x_{2}+\xi_{i,0}^{[3,3]},
\quad
\mu=1,2,\\
&\bar{\xi}_{i}^{[2,2,\mu]}=\sum_{\ell=1}^{2M}\frac{1}{\bar{p}_{i,2}^{[\mu]}-\mathrm{i}k_{\ell}}r_{\ell}+\bar{p}_{i,2}^{[\mu]}x_{1}+\bar{p}_{i,2}^{[\mu]\,2}x_{2}+\bar{\xi}_{i,0}^{[2,2,\mu]},\\
&\bar{\xi}_{i}^{[3,3]}=\sum_{\ell=1}^{2M}\frac{1}{\bar{p}_{i,3}-\mathrm{i}k_{\ell}}r_{\ell}+\bar{p}_{i,3}x_{1}+\bar{p}_{i,3}^{2}x_{2}+\bar{\xi}_{i,0}^{[3,3]},
\quad
\mu=1,2,\\
&\eta_{j}^{[2,2,\nu]}=\sum_{\ell=1}^{2M}\frac{1}{q_{j,2}^{[\nu]}+\mathrm{i}k_{\ell}}r_{\ell}+q_{j,2}^{[\nu]}x_{1}-q_{j,2}^{[\nu]\,2}x_{2}+\eta_{j,0}^{[2,2,\nu]},\\
&\eta_{j}^{[3,3]}=\sum_{\ell=1}^{2M}\frac{1}{q_{j,3}+\mathrm{i}k_{\ell}}r_{\ell}+q_{j,3}x_{1}-q_{j,3}^{2}x_{2}+\eta_{j,0}^{[3,3]},
\quad
\nu=1,2,\\
&\bar{\eta}_{j}^{[2,2,\nu]}=\sum_{\ell=1}^{2M}\frac{1}{\bar{q}_{j,2}^{[\nu]}+\mathrm{i}k_{\ell}}r_{\ell}+\bar{q}_{j,2}^{[\nu]}x_{1}-\bar{q}_{j,2}^{[\nu]\,2}x_{2}+\bar{\eta}_{j,0}^{[2,2,\nu]},\\
&\bar{\eta}_{j}^{[3,3]}=\sum_{\ell=1}^{2M}\frac{1}{\bar{q}_{j,3}+\mathrm{i}k_{\ell}}r_{\ell}+\bar{q}_{j,3}x_{1}-\bar{q}_{j,3}^{2}x_{2}+\bar{\eta}_{j,0}^{[3,3]},
\quad
\nu=1,2,
\end{align*}
The row and column spectral differential operators and the finite
sums preserve the differential and difference relations for the
Gram matrix entries.
Applying the Jacobi identity to the associated bordered
determinants yields \eqref{2d-twris-b-4}.
To evaluate the rational entries, we introduce
\begin{equation*}
G_{1,i}^{[I]}=e^{\sum_{r\ge1}a_{i,r}^{[I]}\kappa^r}e^{\kappa p\partial_p},
\qquad
G_{2,j}^{[J]}=e^{\sum_{r\ge1}b_{j,r}^{[J]}\lambda^r}e^{\lambda q\partial_q},
\qquad G_{3,ij}^{[I,J]}=G_{1,i}^{[I]}G_{2,j}^{[J]}.
\end{equation*}
For an analytic function \(F\), we have
\(e^{\kappa p\partial_p}F(p)=F(e^\kappa p)\).
Let \(K^{[I,J,\mathbf m]}\) denote the undifferentiated kernel
in \(\tilde m_{i,j}^{[I,J,\mathbf m]}\).
Then
\begin{equation}
\label{Schur-polynomials-1-1}
\begin{aligned}
&\frac{G_{3,ij}^{[I,J]}K^{[I,J,\mathbf{m}]}}{K^{[I,J,\mathbf{m}]}}
\Bigg|^{p=p_{i,1}^{[I]}}_{q=q_{j,1}^{[J]}}
=\frac{p_{i,1}^{[I]}+q_{j,1}^{[J]}}{p_{i,1}^{[I]}e^{\kappa}+q_{j,1}^{[J]}e^{\lambda}}
\prod_{\ell=1}^{2M}\left(\frac{p_{i,1}^{[I]}e^{\kappa}- \mathrm{i} k_{\ell}}{p_{i,1}^{[I]}- \mathrm{i} k_{\ell}}\right)^{m_{\ell}}\left(\frac{q_{j,1}^{[J]}e^{\lambda}+\mathrm{i} k_{\ell}}{q_{j,1}^{[J]}+\mathrm{i} k_{\ell}}\right)^{-m_{\ell}}\\
&\qquad{} \mathrm{exp}\left(\sum_{r=1}^{\infty}\left(a_{i,r}^{[I]}\kappa^{r}+b_{j,r}^{[J]}\lambda^{r}\right)\right)\\
&\mathrm{exp}\left(\sum_{\ell=1}^{2M}\left(\frac{1}{p_{i,1}^{[I]}e^{\kappa}-\mathrm{i} k_{\ell}}-\frac{1}{p_{i,1}^{[I]} - \mathrm{i} k_{\ell}}\right) r_{\ell}+\left(p_{i,1}^{[I]}e^{\kappa}-p_{i,1}^{[I]}\right)x_{1}+\left((p_{i,1}^{[I]}e^{\kappa})^{2}-(p_{i,1}^{[I]})^{2}\right)x_{2}\right)\\
&\mathrm{exp}\left(\sum_{\ell=1}^{2M}\left(\frac{1}{q_{j,1}^{[J]}e^{\lambda}+\mathrm{i} k_{\ell}}-\frac{1}{q_{j,1}^{[J]}+\mathrm{i} k_{\ell}}\right)r_{\ell}+\left(q_{j,1}^{[J]}e^{\lambda}-q_{j,1}^{[J]}\right)x_{1}-\left((q_{j,1}^{[J]}e^{\lambda})^{2}-(q_{j,1}^{[J]})^{2}\right)x_{2}\right).
\end{aligned}
\end{equation}
The Cauchy factor in \eqref{Schur-polynomials-1-1}
has the following expansion \cite{YY-2024}:
\begin{equation*}
\begin{aligned}
&\frac{p_{i,1}^{[I]}+q_{j,1}^{[J]}}{p_{i,1}^{[I]}e^{\kappa}+q_{j,1}^{[J]}e^{\lambda}}=\sum_{\nu=0}^{\infty}\left(\frac{p_{i,1}^{[I]}q_{j,1}^{[J]}}{(p_{i,1}^{[I]}+q_{j,1}^{[J]})^{2}}\kappa\lambda\right)^{\nu} \mathrm{exp}\left(\sum_{r=1}^{\infty}(\nu w_{r,1}^{[I,J]}-d_{r,1}^{[I,J]})\kappa^{r}+\sum_{r=1}^{\infty}(\nu w_{r,2}^{[I,J]}-d_{r,2}^{[I,J]})\lambda^{r}\right).
\end{aligned}
\end{equation*}
Extracting the \(\kappa^{n_i}\lambda^{n_j}\) coefficient in
\eqref{Schur-polynomials-1-1} gives
\begin{align*}
\tilde{m}_{i,j}^{[I,J,\mathbf{m}]}\Bigg|^{p=p_{i,1}^{[I]}}_{q=q_{j,1}^{[J]}}=&\frac{1}{p_{i,1}^{[I]}+q^{[J]}_{j,1}}\prod_{\ell=1}^{2M}\left(-\frac{p_{i,1}^{[I]}-z_{\ell}}{q^{[J]}_{j,1}+z_{\ell}}\right)^{m_{\ell}}
e^{\xi_i^{[I,I]}+\eta_j^{[J,J]}}\sum_{\upsilon=0}^{\mathrm{min}\left(n_{i},n_{j}\right)}\left(\frac{p_{i,1}^{[I]}q_{j,1}^{[J]}}{(p_{i,1}^{[I]}+q^{[J]}_{j,1})^{2}}\right)^{\upsilon}
\\
&S_{n_{i}-\upsilon}\left(\mathbf{x}^{+\,[I]}_{\mathbf{m}}-\mathbf{d_{1}}^{[I,J]}+\upsilon\mathbf{w_{1}}^{[I,J]}\right)
 S_{n_{j}-\upsilon}\left(\mathbf{x}^{-\,[J]}_{\mathbf{m}}-\mathbf{d_{2}}^{[I,J]}+\upsilon\mathbf{w_{2}}^{[I,J]}\right),
\end{align*}
Here \(\xi_i^{[I,I]}=\xi^{[I,I]}|_{p=p_{i,1}^{[I]}}\) and \(\eta_j^{[J,J]}=\eta^{[J,J]}|_{q=q_{j,1}^{[J]}}\).

Similarly, we obtain:
\begin{align*}
\tilde{m}_{i,j}^{[I,3,\mathbf{m}]}\Big|_{p=p_{i,1}^{[I]}}
&=\sum_{\nu=1}^{2}\frac{1}{p_{i,1}^{[I]}+q_{j,2}^{[\nu]}}
\prod_{\ell=1}^{2M}\left(-\frac{p_{i,1}^{[I]}-\mathrm{i}k_{\ell}}{q_{j,2}^{[\nu]}+\mathrm{i}k_{\ell}}\right)^{\!m_{\ell}}
e^{\xi_i^{[I,I]}+\eta_{j}^{[2,2,\nu]}}
S_{n_{i}}\!\left(\mathbf{x}^{+\,[I]}_{\mathbf{m}}-\mathbf{d}^{[I,3,\nu]}\right),
\\
\tilde{m}_{i,j}^{[I,4,\mathbf{m}]}\Big|_{p=p_{i,1}^{[I]}}
&=\sum_{\nu=1}^{2}\frac{1}{p_{i,1}^{[I]}+\bar{q}_{j,2}^{[\nu]}}
\prod_{\ell=1}^{2M}\left(-\frac{p_{i,1}^{[I]}-\mathrm{i}k_{\ell}}{\bar{q}_{j,2}^{[\nu]}+\mathrm{i}k_{\ell}}\right)^{\!m_{\ell}}
e^{\xi_i^{[I,I]}+\bar{\eta}_{j}^{[2,2,\nu]}}
S_{n_{i}}\!\left(\mathbf{x}^{+\,[I]}_{\mathbf{m}}-\mathbf{d}^{[I,4,\nu]}\right),
\\
\tilde{m}_{i,j}^{[I,5,\mathbf{m}]}\Big|_{p=p_{i,1}^{[I]}}
&=\frac{1}{p_{i,1}^{[I]}+q_{j,3}}
\prod_{\ell=1}^{2M}\left(-\frac{p_{i,1}^{[I]}-\mathrm{i} k_{\ell}}{q_{j,3}+\mathrm{i} k_{\ell}}\right)^{\!m_{\ell}}
e^{\xi_i^{[I,I]}+\eta_{j}^{[3,3]}}
S_{n_{i}}\!\left(\mathbf{x}^{+\,[I]}_{\mathbf{m}}-\mathbf{d}^{[I,5]}\right),
\\
\tilde{m}_{i,j}^{[I,6,\mathbf{m}]}\Big|_{p=p_{i,1}^{[I]}}
&=\frac{1}{p_{i,1}^{[I]}+\bar{q}_{j,3}}
\prod_{\ell=1}^{2M}\left(-\frac{p_{i,1}^{[I]}-\mathrm{i} k_{\ell}}{\bar{q}_{j,3} + \mathrm{i} k_{\ell}}\right)^{\!m_{\ell}}
e^{\xi_i^{[I,I]}+\bar{\eta}_{j}^{[3,3]}}
S_{n_{i}}\!\left(\mathbf{x}^{+\,[I]}_{\mathbf{m}}-\mathbf{d}^{[I,6]}\right),
\\
\tilde{m}_{i,j}^{[3,J,\mathbf{m}]}\Big|_{q=q_{j,1}^{[J]}}
&=\sum_{\mu=1}^{2}\frac{1}{p_{i,2}^{[\mu]}+q_{j,1}^{[J]}}
\prod_{\ell=1}^{2M}\left(-\frac{p_{i,2}^{[\mu]}-\mathrm{i}k_{\ell}}{q_{j,1}^{[J]}+\mathrm{i}k_{\ell}}\right)^{\!m_{\ell}}
e^{\xi_{i}^{[2,2,\mu]}+\eta_j^{[J,J]}}
S_{n_{j}}\!\left(\mathbf{x}^{-\,[J]}_{\mathbf{m}}-\mathbf{d}^{[3,J,\mu]}\right),
\\
\tilde{m}_{i,j}^{[4,J,\mathbf{m}]}\Big|_{q=q_{j,1}^{[J]}}
&=\sum_{\mu=1}^{2}\frac{1}{\bar{p}_{i,2}^{[\mu]}+q_{j,1}^{[J]}}
\prod_{\ell=1}^{2M}\left(-\frac{\bar{p}_{i,2}^{[\mu]}- \mathrm{i} k_{\ell}}{q_{j,1}^{[J]} + \mathrm{i} k_{\ell}}\right)^{\!m_{\ell}}
e^{\bar{\xi}_{i}^{[2,2,\mu]}+\eta_j^{[J,J]}}
S_{n_{j}}\!\left(\mathbf{x}^{-\,[J]}_{\mathbf{m}}-\mathbf{d}^{[4,J,\mu]}\right),
\\
\tilde{m}_{i,j}^{[5,J,\mathbf{m}]}\Big|_{q=q_{j,1}^{[J]}}
&=\frac{1}{p_{i,3}+q_{j,1}^{[J]}}
\prod_{\ell=1}^{2M}\left(-\frac{p_{i,3}- \mathrm{i} k_{\ell}}{q_{j,1}^{[J]} + \mathrm{i} k_{\ell}}\right)^{\!m_{\ell}}
e^{\xi_{i}^{[3,3]}+\eta_j^{[J,J]}}
S_{n_{j}}\!\left(\mathbf{x}^{-\,[J]}_{\mathbf{m}}-\mathbf{d}^{[5,J]}\right),
\\
\tilde{m}_{i,j}^{[6,J,\mathbf{m}]}\Big|_{q=q_{j,1}^{[J]}}
&=\frac{1}{\bar{p}_{i,3}+q_{j,1}^{[J]}}
\prod_{\ell=1}^{2M}\left(-\frac{\bar{p}_{i,3}- \mathrm{i} k_{\ell}}{q_{j,1}^{[J]} + \mathrm{i} k_{\ell}}\right)^{\!m_{\ell}}
e^{\bar{\xi}_{i}^{[3,3]}+\eta_j^{[J,J]}}
S_{n_{j}}\!\left(\mathbf{x}^{-\,[J]}_{\mathbf{m}}-\mathbf{d}^{[6,J]}\right),
\end{align*}
As in Section~2, the boldface vectors collect the corresponding
coefficients.
For \(I,J=1,2\) and \(\mu,\nu=1,2\), the coefficients
\(w_{r,1}^{[I,J]},\allowbreak w_{r,2}^{[I,J]},\allowbreak
d_{r,1}^{[I,J]},\allowbreak d_{r,2}^{[I,J]},\allowbreak
d_r^{[I,3,\nu]},\allowbreak d_r^{[I,4,\nu]},\allowbreak
d_r^{[I,5]},\allowbreak d_r^{[I,6]},\allowbreak
d_r^{[3,J,\mu]},\allowbreak d_r^{[4,J,\mu]},\allowbreak
d_r^{[5,J]}\), and \(d_r^{[6,J]}\)
are defined by the following generating series:
{\small
\begin{equation*}
\renewcommand{\arraystretch}{2.2}
\begin{array}{@{}l@{\quad}l@{}}
\ln\!\left[\dfrac{e^{\kappa}-1}{\kappa}\,
\dfrac{p_{i,1}^{[I]}+q_{j,1}^{[J]}}{p_{i,1}^{[I]}e^{\kappa}+q_{j,1}^{[J]}}\right]
=\displaystyle\sum_{r=1}^{\infty}w_{r,1}^{[I,J]}\kappa^{r}, & \ln\!\left[\dfrac{e^{\lambda}-1}{\lambda}\,
\dfrac{p_{i,1}^{[I]}+q_{j,1}^{[J]}}{q_{j,1}^{[J]}e^{\lambda}+p_{i,1}^{[I]}}\right]
=\displaystyle\sum_{r=1}^{\infty}w_{r,2}^{[I,J]}\lambda^{r},\\[6pt]
\ln\!\left[\dfrac{p_{i,1}^{[I]}e^{\kappa}+q_{j,1}^{[J]}}{p_{i,1}^{[I]}+q_{j,1}^{[J]}}\right]
=\displaystyle\sum_{r=1}^{\infty}d_{r,1}^{[I,J]}\kappa^{r}, & \ln\!\left[\dfrac{q_{j,1}^{[J]}e^{\lambda}+p_{i,1}^{[I]}}{q_{j,1}^{[J]}+p_{i,1}^{[I]}}\right]
=\displaystyle\sum_{r=1}^{\infty}d_{r,2}^{[I,J]}\lambda^{r},\\[6pt]
\ln\!\left[\dfrac{p_{i,1}^{[I]}e^{\kappa}+q_{j,2}^{[\nu]}}{p_{i,1}^{[I]}+q_{j,2}^{[\nu]}}\right]
=\displaystyle\sum_{r=1}^{\infty}d_{r}^{[I,3,\nu]}\kappa^{r}, & \ln\!\left[\dfrac{q_{j,1}^{[J]}e^{\lambda}+p_{i,2}^{[\mu]}}{q_{j,1}^{[J]}+p_{i,2}^{[\mu]}}\right]
=\displaystyle\sum_{r=1}^{\infty}d_{r}^{[3,J,\mu]}\lambda^{r},\\[6pt]
\ln\!\left[\dfrac{p_{i,1}^{[I]}e^{\kappa}+\bar{q}_{j,2}^{[\nu]}}{p_{i,1}^{[I]}+\bar{q}_{j,2}^{[\nu]}}\right]
=\displaystyle\sum_{r=1}^{\infty}d_{r}^{[I,4,\nu]}\kappa^{r}, & \ln\!\left[\dfrac{q_{j,1}^{[J]}e^{\lambda}+\bar{p}_{i,2}^{[\mu]}}{q_{j,1}^{[J]}+\bar{p}_{i,2}^{[\mu]}}\right]
=\displaystyle\sum_{r=1}^{\infty}d_{r}^{[4,J,\mu]}\lambda^{r},\\[6pt]
\ln\!\left[\dfrac{p_{i,1}^{[I]}e^{\kappa}+q_{j,3}}{p_{i,1}^{[I]}+q_{j,3}}\right]
=\displaystyle\sum_{r=1}^{\infty}d_{r}^{[I,5]}\kappa^{r}, & \ln\!\left[\dfrac{q_{j,1}^{[J]}e^{\lambda}+p_{i,3}}{q_{j,1}^{[J]}+p_{i,3}}\right]
=\displaystyle\sum_{r=1}^{\infty}d_{r}^{[5,J]}\lambda^{r},\\[6pt]
\ln\!\left[\dfrac{p_{i,1}^{[I]}e^{\kappa}+\bar{q}_{j,3}}{p_{i,1}^{[I]}+\bar{q}_{j,3}}\right]
=\displaystyle\sum_{r=1}^{\infty}d_{r}^{[I,6]}\kappa^{r}, & \ln\!\left[\dfrac{q_{j,1}^{[J]}e^{\lambda}+\bar{p}_{i,3}}{q_{j,1}^{[J]}+\bar{p}_{i,3}}\right]
=\displaystyle\sum_{r=1}^{\infty}d_{r}^{[6,J]}\lambda^{r}.
\end{array}
\end{equation*}
}
The components of \(\mathbf x_{\mathbf m}^{+\,[I]}\) and
\(\mathbf x_{\mathbf m}^{-\,[J]}\) are given by
\begin{equation}\label{Schur-polynomials-1-5}
\begin{aligned}
x_{r,\mathbf m}^{+\,[I]}
&=
\frac{p_{i,1}^{[I]}}{r!}x_1
+
\frac{2^r(p_{i,1}^{[I]})^2}{r!}x_2
+
\sum_{\ell=1}^{2M}\theta_{r,\ell}^{+\,[I]}r_\ell
+
\sum_{\ell=1}^{2M}h_{r,\ell}^{+\,[I]}m_\ell
+
a_{i,r}^{[I]},
\\
x_{r,\mathbf m}^{-\,[J]}
&=
\frac{q_{j,1}^{[J]}}{r!}x_1
-
\frac{2^r(q_{j,1}^{[J]})^2}{r!}x_2
+
\sum_{\ell=1}^{2M}\theta_{r,\ell}^{-\,[J]}r_\ell
-
\sum_{\ell=1}^{2M}h_{r,\ell}^{-\,[J]}m_\ell
+
b_{j,r}^{[J]}.
\end{aligned}
\end{equation}
With \(z_\ell=\mathrm{i}k_\ell\), the \(\theta\)- and
\(h\)-coefficients are defined by the following generating series:
\begin{align*}
&\frac{1}{p_{i,1}^{[I]}e^{\kappa}-z_{\ell}}-\frac{1}{p_{i,1}^{[I]}-z_{\ell}}=\sum_{r=1}^{\infty}\theta_{r,\ell}^{+\,[I]}\kappa^{r},
\quad
\ln{\left(\frac{p_{i,1}^{[I]}e^{\kappa}-z_{\ell}}{p_{i,1}^{[I]}-z_{\ell}}\right)}=\sum_{r=1}^{\infty}h_{r,\ell}^{+\,[I]}\kappa^{r},
\\
&\frac{1}{q_{j,1}^{[J]}e^{\lambda}+z_{\ell}}-\frac{1}{q_{j,1}^{[J]}+z_{\ell}}=\sum_{r=1}^{\infty}\theta_{r,\ell}^{-\,[J]}\lambda^{r},
\quad
\ln{\left(\frac{q_{j,1}^{[J]}e^{\lambda}+z_{\ell}}{q_{j,1}^{[J]}+z_{\ell}}\right)}=\sum_{r=1}^{\infty}h_{r,\ell}^{-\,[J]}\lambda^{r}.
\end{align*}
\subsection{Complex-conjugation condition}
The complex-conjugation condition
$
\tilde{\gamma}_{\mathbf m}^{*}
=
\tilde{\gamma}_{-\mathbf m}
$
is ensured by the following parameter restrictions:
\begin{equation}
\label{restriction-11}
\begin{aligned}
&p_{i,1}^{*[I]}=q_{i,1}^{[I]}, \quad q_{j,2}^{[\nu]}=p_{j,2}^{*[\nu]},\quad \bar{q}_{j,2}^{[\nu]}=\bar{p}_{j,2}^{*[\nu]},  \quad q_{j,3}=p_{j,3}^{*},\quad \bar{q}_{j,3}=\bar{p}_{j,3}^{*},  \quad \eta_{j,0}^{[2,2,\nu]}=\xi_{j,0}^{*[2,2,\nu]},\\
&\bar{\eta}_{j,0}^{[2,2,\nu]}=\bar{\xi}_{j,0}^{*[2,2,\nu]},\quad
\eta_{j,0}^{[3,3]}=\xi_{j,0}^{*[3,3]}, \quad
\bar{\eta}_{j,0}^{[3,3]}=\bar{\xi}_{j,0}^{*[3,3]}, \quad
\eta_0^{[I,I]}=\xi_0^{*\,[I,I]},\quad b_{i,r}^{[I]}=a_{i,r}^{*\,[I]},  \quad
\mu,\nu=1,2.
\end{aligned}
\end{equation}
Since \(x_1\) and \(r_\ell\) are real and \(x_2\) is purely
imaginary, the above parameter restrictions imply \(\eta=\xi^*\).
The generating series then yield
\begin{equation*}
\theta_{r,\ell}^{-\,[I]}=(\theta_{r,\ell}^{+\,[I]})^*,
\qquad h_{r,\ell}^{-\,[I]}=(h_{r,\ell}^{+\,[I]})^*,
\qquad (\mathbf x_{\mathbf m}^{+\,[I]})^*
=\mathbf x_{-\mathbf m}^{-\,[I]}.
\end{equation*}
With the dependence on the row and column indices made explicit,
we obtain
\begin{equation*}
\begin{gathered}
\bigl(d_{r,1}^{[I,J]}(i,j)\bigr)^*
=d_{r,2}^{[J,I]}(j,i),\qquad
\bigl(w_{r,1}^{[I,J]}(i,j)\bigr)^*
=w_{r,2}^{[J,I]}(j,i),\\
\bigl(d_r^{[I,3,\nu]}(i,j)\bigr)^*
=d_r^{[3,I,\nu]}(j,i).
\end{gathered}
\end{equation*}
The coefficients in the remaining mixed blocks satisfy analogous
conjugation relations, with the row and column indices interchanged.
For the Cauchy and lattice factors, we use the identity
\begin{equation*}
\left[\frac{1}{p+q}\prod_{\ell=1}^{2M}
\left(-\frac{p-\mathrm{i}k_\ell}{q+\mathrm{i}k_\ell}\right)^{m_\ell}\right]^{*}
=\frac{1}{q^*+p^*}\prod_{\ell=1}^{2M}
\left(-\frac{q^*-\mathrm{i}k_\ell}{p^*+\mathrm{i}k_\ell}\right)^{-m_\ell}.
\end{equation*}
In particular, the rational--breather block satisfies
\begin{equation*}
\begin{aligned}
\bigl(\tilde m_{i,j}^{[1,3,\mathbf m]}\bigr)^{*}
&=\sum_{\nu=1}^{2}
\frac{e^{\eta_i^{[1,1]}+\xi_j^{[2,2,\nu]}}}
{q_{i,1}^{[1]}+p_{j,2}^{[\nu]}} \prod_{\ell=1}^{2M}
\left(-\frac{p_{j,2}^{[\nu]}-\mathrm{i}k_\ell}
{q_{i,1}^{[1]}+\mathrm{i}k_\ell}\right)^{-m_\ell}
S_{n_i}\!\left(\mathbf x_{-\mathbf m}^{-\,[1]}
-\mathbf d^{[3,1,\nu]}\right)
=\tilde m_{j,i}^{[3,1,-\mathbf m]}.
\end{aligned}
\end{equation*}
The same calculation applies to the remaining blocks and yields
\begin{equation*}
\bigl(\tilde m_{i,j}^{[I,J,\mathbf m]}\bigr)^{*}
=\tilde m_{j,i}^{[J,I,-\mathbf m]},\qquad I,J=1,\ldots,6.
\end{equation*}
The terms \(\pm\delta_{ij}\) in the dark-soliton blocks are real
and symmetric, so they satisfy the same relation.
Taking determinants yields
\(\tilde\gamma_{\mathbf m}^*=\tilde\gamma_{-\mathbf m}\).
In particular, the matrix at \(\mathbf m=\mathbf0\) is Hermitian,
so \(\tilde\gamma_{\mathbf0}\) is real.

\subsection{Spatial-reflection condition}
To implement the spatial-reflection reduction
\eqref{nonlocal-conditions}, we impose the following
parameter constraints:
\begin{equation}
\label{restriction-1}
\begin{aligned}
&p_{i,1}^{[1]}=p_{i,1},\quad p_{i,1}^{[2]}=\bar{p}_{i,1},\quad \bar{p}_{i,1}=-p_{i,1}, \quad \bar{p}_{i,2}^{[\nu]}=-p_{i,2}^{[\nu]},\quad \bar{q}_{j,2}^{[\nu]}=-q_{j,2}^{[\nu]},\\
&\bar{\xi}_{i,0}^{[2,2,\nu]}=\xi_{i,0}^{[2,2,\nu]}, \quad \xi_{0}^{[1,1]}
=\xi_{0}^{[2,2]},\quad
\bar{p}_{i,3}=-p_{i,3},\quad
\bar{\xi}_{i,0}^{[3,3]}=\xi_{i,0}^{[3,3]}, \quad a_{i,r}^{[2]}=a_{i,r}^{[1]}=a_{i,r}.
\end{aligned}
\end{equation}
These constraints, together with the background parameter relations,
yield the following identities for \(\ell=1,\ldots,M\):
\begin{equation*}
\theta_{r,\ell}^{+\,[2]}=-\theta_{r,M+\ell}^{+\,[1]},\qquad
h_{r,\ell}^{+\,[2]}=h_{r,M+\ell}^{+\,[1]},\qquad
\mathbf x_{\mathbf m}^{+\,[1]}(-x)
=\mathbf x_{\mathbf m}^{+\,[2]}(x)\quad(\mathbf m=\mathbf0),
\end{equation*}
and the corresponding relations for the column vectors and coefficients.
The phase functions are paired under \(x\mapsto-x\).
Under the spectral reflection \((p,q)\mapsto(-p,-q)\),
the Cauchy factor changes sign, while the operators
\(p\partial_p\) and \(q\partial_q\) remain invariant.
For the rational blocks, \eqref{restriction-11} and
\eqref{restriction-1} imply
\(q_{j,1}^{[2]}=-q_{j,1}^{[1]}\).
The generating series then yield
\begin{equation*}
\mathbf d_a^{[2,2]}=\mathbf d_a^{[1,1]},\qquad
\mathbf w_a^{[2,2]}=\mathbf w_a^{[1,1]},\qquad a=1,2.
\end{equation*}
Using these identities and the reflection relations for the
phase functions and Schur vectors, we obtain
\begin{equation*}
\begin{aligned}
\tilde m_{i,j}^{[1,1,\mathbf0]}(-x)
=&-\frac{e^{\xi_i^{[2,2]}(x)+\eta_j^{[2,2]}(x)}}
{p_{i,1}^{[2]}+q_{j,1}^{[2]}}
\sum_{\nu=0}^{\min(n_i,n_j)}
\left[\frac{p_{i,1}^{[2]}q_{j,1}^{[2]}}
{(p_{i,1}^{[2]}+q_{j,1}^{[2]})^2}\right]^\nu S_{n_i-\nu}\!\left(
\mathbf x_{\mathbf0}^{+\,[2]}(x)-\mathbf d_1^{[2,2]}
+\nu\mathbf w_1^{[2,2]}\right)\\
& S_{n_j-\nu}\!\left(
\mathbf x_{\mathbf0}^{-\,[2]}(x)-\mathbf d_2^{[2,2]}
+\nu\mathbf w_2^{[2,2]}\right)
=-\tilde m_{i,j}^{[2,2,\mathbf0]}(x).
\end{aligned}
\end{equation*}
For the dark-soliton block, the constant term gives
\begin{equation*}
\begin{aligned}
\tilde m_{i,j}^{[5,5,\mathbf0]}(-x)
&=\delta_{ij}+\frac{e^{\bar\xi_i^{[3,3]}(x)+\bar\eta_j^{[3,3]}(x)}}{p_{i,3}+q_{j,3}}=\delta_{ij}-\frac{e^{\bar\xi_i^{[3,3]}(x)+\bar\eta_j^{[3,3]}(x)}}{\bar p_{i,3}+\bar q_{j,3}}
=-\tilde m_{i,j}^{[6,6,\mathbf0]}(x).
\end{aligned}
\end{equation*}
The matrix entries satisfy the following reflection relations:
\begin{equation*}
\begin{aligned}
\tilde m_{i,j}^{[I,J,\mathbf0]}(-x,y,t)
&=-\tilde m_{i,j}^{[I',J',\mathbf0]}(x,y,t),\\
\tilde m_{i,j}^{[I,J,\mathbf e_\ell]}(-x,y,t)
&=-\tilde m_{i,j}^{[I',J',\mathbf e_{M+\ell}]}(x,y,t),
\qquad \ell=1,\ldots,M,
\end{aligned}
\end{equation*}
where \(I'\) and \(J'\) are paired with \(I\) and \(J\), respectively, according to \(1\leftrightarrow2\), \(3\leftrightarrow4\), and \(5\leftrightarrow6\).

Since the matrix order \(2(N_1+N_2+N_3)\) is even,
the common minus sign does not affect the determinant.
Applying the same permutation to the rows and columns
also preserves the determinant. Hence
\begin{equation*}
\begin{aligned}
\tilde\gamma_{\mathbf0}(-x,y,t)
&=\tilde\gamma_{\mathbf0}(x,y,t),\quad
\tilde\gamma_{\mathbf e_\ell}(-x,y,t)
&=\tilde\gamma_{\mathbf e_{M+\ell}}(x,y,t),
\qquad \ell=1,\ldots,M.
\end{aligned}
\end{equation*}
These identities imply the reflection constraint
\eqref{nonlocal-constraint}.

Extracting the exponential factors from the rational rows and columns
of the determinant \eqref{tau-4-1} yields
\begin{equation}\label{tau-relation}
\tilde\gamma_{\mathbf m}
=
\exp\!\left[
\sum_{i=1}^{N_1}
\left(
\xi_i^{[1,1]}+\xi_i^{*\,[1,1]}
+\xi_i^{[2,2]}+\xi_i^{*\,[2,2]}
\right)
\right]\gamma_{\mathbf m}.
\end{equation}
The exponential factor in \eqref{tau-relation} is real,
strictly positive, and independent of \(\mathbf m\).
In view of \(p_{i,1}^{[2]}=-p_{i,1}^{[1]}\), it is also
independent of \(x\).
Hence \(\tilde\gamma_{\mathbf m}\) and \(\gamma_{\mathbf m}\)
yield the same fields under the transformation
\eqref{12d-M-semi-rational-solutions-1-1}.
Since \(\tilde\gamma_{\mathbf0}\) is real,
the denominator \(f^{[N_1,N_2,N_3]}=\gamma_{\mathbf0}\)
is real-valued and has the same zero set as
\(\tilde\gamma_{\mathbf0}\).
This completes the proof of Theorem~1.

\section{Conclusion}

We have constructed higher-order semi-rational solutions for the
(2+1)-dimensional nonlocal multi-component Yajima--Oikawa system
using Hirota's bilinear method and the KP hierarchy reduction.
These solutions admit a unified \(6\times6\) block Gram-determinant
representation.
Its entries involve Schur polynomials and exponential functions,
with separate spectral parameter sets assigned to the rational,
breather, and dark-soliton blocks.
Algebraic constraints on the spectral parameters and determinant
blocks implement the complex-conjugation and spatial-reflection
reductions.
The complex-conjugation reduction also ensures that the denominator
tau function is real-valued.
Varying the block sizes \(N_1,N_2,N_3\) yields the pure and mixed
solution families.

Under distinct spectral conditions, the fundamental rational
solutions describe pairs of lumps or line rogue waves.
Their separated asymptotic profiles and plane-wave background
limits have been obtained.
For the fundamental breather and dark-soliton pairs,
asymptotic analysis determines the position and phase shifts.
Their amplitudes, widths, and velocities are preserved,
confirming the elastic nature of their interactions.
Representative mixed solutions illustrate the interactions
of lumps or line rogue waves with breathers, dark solitons,
or both. Future work will investigate resonant interactions and bound
states in the mixed semi-rational solutions.
The associated spectral conditions and asymptotic dynamics
will be examined.

\bigskip

\frenchspacing
\noindent{\bf acknowledgments.}
 This work is supported by the National Natural Science Foundation of China under Grant Nos. 12171132, 12171133, 12501330 and 12271136.

\frenchspacing
\noindent{\bf Author Contributions.}
Wenqiang Lu: Writing -- review \& editing,
Writing -- original draft, Visualization, Formal analysis;
Kelei Tian: Writing -- review \& editing, Supervision;
Ying Xu: Writing -- review \& editing;
Ge Yi: Writing -- review \& editing;
Xiaoming Zhu: Writing -- review \& editing, Supervision.
\frenchspacing

\noindent{\bf Conflict of interest statement.} The authors have no conflicts to disclose.
\frenchspacing

\noindent{\bf Data availability.} The data that support the findings of this study are available within the article.

\end{document}